\documentclass[prc,aps,12pt,nofootinbib,floatfix]{revtex4}
\usepackage{amsfonts}
\usepackage{amsmath}
\usepackage{amssymb}
\usepackage{epsfig}
\usepackage{slashed}
\usepackage{graphicx}
\usepackage{color}
\usepackage{multirow}

\newcommand{\be}{\begin{equation}}
\newcommand{\ee}{\end{equation}}
\newcommand{\ba}{\begin{eqnarray}}
\newcommand{\ea}{\end{eqnarray}}
\newcommand{\pa}{\partial}
\newcommand{\unit}{\mathbb{I}}
\newcommand{\nn}{\nonumber}

\newcommand{\pvec}{{\bf p}}
\newcommand{\kvec}{{\bf k}}

\newcommand{\dtilde}[1]{\frac{d^3 #1}{(2\pi)^3}}

\begin{document}
\title{Equation of state of a quark-meson mixture \\ in the improved-PNJL model at finite chemical potential}
\author{Juan M. Torres-Rincon}
\affiliation{Frankfurt Institute for Advances Studies. Johann Wolfgang Goethe University, Ruth-Moufang-Str. 1,
60438, Frankfurt am Main, Germany}
\author{Joerg Aichelin}
\affiliation{Subatech, UMR 6457, IN2P3/CNRS, Universit\'e de Nantes, \'Ecole de Mines de Nantes, 4 rue Alfred Kastler 44307,
Nantes, France}
\affiliation{Frankfurt Institute for Advances Studies. Johann Wolfgang Goethe University, Ruth-Moufang-Str. 1,
60438, Frankfurt am Main, Germany}
\date{\today}
\begin{abstract}

   We study the equation of state of QCD using an improved version of the three-flavor Polyakov-Nambu-Jona-Lasinio model beyond 
the mean-field approximation. It incorporates the effects of unquenched quarks into the Polyakov-loop effective potential, as well as mesonic 
contributions to the grand-canonical potential. We study in full detail the calculation of the thermodynamical potential in this approach and compare the resulting 
pressure and entropy density with the most-recent lattice-QCD calculations at zero baryochemical potential. Finally, we present some exploratory results
at finite chemical potential which include the phase diagram of the model, the quark and meson masses, and finally, the thermodynamical pressure.
   
\end{abstract}
\maketitle

\tableofcontents

\section{Introduction}

    One of the most widely used low-energy realizations of quantum chromodynamics (QCD) for the quark degrees of freedom is
the Nambu--Jona-Lasinio (NJL) model~\cite{Nambu:1961tp,Vogl:1991qt,Klevansky:1992qe,Hatsuda:1994pi,Alkofer:1995mv,Buballa:2003qv}.
In this effective model one only considers dynamical quarks which exchange small momenta in their interaction, whereas gluons are integrated 
out. This model shares the global symmetries of QCD, and it is able to describe the phenomenon 
of spontaneous symmetry breaking and its high-temperature/density restoration. In particular, the local version of
the model (where quarks interact locally in space through a four-point interaction) has the property of simplicity
and transparency, as opposed to other more complicated approximations to QCD. In addition, thanks to its straightforward extension to
finite temperatures and densities, it allows for the study of the QCD phase diagram.

    Static properties of gluons can be implemented in the model by a coupling to a Polyakov-loop effective
potential~\cite{Fukushima:2003fw,Megias:2004hj,Ratti:2005jh,Hansen:2006ee,Fukushima:2008wg}. Such static properties,
like the equation of state, are nicely described when the effective potential is conveniently parametrized by using 
lattice-QCD calculations at finite temperature. However, as opposed to QCD on a lattice, the Polyakov-Nambu-Jona-Lasinio
(PNJL) model allows for a direct analysis of the phase diagram in the whole temperature ($T$)-baryochemical potential ($\mu_B$) plane. At low baryonic densities one can compare to
lattice-QCD calculations that use a Taylor expansion around $\mu_B=0$~\cite{Ratti:2005jh}.

    Old tensions in the equation of state at $\mu_B=0$, computed by two different lattice-QCD collaborations, have been solved by newer results and
a nice agreement has been achieved. It has also been confirmed that for physical $N_f=2+1$ quark masses the equation of state at
vanishing chemical potential presents a crossover phase transition~\cite{Aoki:2006we}, with a ``critical temperature'' (the temperature at the inflection point of the chiral quark condensate)
around $T_c = 155$ MeV~\cite{Borsanyi:2010bp,Bazavov:2011nk}. In addition, the temperature-dependence of the QCD
pressure (as well as other thermodynamical quantities) turns out to be very similar for both groups
within a wide range of temperatures~\cite{Borsanyi:2013bia,Bazavov:2014pvz}.

   Given this robust result for the equation of state on the lattice, one would like to check whether the thermodynamics of the PNJL model
is consistent with it. Unfortunately, when one compares the lattice-QCD results with the classical mean-field calculations 
of the pressure in the PNJL model~\cite{Ratti:2005jh} one realizes several deficiencies. At low temperature the pressure is well below the lattice-QCD results, and above $T_c$ it increases too fast towards
the ideal gas expectations, as opposed to what is seen in lattice results (see Ref.~\cite{Marty:2013ita} for an illustration).

   In this paper we revisit the thermodynamics of the PNJL model and show
that these two discrepancies with respect to the lattice-QCD results have two natural origins, and can be cured using
already-established approaches. 
First, at low temperature, where hadronic degrees of freedom dominate, the pressure is
mainly produced by the lightest mesonic states. The calculation of the thermodynamical potential beyond
mean field includes meson-like fluctuations~\cite{Quack:1993ie,Hufner:1994ma} and brings the hadronic pressure much
closer to the lattice data. Second, at moderate temperatures the interaction of quarks in the gluonic environment modifies the Polyakov loop effective potential.
Introducing an effective potential which includes the quark back reaction to the gluons~\cite{Haas:2013qwp} we show that quarks are able to control the pressure of the glue 
sector (in a dual way in which the Polyakov loop regulates the quark pressure below the critical temperature) and
this interaction lowers the pressure around $T/T_c \simeq 1.5-2$ to lattice-QCD values.

    Therefore, we implement for the first time the quark back reaction into the Polyakov-loop effective potential in a NJL-like model (our results at $\mu_B=0$ have
been presented in a short note in~\cite{Torres-Rincon:2016ahl}), determining the grand canonical potential---and the phase boundary---beyond mean field. To the best of our knowledge, this
is the first time that these techniques have been jointly implemented into a NJL model at finite density.
    
    In Sec.~\ref{sec:mf} we introduce the PNJL model with scalar and pseudoscalar interactions, and compute the grand-canonical potential at mean-field level.
We introduce the ``glue'' potential that accounts for the back reaction of quarks onto the Polyakov loop effective potential. We will show that at high temperatures the pressure is
consistent with lattice-QCD data. In Sec.~\ref{sec:fluc} we add mesonic fluctuations to the thermodynamic potential which
enter as collective modes of the propagating quarks. This improves the description of the low-temperature pressure, bringing it closer to the lattice-QCD results.
After achieving a reasonable agreement with lattice-QCD calculations at $\mu_B=0$, we will compute results at finite chemical potential in Sec.~\ref{sec:mu}. We will determine the phase diagram of
the model as a function of $T$ and $\mu_B$, show some examples of the behavior of quark and meson masses as functions of the baryochemical potential, and finally, compute the pressure of the model at finite $\mu_B$.
In Sec.~\ref{sec:conc} we present our conclusions and describe the possible extensions of this work to reach a finer description of the QCD thermodynamics.

\section{\label{sec:mf} Thermodynamics at $\mu_q=0$: the mean field approximation}
  
  In this section we will review the effective Lagrangian and the basic properties of the PNJL model at finite
temperature and finite chemical potential~\cite{Klevansky:1992qe,Buballa:2003qv,Fukushima:2003fw,Fukushima:2008wg,Costa:2010zw}.
We will detail the grand-canonical potential $\Omega (T,\mu_q)$ at mean-field level and explain the role of the
effective potential for the Polyakov loop. Then, we will argue for the need to account for modifications
of the Polyakov loop effective potential due to the presence of quarks, and adopt the findings of Ref.~\cite{Haas:2013qwp} to improve the description
of the gluonic sector of the theory. Finally, we will compute the equation of state of the improved
PNJL model at mean-field level and $\mu_q=0$.

\subsection{PNJL Lagrangian and grand-canonical potential at ${\cal O} (N_c)$~\label{sec:lagrangian}}

  We consider the Lagrangian of the PNJL model~\cite{Fukushima:2003fw,Megias:2004hj,Ratti:2005jh,Hansen:2006ee,Fukushima:2008wg,Torres-Rincon:2015rma} 
with (color neutral) pseudoscalar and scalar interactions (neglecting the vector and axial-vector vertices for simplicity),
\ba \label{eq:lagPNJL} {\cal L}_{PNJL} &=& \sum_i \bar{\psi}_i (i \slashed{D}-m_{0i}+\mu_{i} \gamma_0) \psi_i \nn \\
&+& G \sum_{a} \sum_{ijkl} \left[ (\bar{\psi}_i \ i\gamma_5 \tau^{a}_{ij} \psi_j) \ 
(\bar{\psi}_k \ i \gamma_5 \tau^{a}_{kl} \psi_l)
+ (\bar{\psi}_i \tau^{a}_{ij} \psi_j) \ 
(\bar{\psi}_k  \tau^{a}_{kl} \psi_l) \right] \nn \\
& -&    H \det_{ij} \left[ \bar{\psi}_i \ ( \unit - \gamma_5 ) \psi_j \right] - H \det_{ij} \left[ \bar{\psi}_i \ ( \unit + \gamma_5 ) \psi_j \right]  \nn \\ 
&-& {\cal U} (T;\Phi,\bar{\Phi})\ , \ea
where the flavor indices $i,j,k,l=1,2,3$ and $\tau^{a}$ ($a=1,...,8$) being the $N_f=3$ flavor generators with
normalization 
\be \textrm{tr}_f \  (\tau^{a} \tau^{b}) = 2\delta^{ab}  \ , \ee
with $\textrm{tr}_f$ denoting the trace in flavor space. 

In the Lagrangian~(\ref{eq:lagPNJL}) the bare quark masses are represented by $m_{0i}$ and their chemical potential by
$\mu_{i}$. The covariant derivative in the Polyakov gauge reads $D^\mu=\pa^\mu - i \delta^{\mu 0} A^0$, with $A^0=-iA_4$ being the temporal component of the gluon field in
Euclidean space (we denote $A^\mu = g_s A_{a}^\mu T_{a}$). The coupling constant for the scalar and pseudoscalar interaction $G$ is taken as
a free parameter (fixed e.g. by the pion mass in vacuum).

The third term of Eq.~(\ref{eq:lagPNJL}) is the so-called 't Hooft Lagrangian. It mimics the effect of the axial $U(1)$ anomaly, accounting for
the physical splitting between the $\eta$ and the $\eta'$ mesons. $H$ is a coupling constant (fixed by the value
of $m_{\eta'}-m_{\eta}$) and $\unit$ is the identity matrix in Dirac space.

  Finally, ${\cal U} (T;\Phi,\bar{\Phi})$ is the so-called Polyakov-loop effective potential used to account for
static gluonic contributions to the pressure. The Polyakov line and the Polyakov loop are respectively defined as
\be  L({\bf x}) = {\cal P} \exp \left( i \int_0^{1/T} d\tau A_4 (\tau,{\bf x}) \right) \ , \quad 
\Phi ({\bf x})= \frac{1}{N_c} {\rm tr}_c L({\bf x}) \ , \ee
where ${\cal P}$ is the path-integral ordering operator, and the trace ${\rm tr}_c$ is taken in the color space.

\begin{figure}[ht]
\begin{center}
\includegraphics[scale=0.4]{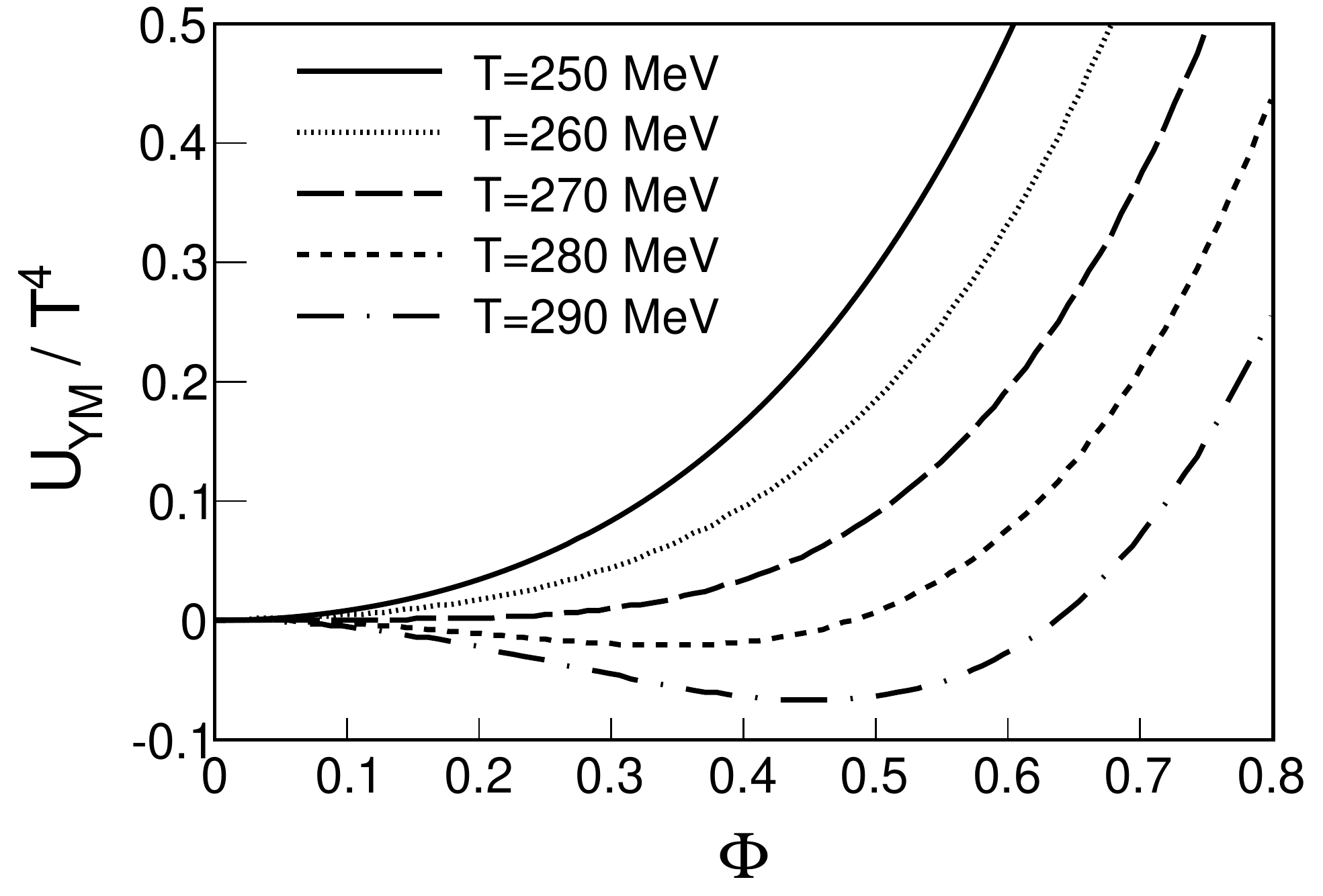}
\includegraphics[scale=0.4]{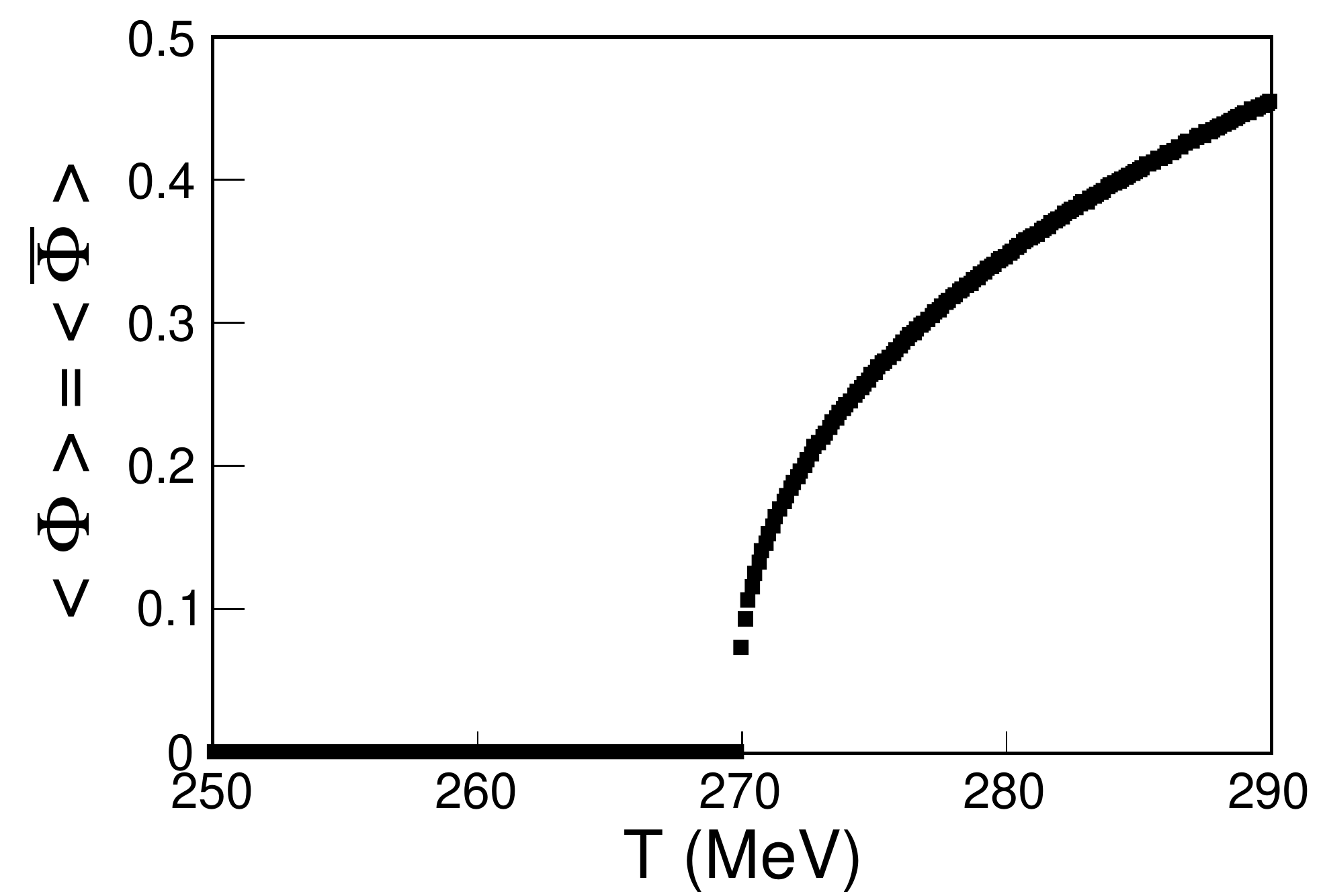}
\end{center} 
\caption{\label{fig:YMeffpot} (Left) ${\cal U}_{\rm{YM}}/T^4$ of Eq.~(\ref{eq:effU}) for several temperatures around $T_0=270$ MeV, 
where a first-order phase transition occurs. (Right) Expectation value of the Polyakov loop as a function of the temperature for the case $N_f=0$ (pure gauge).}
\end{figure}

  In the limit of infinitely heavy quarks ---i.e. for a Yang-Mills (YM) theory--- the expectation value (minimum of the effective potential) of the
Polyakov loop (EVPL) $\langle \Phi \rangle (T)$ is a true order parameter of the deconfinement transition. That means that the
EVPL is strictly zero (nonzero) in the confined (deconfined) phase
(cf. right panel of Fig.~\ref{fig:YMeffpot}). In the standard approach, the EVPL is taken to be a real function, but independent of its complex conjugated~\cite{Dumitru:2005ng}. 

Following \cite{Ratti:2005jh} we take an homogeneous Polyakov loop field $\Phi({\bf x})=\Phi=$const., and calculate the 
expectation values $\langle \Phi \rangle (T),\langle \bar{\Phi} \rangle (T)$ that minimize the effective potential ${\cal U} (T; \Phi,\bar{\Phi})$ at a given temperature
\be \label{eq:minU} \left. \frac{\partial {\cal U} (T;\Phi,\bar{\Phi})}{\partial \Phi} \right|_{\Phi= \langle \Phi \rangle (T),\bar{\Phi}= \langle \bar{\Phi} \rangle (T)}
 =0 \quad , \quad \left. \frac{\partial {\cal U} (T;\Phi,\bar{\Phi})}{\partial \bar{\Phi}}
\right|_{\Phi=\langle \Phi \rangle (T), \bar{\Phi}=\langle \bar{\Phi} \rangle (T)} =0  \ . \ee

  At this point it is convenient to remind that for a given temperature, although the effective potential is a function of the Polyakov loop (and its conjugate), the only
physical information is carried by the EVPL, where the effective potential takes its minimum. The reasoning is in the spirit of the Landau-Ginzburg functional for an homogeneous order parameter.
In addition, the value of the effective potential at the EVPL gives the physical pressure of the YM theory, which is a function of the temperature only,
$P(T)=-{\cal U} (T; \langle \Phi \rangle (T), \langle \bar{\Phi} \rangle (T))$. The EVPL and the pressure of the $SU(3)$ YM system can 
be computed by lattice QCD as functions of the temperature~\cite{Borsanyi:2012ve}. Therefore, lattice-QCD results can be used to constrain the minima of
${\cal U}$ as a function of the temperature; and eventually, to parametrize the effective potential as done in Ref.~\cite{Ratti:2005jh}.

  However it is important to notice that the knowledge of $\langle \Phi \rangle(T)$ and $P(T)$ can only help to fix the effective potential {\it at} its minimum. The complete dependence of the 
effective potential on $\Phi$ (and ${\bar \Phi}$) is not determined by these functions. For this reason, several {\it ans\"atze} for ${\cal U}$ have been introduced in the literature (compare for
example the different parametrizations considered in Ref.~\cite{Dutra:2013lya}), with the requirement of being compatible with the breaking pattern of the ${\cal Z}_3$ center symmetry of QCD.

Following our previous work~\cite{Torres-Rincon:2015rma} we will take the so-called ``polynomial parametrization'' for
the Yang-Mills potential, introduced in Ref.~\cite{Ratti:2005jh},
\be \label{eq:effU}  \frac{{\cal U}_{\rm{YM}} (T;\Phi,\bar{\Phi})}{T^4} = - \frac{b_2(T)}{2} \bar{\Phi} \Phi - \frac{b_3}{6}\left( \Phi^3 + 
\bar{\Phi}^3 \right) + \frac{b_4}{4} \left(  \bar{\Phi} \Phi \right)^2 \ , \ee
where 
\be \label{eq:effU2} b_2(T) = a_0 + a_1 \frac{T_0}{T} + a_2  \left( \frac{T_0}{T} \right)^2 + a_3  \left( \frac{T_0}{T} \right)^3 \ , \ee
and the parameters $a_0,a_1,a_2,a_3,b_3,b_4$ and $T_0$ are given in Ref.~\cite{Ratti:2005jh} (see Table~\ref{tab:param}). The effective potential is plotted in the left panel of
Fig.~\ref{fig:YMeffpot} for several temperatures around $T_0$.

\begin{table}[ht]
\begin{center}
\begin{ruledtabular}
\begin{tabular}{cccccccc}
Parameter & $a_0$ & $a_1$ & $a_2$ & $a_3$ & $b_3$ & $b_4$ & $T_0$ \\ 
\hline
\multirow{2}{*}{Value} & \multirow{2}{*}{6.75} & \multirow{2}{*}{-1.95} & \multirow{2}{*}{2.625} & \multirow{2}{*}{-7.44} & \multirow{2}{*}{0.75} & \multirow{2}{*}{7.5} & 270 MeV ($N_f=0$) \\ 
  & & & & & & & 190 MeV ($N_f=2+1$) \\ 

\end{tabular}
\caption{\label{tab:param} Parameters for ${\cal U}_{\rm{YM}} (T;\Phi,\bar{\Phi})$ in Eq.~(\ref{eq:effU}), taken from Ref.~\cite{Ratti:2005jh}.}
\end{ruledtabular}
\end{center}
\end{table}

 The EVPL is obtained by performing the minimization~(\ref{eq:minU}) of the (Yang-Mills) effective potential in~(\ref{eq:effU}). The phase transition is of first order with a transition
temperature of $T_0=270$ MeV (see right panel of Fig.~\ref{fig:YMeffpot}).

  In the presence of dynamical quarks, the effective potential ${\cal U}_{\rm{YM}} (T;\Phi,\bar{\Phi})$ complements the Lagrangian of the quark sector in Eq.~(\ref{eq:lagPNJL}). The associated grand-canonical potential
$\Omega_{\rm{PNJL}}$ accounts for the thermodynamical properties of a gas of quarks and ``static'' gluons,
\be \label{eq:Omega} \Omega_{\rm{PNJL}} (T,\mu_i;\langle \bar{\psi}_i \psi_i \rangle,\Phi,\bar{\Phi}) = \Omega_{q} (T,\mu_i;\langle \bar{\psi}_i \psi_i \rangle,\Phi,\bar{\Phi})+ {\cal U}_{YM} (T;\Phi,\bar{\Phi}) \ , \ee
where $\langle \bar{\psi}_i \psi_i \rangle$ is the scalar quark bilinear for each flavor $i$. It should not be confused with its expectation value (to be defined later), for which we will use a double bracket. The quark bilinear is merely a variable of the grand-canonical potential, not yet fixed to its physical value. We use a single bracket notation to distinguish it from the $\bar{\psi}_i(x) \psi_i (x)$ operator.

The quark potential will be eventually expanded in inverse powers of $N_c$,
\be \Omega_{q} = \Omega_{q}^{(-1)} + \Omega_q^{(0)} + \cdots \ , \ee
where the exponent $(a)$ denotes the term at order ${\cal O} (1/N_c^a)$. 

For the time being we consider the mean-field result~\cite{Hufner:1994ma}, which is equivalent to the leading-order term ${\cal O}(N_c)$ and reads~\cite{Hansen:2006ee,Torres-Rincon:2015rma}
\ba \label{eq:OmegaMF} \Omega_q^{(-1)} (T,\mu_i;\langle \bar{\psi}_i \psi_i \rangle,\Phi,\bar{\Phi})& =&  2G \sum_i \langle \bar{\psi}_i \psi_i \rangle^2
 -4H \prod_i  \langle \bar{\psi}_i \psi_i \rangle  - 2N_c \sum_i \int \frac{d^3 k}{(2\pi)^3} E_i \nn \\
&-& 2 T N_c \sum_i \int \frac{d^3 k}{(2\pi)^3} \left[ \frac{1}{N_c}\textrm{ tr}_c \log \left(1+L e^{-(E_i-\mu_i)/T} \right)
\right. \nn \\
& & \left. + \frac{1}{N_c} \textrm{ tr}_c \log \left(1+L^\dag e^{-(E_i+\mu_i)/T} \right)  \right] \ ,  \ea
where $E_i=\sqrt{k^2+m_i^2}$. Notice that $G \sim {\cal O}(1/N_c), H \sim {\cal O} (1/N_c^2)$ and $\langle \bar{\psi}_i \psi_i \rangle \sim {\cal O}(N_c)$, making all terms in Eq.~(\ref{eq:OmegaMF}) to be ${\cal O} (N_c)$.

\begin{figure}[ht]
 \begin{center}
 \includegraphics[scale=0.65]{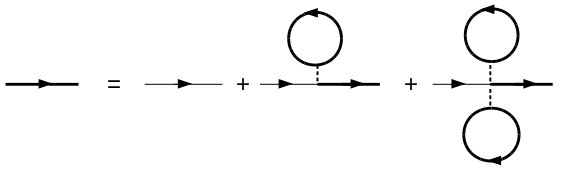}
 \end{center} 
 \caption{\label{fig:gap} Gap equation for the quark mass in Eq.~(\ref{eq:gap}). We have explicitly included a finite spatial range for the 
 four-point and six-point contact interactions to detail the topology of the diagrams.}
 \end{figure}

In Eq.~(\ref{eq:OmegaMF}) it is important to notice that due to interactions, the constituent quark mass $m_i$ depends on the quark bilinears $\langle {\bar \psi}_j \psi_j \rangle$ via a ``gap equation''.
In a diagrammatic way, this equation is sketched in Fig.~\ref{fig:gap}, where we have added to the 4-point and 6-point interactions a finite spatial range (in dashed line) to detail the topology of the diagrams.

In the mean-field approximation the gap equation reads~\cite{Buballa:2003qv,Hansen:2006ee}
\be \label{eq:gap} m_i = m_{i0} - 4 G \langle  \bar{\psi}_i \psi_i \rangle + 2 H  \langle \bar{\psi}_j \psi_j \rangle  \langle \bar{\psi}_k \psi_k  \rangle \ ,
\quad j,k\neq i; j\neq k \ . \ee

In addition, the color traces ($\textrm{tr}_c$) of Eq.~(\ref{eq:OmegaMF}) can be performed for $N_c=3$ using a mean-field approximation for the Polyakov line $L = \exp \left( i A_4 /T \right) \in SU (3)$~\cite{Fukushima:2003fw}. In the Polyakov gauge, it
can be represented by~\cite{Fukushima:2003fw}
\be L = \textrm{diag } (e^{i\phi},e^{i\phi'}, e^{-i(\phi+\phi')}) \ . \ee

Using this diagonal representation one finds
\be \textrm{ tr}_c \log \left(1+L e^{-(E_i-\mu_i)/T} \right) = \sum_{j=1}^3 \log \left( 1 +  e^{-(E_i-\mu_i)/T} L_{jj} \right) = \log \prod_{j=1}^3 \left( 1 +  e^{-(E_i-\mu_i)/T} L_{jj} \right) \ . \ee
Expanding the last product, and writing the result in terms of the Polyakov loop (and its conjugate):
\be \Phi= \frac{1}{3} \textrm{ tr}_c L =\frac{1}{3} \left( e^{i\phi} + e^{i \phi'} + e^{-i(\phi+\phi')} \right) \ , \ee
it is straightforward to get
\ba 
 \textrm{ tr}_c \log \left(1+L e^{-(E_i-\mu_i)/T} \right) &=& \log \left[ 1 + 3(\Phi + \bar{\Phi} e^{-(E_i-\mu_i)/T}) e^{-(E_i-\mu_i)/T}
+e^{-3(E_i-\mu_i)/T} \right] \ , \qquad \\
 \textrm{ tr}_c \log \left(1+L^\dag e^{-(E_i+\mu_i)/T} \right) &=& \log \left[ 1 + 3(\bar{\Phi} + \Phi e^{-(E_i+\mu_i)/T}) e^{-(E_i+\mu_i)/T}
+e^{-3(E_i+\mu_i)/T} \right] \ . \qquad  \label{ctrace}
\ea

Similarly to Eq.~(\ref{eq:minU}), the grand-canonical potential (\ref{eq:Omega}) should be minimized with respect to $\langle \bar{\psi}_i \psi_i \rangle, \Phi$ and $\bar{\Phi}$,
\be \label{eq:min}  \frac{\pa \Omega_{PNJL}}{\pa \langle {\bar \psi}_i \psi_i \rangle} =0 \ , \quad \frac{\pa \Omega_{PNJL}}{\pa \Phi} =0 \ , \quad \frac{\pa \Omega_{PNJL}}{\pa \bar{\Phi}} =0  \ , \ee
to get the expectation value of the quark bilinear---the quark condensate---, the EVPL and its conjugate, denoted respectively as $\langle \langle \bar{\psi}_i \psi_i \rangle \rangle$, 
$\langle \Phi \rangle$ and $\langle \bar{\Phi} \rangle$. 

The first equation of Eq.~(\ref{eq:min}) provides the expression for the quark condensate (recall that the derivative with respect to the quark bilinear also affects the quark constituent mass)
\be \label{eq:condenpnjl} \langle \langle \bar{\psi}_i \psi_i \rangle \rangle = - 6 \int  \frac{d^3 k}{(2\pi)^3}
\frac{m_i}{E_i} \left[1 - f_\Phi^+(E_i-\mu_i) - f_\Phi^- (E_i+\mu_i)\right]  \ , \ee
which can be written as $\langle \langle \bar{\psi}_i \psi_i \rangle \rangle = 3 m_i A_i /4 \pi^2$ in terms of the function $A_q$ defined in
Eq.~(\ref{eq:A}) of App.~\ref{app:polfunc}.

The Fermi-Dirac distribution functions are modified by the Polyakov loop,
\ba \label{eq:fpol1} f_\Phi^+ (E_i-\mu_i) &=& \frac{ ( \Phi + 2 \bar{\Phi} e^{- (E_i-\mu_i)/T} ) e^{- (E_i-\mu_i)/T} + e^{-3  (E_i-\mu_i)/T}}{1+
 3( \Phi + \bar{\Phi} e^{- (E_i-\mu_i)/T}) e^{- (E_i-\mu_i)/T } + e^{-3 (E_i-\mu_i)/T}} \ , \\
\label{eq:fpol2} f_\Phi^- (E_i+\mu_i) & =& \frac{ ( \bar{\Phi} + 2 \Phi e^{- (E_i+\mu_i)/T} ) e^{- (E_i+\mu_i)/T } + e^{-3 (E_i+\mu_i)/T}}{1+ 
3( \bar{\Phi} + \Phi e^{- (E_i+\mu_i)/T}) e^{- (E_i+\mu_i)/T } + e^{-3 (E_i+\mu_i) /T}} \  .  \ea

For $N_f=3$ the gap equation~(\ref{eq:gap}), together with Eq.~(\ref{eq:condenpnjl}), and the last two equations in~(\ref{eq:min}) form a system of five coupled equations. We solve this system numerically to obtain the value of 
$\langle \langle \bar{\psi}_i \psi_i \rangle \rangle$ (or equivalently $m_i$), $\langle \Phi \rangle$ and $\langle \bar{ \Phi} \rangle$.
In the isospin limit, two gap equations are degenerate giving $m_u=m_d$ (and $\langle \langle {\bar \psi}_u \psi_u \rangle \rangle=\langle \langle \bar{\psi}_d \psi_d \rangle \rangle$).
In addition, at vanishing chemical potential one has $\Phi=\bar{\Phi}$, which is evident from our equations.

  When minimizing (\ref{eq:Omega}) using the last two equations in (\ref{eq:min}), it is clear that the presence of quarks will affect the determination of
the EVPL (simply because $\Omega_q$ also depends on $\Phi,\bar{\Phi}$). The quark sector is affected as well by
the EVPL, cf. Eq.~(\ref{eq:condenpnjl}). The most drastic effect is the transformation of the deconfinement transition
from a first-order transition in the pure gauge case to a crossover in the unquenched theory, with an
approximate transition temperature at $T_0=270$ MeV. 

\begin{figure*}[htp]
\begin{center}
\includegraphics[scale=0.4]{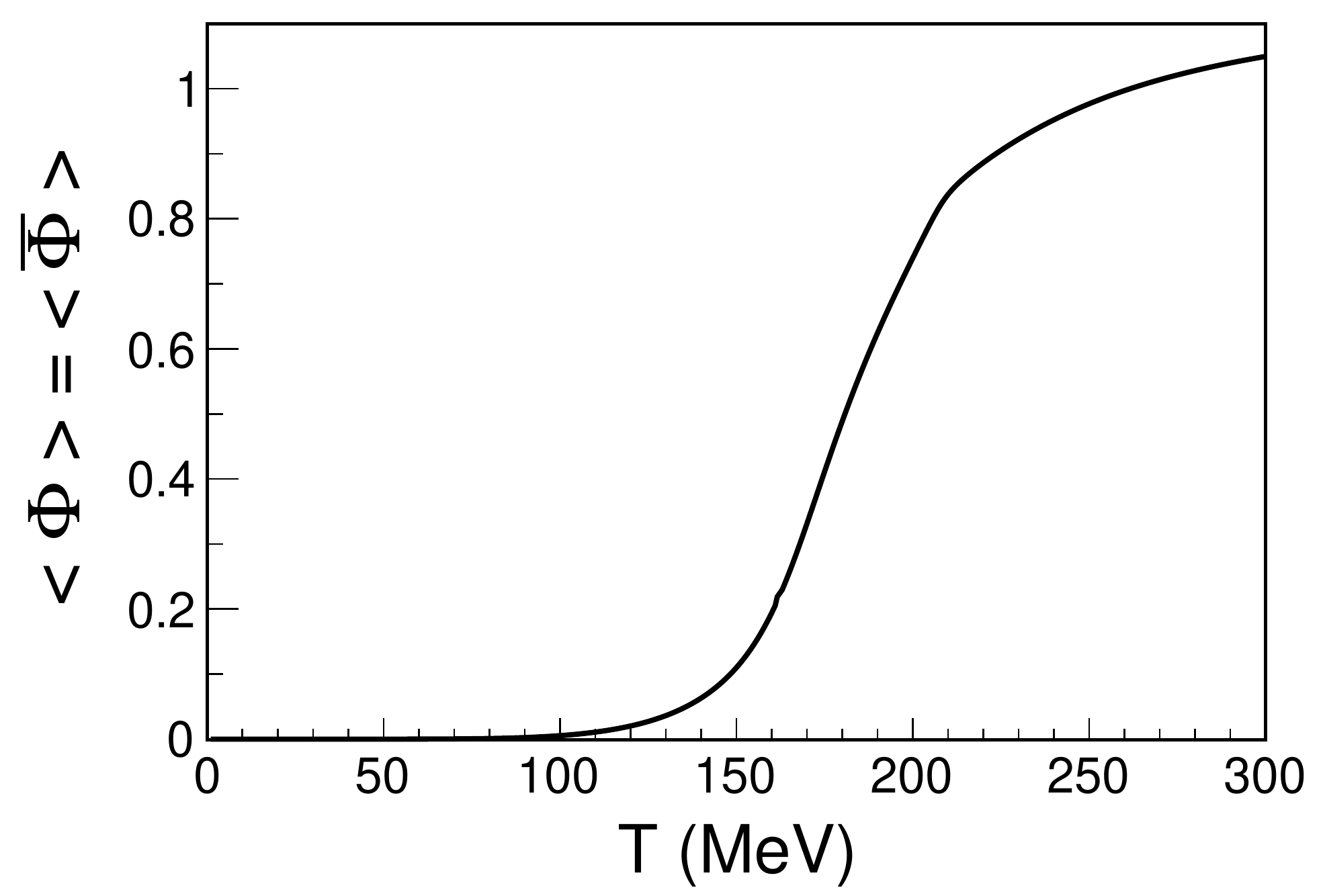}
\includegraphics[scale=0.4]{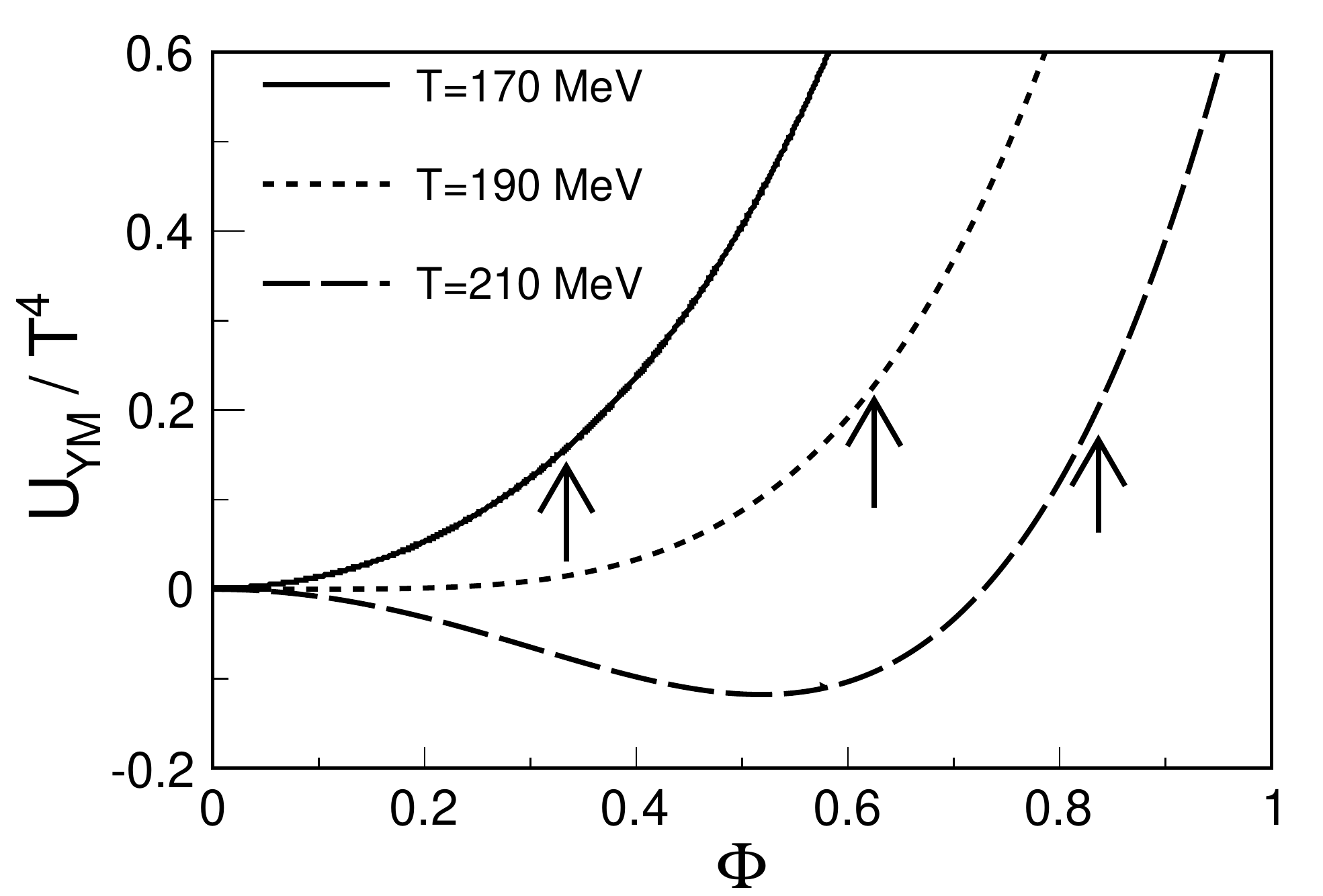}
\caption{\label{fig:EVPL} Left panel: Expectation value of the Polyakov loop as a function of the temperature for the PNJL model. 
We assume the polynomial parametrization of the Polyakov-loop effective potential in Table~\ref{tab:param}, with $T_0=190$ MeV
as suggested in Ref.~\cite{Schaefer:2007pw}. Right panel: Effective potential ${\cal U}_{\rm{YM}}/T^4$ as a function of the Polyakov loop 
for three different temperatures around $T_0=190$ MeV. For each potential we mark the position of the EVPL with an arrow.}
\end{center} 
\end{figure*}

  As argued in Ref.~\cite{Schaefer:2007pw} on the basis of renormalization group arguments, the presence of dynamical quarks should also be accompanied by
a reduction of the $T_0$ parameter of the gluon potential (\ref{eq:effU2}). This induces a decrease of the transition temperature. Following our last
work~\cite{Torres-Rincon:2015rma} we assume a value of $T_0=190$ MeV for $N_f=2+1$ flavors. The EVPL as a function
of the temperature can be seen in the left panel of Fig.~\ref{fig:EVPL}. In particular note
that $\Phi$ becomes larger than one at high temperatures, as already noticed in Ref.~\cite{Schaefer:2007pw}.  
When the Polyakov-loop effective potential is coupled to the quark sector, the restriction $\Phi,{\bar \Phi} \le 1$ in the Yang-Mills case does not necessarily hold anymore.

  It is important to note that the change of $T_0$ does not modify the functional form of the Polyakov-loop effective
potential. Therefore, at a given temperature, the EVPL does not coincide anymore with the minimum 
of the effective potential, simply because $\langle \Phi \rangle (T)$ is obtained by the minimization of the grand canonical potential,
and not of ${\cal U}_{YM}$. We illustrate this in the right panel of Fig.~\ref{fig:EVPL} where we plot ${\cal U}_{\rm{YM}}/T^4$
(with $T_0=190$ MeV) for three different temperatures $T=170, 190, 210$ MeV, together with the location
(marked with arrows) of the EVPL for each temperature. One can see that the EVPL might be considerably away
from the minimum of the effective potential.

  For this reason, to obtain the ``gluonic'' contribution to the pressure $P_{\rm{gluon}}=-{\cal U}_{YM}(T;\langle \Phi \rangle(T),\langle \bar{\Phi} \rangle(T))$, 
one should evaluate the effective potential in a region not modified by the presence of dynamical quarks, i.e. away from its minimum (points marked with arrows in Fig.~\ref{fig:EVPL}). 
In view of this situation, it is legitimate to ask for a possible effect of quarks on the Polyakov-loop effective potential out of its minima. Any modification of ${\cal U}_{\textrm{YM}}$ due to quarks would result
critical for the correct computation of the gluonic pressure.

\subsection{Effect of quarks into the Polyakov-loop effective potential}

   To pursue such a study, we introduce the quark back reaction onto ${\cal U}$, not limited to their effect at the minimum of the potential. This has been considered in Ref.~\cite{Haas:2013qwp} in the context of the functional renormalization group (FRG) 
equations applied to QCD.

   In Ref.~\cite{Haas:2013qwp} the pure gauge effective potential ${\cal U}_{YM}$ is compared
to the ``glue'' potential ${\cal U}_{glue}$, where quark-antiquark excitations are allowed in the gluon propagator.
The technical details of the computation within the framework of the FRG are explained in detail in that reference.
The authors found remarkable differences between the two effective potentials. However, the study
shows that the glue potential turns out to be very approximately related to the original ${\cal U}_{YM}$ but with a rescaling of the temperature. The glue potential is
related to the YM one as~\cite{Haas:2013qwp}:
\be \label{eq:paw} \frac{{\cal U}_{glue}}{T^4} (T;\Phi, \bar{\Phi}) = \frac{{\cal U}_{YM}}{T_{YM}^4} (T_{YM};\Phi, \bar{\Phi}) \ , \ee
where the relation between $T_{YM}$ and the temperature $T$ is given by
\be \label{eq:rescaling} \frac{T_{YM} - T_{YM}^{cr}}{T_{YM}^{cr}} = 0.57 \ \frac{T-T_{glue}^{cr}}{T_{glue}^{cr}} \ , \ee
with $T_{YM}^{cr}$ playing the role of the deconfinement temperature in the YM case (and fixed to $T_{YM}^{cr}=270$ MeV),
and $T_{glue}^{cr}$ the transition temperature in the unquenched case. The numerical
coefficient 0.57 is the outcome from the comparison of the two effective potentials. 

  The value $T^{cr}_{glue}=210$ MeV is used in Ref.~\cite{Haas:2013qwp} in the context of the quark-meson model, 
fixed to reproduce lattice-QCD data. In our case, we simply fix this value to $T_{glue}^{cr}=190$ MeV, reminiscent 
of the $T_0$ parameter that we used in Ref.~\cite{Torres-Rincon:2015rma}. As we will check later, this value will eventually
provide a reasonable result when comparing to lattice-QCD data. In addition, this choice lies within the uncertainties
quoted in~\cite{Haas:2013qwp}.

  Before presenting our results we will comment the choice of parameters and the conventions used. For the
NJL sector we will use the same parameters of our Ref.~\cite{Torres-Rincon:2015rma}, chosen to
reproduce experimental results of meson and baryons masses, as well as the pion decay constant and the value of the
quark condensate in vacuum. As we do not treat baryons here, nor vector mesons, we only have 5 parameters, viz.
$m_{q0}$, $m_{s0}$, $\Lambda$, $G$ and $H$. They are fixed in vacuum with the help of $m_\pi$, $m_K$, $\langle \langle \bar{\psi}_q \psi_q \rangle \rangle$, $f_\pi$ and $m_\eta-m_{\eta'}$. At finite temperature we parametrize the YM
Polyakov loop effective potential as described in Sec.~\ref{sec:lagrangian} with the parameters given in Table~\ref{tab:param}.

  As a low-energy realization of the QCD dynamics, the PNJL Lagrangian is not renormalizable. We use an ultraviolet (UV) regulator $\Lambda$ that cuts
  off the momentum integrals. This is consistent with the idea that the model is only valid at low momenta (below $\Lambda$). Nevertheless, at finite temperature the high-momentum modes are automatically suppressed by the distribution function inside the integrals. For this reason, we regularize our momentum integrals in such a way that $\Lambda$ is set to infinity in all UV-finite integrals, but we keep it finite in the divergent integrals (like vacuum terms). With this prescription one can reproduce the Stefan-Boltzmann limit of the thermodynamic functions in the asymptotic $T\rightarrow \infty$ limit~\cite{Zhuang:1994dw,Costa:2009ae,Costa:2010zw}.

\begin{figure*}[htp]
\begin{center}
\includegraphics[scale=0.4]{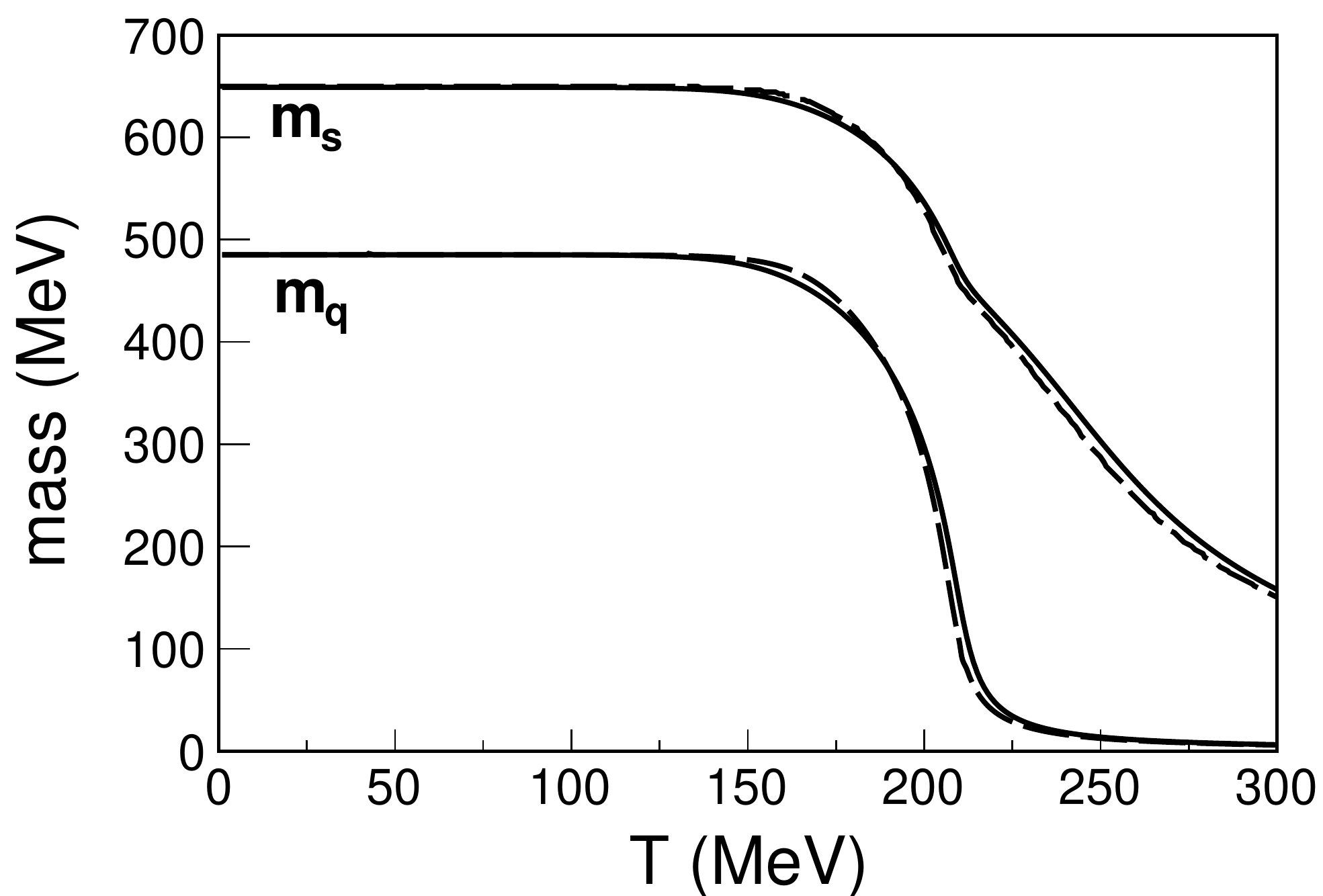}
\includegraphics[scale=0.4]{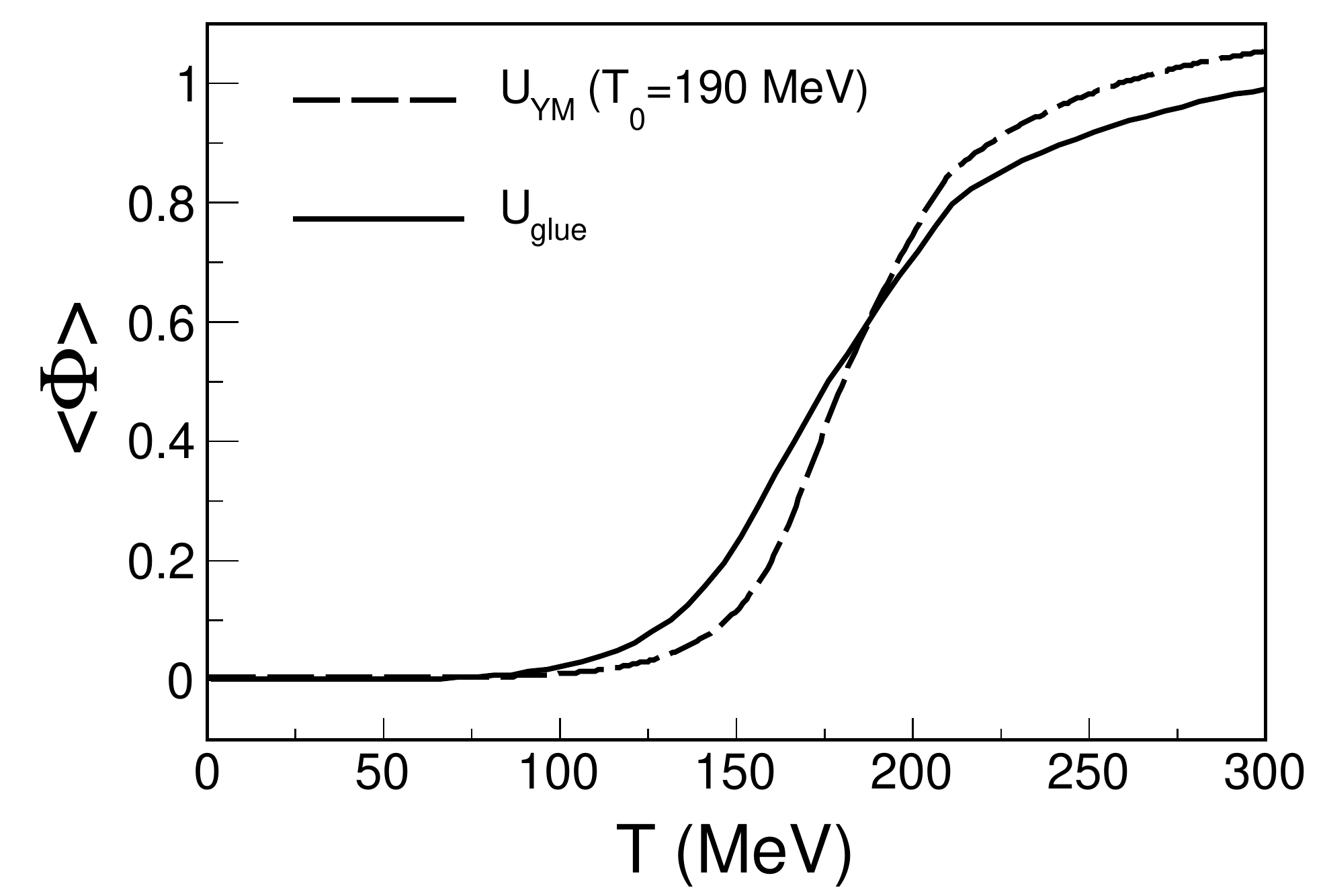}
\caption{\label{fig:quarkmass} Quark masses (left panel) and EVPL (right panel) 
in the PNJL model as a function of the temperature at vanishing chemical potential for
${\cal U}_{\rm{YM}}$ with $T_0=190$ MeV (dashed lines) and 
for ${\cal U}_{\rm{glue}}$ with $T^{cr}_{glue}=190$ MeV (solid lines).}
\end{center} 
\end{figure*}

  Our first observation is that the use of ${\cal U}_{\rm{glue}}$ does not appreciably change the mass of quarks as compared
to the YM potential. We compare the two sets of results in the left panel of Fig.~\ref{fig:quarkmass}. As being 
directly related to $m_i$, the quark condensates do not present sensible differences. However, in the 
right panel of Fig.~\ref{fig:quarkmass} we see some changes in the EVPL, $\langle \Phi \rangle$, calculated by Eq.~(\ref{eq:min}). The glue effective
potential provides a reduction of the EVPL at high temperatures to values smaller than 1 for the
temperatures considered here. A similar reduction in the Polyakov loop is seen in the context of the quark-meson model in Ref.~\cite{Haas:2013qwp}.

 The chiral transition temperature $T_c$ can be defined as the inflection point of the quark condensate. Using the results in~\cite{Borsanyi:2010bp} 
(with a $T_{c}=157 \pm 6 $ MeV) we compute the renormalized chiral condensate $\Delta_{l,s}$ defined as
 \be \label{eq:rencond} \Delta_{l,s} (T) = \frac{\langle \langle \bar{\psi}_q \psi_q \rangle \rangle (T) - \frac{m_{q0}}{m_{s0}} \langle \langle \bar{\psi}_s \psi_s \rangle \rangle (T) }
{\langle \langle \bar{\psi}_q \psi_q \rangle \rangle (0) - \frac{m_{q0}}{m_{s0}} \langle \langle \bar{\psi}_s \psi_s \rangle \rangle (0) } \ . \ee

\begin{figure}[htp]
\begin{center}
\includegraphics[scale=0.45]{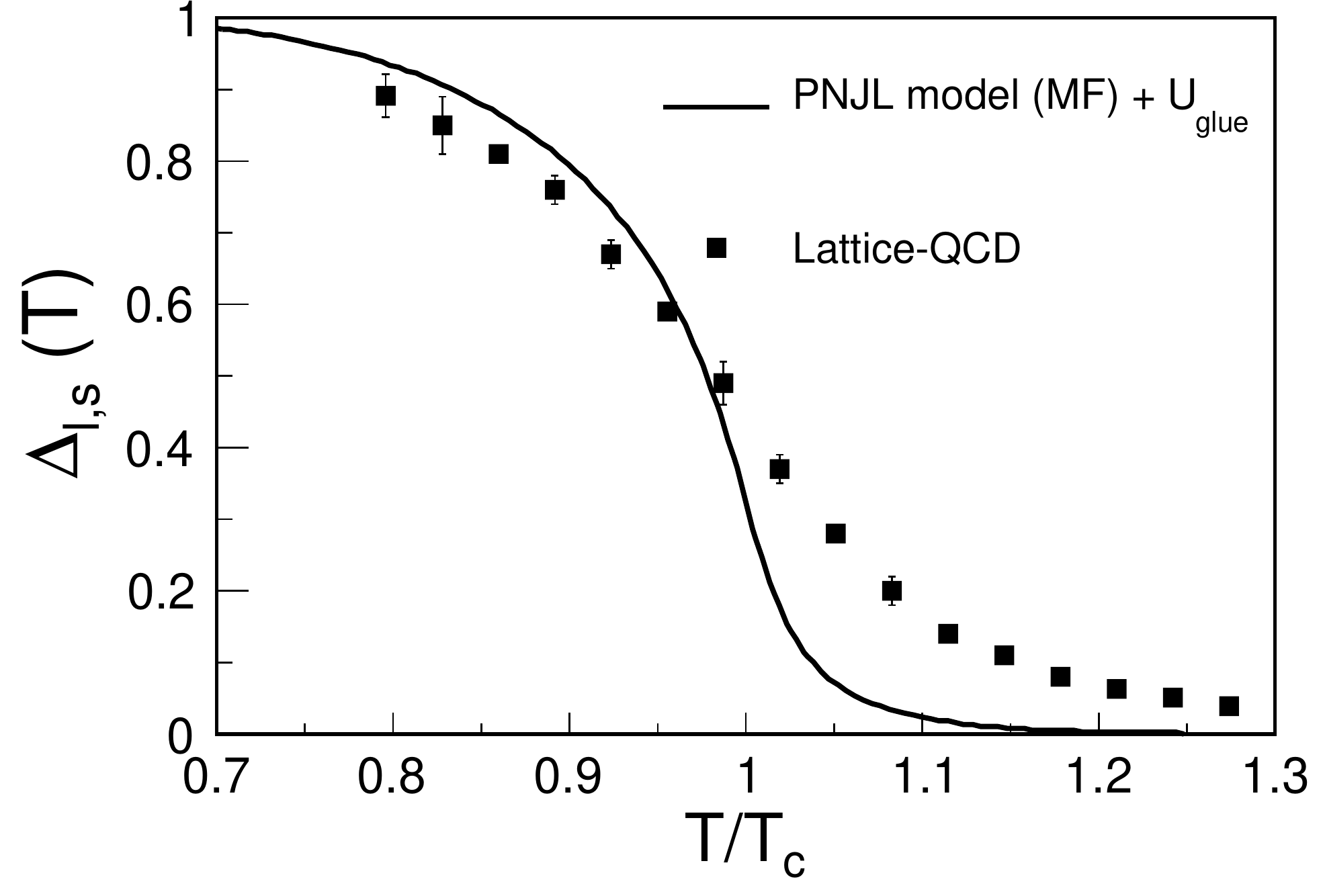}
\end{center} 
\caption{\label{fig:condlattice} Renormalized quark condensate (\ref{eq:rencond}) as a function of the temperature versus $T/T_{c}$ for
the PNJL model using ${\cal U}_{\rm{glue}}$ effective potential. The lattice-QCD result is taken from Ref.~\cite{Borsanyi:2010bp}.}
\end{figure}

  In Fig.~\ref{fig:condlattice} we show the close correspondence of $\Delta_{l,s}$ with lattice results when plotted as a function of $T/T_{c}$.
In the PNJL model the transition temperature is typically shifted to higher temperatures. In this case we get $T_{c,l}=209$ MeV for the light sector, and 
$T_{c,s}=208$ MeV for the strange one. For this reason, to compare with lattice-QCD results we opt to plot this condensate as a function of $T/T_{c}$.

  Finally, we present the result for the mean-field pressure at $\mu_q=0$. It is computed as,
\be \label{eq:pressure} P(T)= - [ \Omega_{\rm{PNJL}} (T,\mu_q=0) - \Omega_{\rm{PNJL}} (T=0,\mu_q=0)] \ , \ee
where we subtract the vacuum pressure to have $P(T=0)=0$. We remind that at mean-field level the grand-canonical potential of the model is $\Omega_{\rm{PNJL}}=\Omega_q^{(-1)} + {\cal U}$.

 \begin{figure*}[htp]
\begin{center}
\includegraphics[scale=0.45]{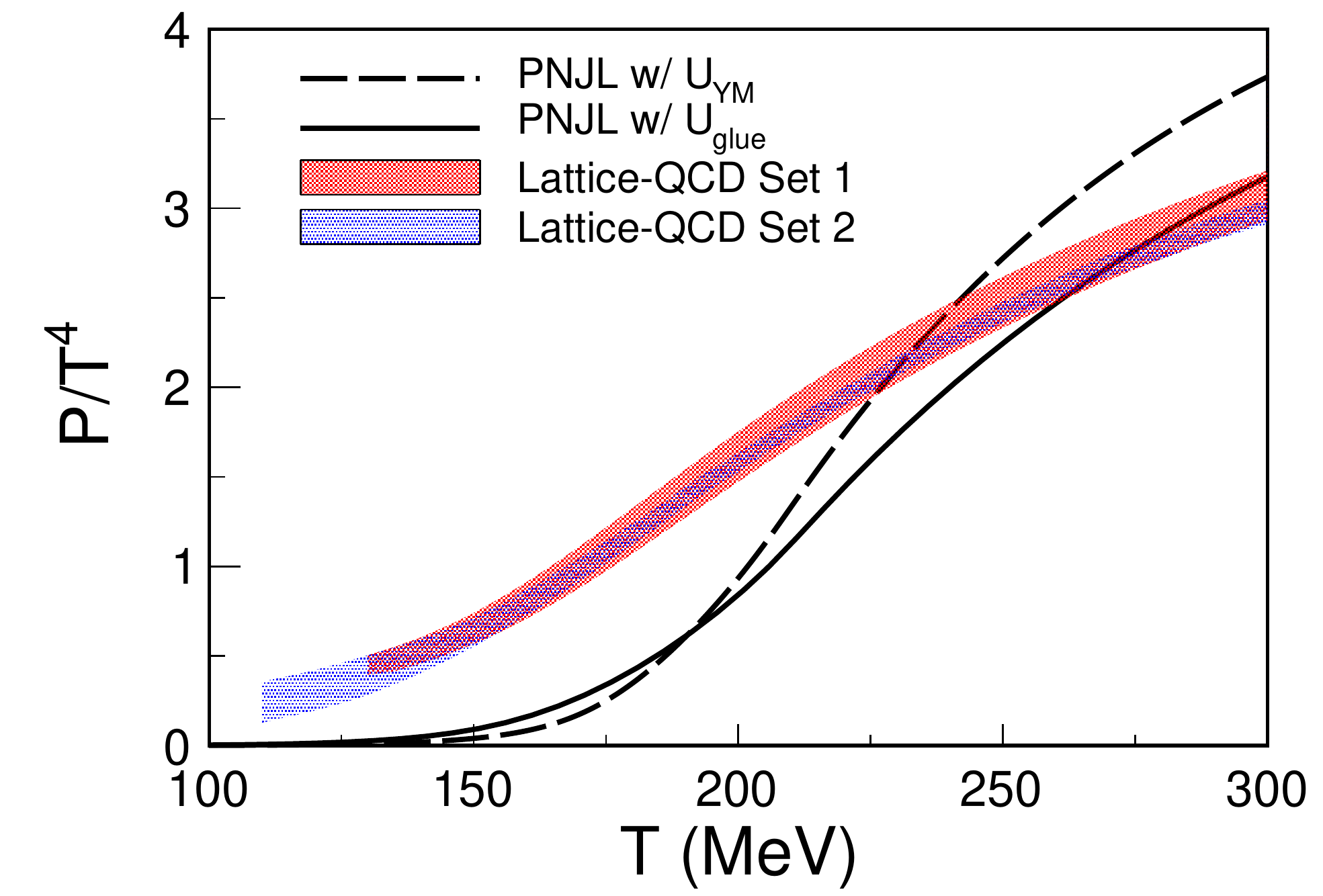}
\caption{\label{fig:pressureMF} Pressure of the PNJL model at mean-field level using ${\cal U}_{\rm{YM}}$ with $T_0=190$ MeV,
and using ${\cal U}_{\rm{glue}}$. We compare with lattice-QCD results denoted as Set 1 (taken from Ref.~\cite{Borsanyi:2013bia}) and Set 2 (from Ref.~\cite{Bazavov:2014pvz}).}
\end{center} 
\end{figure*}

  In Fig.~\ref{fig:pressureMF} we plot the pressure of the PNJL model at mean-field level using ${\cal U}={\cal U}_{YM}$ with
$T_0=190$ MeV (dashed line), and using ${\cal U}={\cal U}_{glue}$ (solid line). The YM potential provides a pressure which increases fast
to the gluon SB limit, whereas the glue potential limits this increase and gives a lower pressure at high temperatures. 
We also include the lattice-QCD results from Refs.~\cite{Borsanyi:2013bia} (Set 1) and~\cite{Bazavov:2014pvz} (Set 2).
We make note that the fast overshooting of the lattice-QCD results by the YM potential cannot be cured by modifying 
the parameters of the PNJL model alone, as we have checked with several parametrizations and regularization procedures 
in the literature. It is natural that the PNJL model cannot explain the QCD pressure 
at some intermediate range of temperatures because it lacks from attractive gluon interactions, which should contribute to the pressure
above $T \simeq 2-2.5T_c$~\cite{Ratti:2005jh} (lattice-QCD data at higher temperatures can be described by the hard-thermal-loop perturbation theory~\cite{Haque:2014rua}). At 
asymptotically large temperatures the Polyakov-loop effective potential approaches the SB limit by construction.

  We end this section by pointing out the interesting dual role of the coupling between the NJL Lagrangian and
the glue effective potential. On the one hand, it is known that the Polyakov loop potential helps to suppress
non-physical quark pressure at low temperature, contributing to the effective confinement of the system. On the other
hand, we have seen that quarks also play a similar role at high temperature, helping to reduce the gluonic contribution 
to the pressure. We find this observation particularly interesting. It should be mentioned that the glue potential 
has been recently used in the context of the $N_f=2+1$ quark-meson model in Refs.~\cite{Kovacs:2016juc,Stiele:2016cfs}
providing an analogous gluon pressure reduction, and a similar comparison to lattice-QCD data.

  In the next sections we will fix ${\cal U} = {\cal U}_{glue}$ given by Eq.~(\ref{eq:paw}), and analyze the
low-temperature regime by computing a systematic correction to the pressure beyond the mean-field level.

\section{\label{sec:fluc} Beyond mean field: Mesonic fluctuations at $\mu_q=0$}

   The modification of the Polyakov-effective potential in the PNJL model provides a reasonable
approximation to the pressure computed in lattice QCD at moderate temperatures around $T\sim (1.5-2) T_c$, as can be seen in Fig.~\ref{fig:pressureMF}. 
However, at low temperatures the pressure of the system is underestimated by the mean-field calculation.

   Hadronic states should also contribute to the pressure of the system, especially at $T<T_c$, where
they represent the main degrees of freedom (quarks and gluon remain confined at low temperatures and their pressure should be suppressed). They should be responsible for the pressure deficit
below $T_c$ as compared to lattice data in Fig.~\ref{fig:pressureMF}. On the other hand, the addition of the hadronic states cannot
spoil the current agreement at high temperature. One expects that hadrons deconfine around $T_c$ and become part of the partonic medium. 
   
   To include the effect of the lightest hadrons on the thermodynamic potential, we need to go beyond the mean-field approximation, by accounting for next-order diagrams in the large-$N_c$ expansion.

   We will illustrate the large-$N_c$ expansion of the thermodynamic potential by the diagrammatic picture adapted from Refs.~\cite{Quack:1993ie,Hufner:1994ma}.
This expansion is also sketched in Ref.~\cite{Zhuang:1994dw}. In practice---due to technical simplifications~\cite{abrikosov,fetter}---the
$1/N_c$-corrections are computed, not directly on the grand-canonical potential, but at the level of the quark self-energy, as we will detail later. We believe, however, that a more pedagogical understanding of the
large-$N_c$ expansion is provided in terms of the grand-canonical potential.

   \subsection{Grand-canonical potential at ${\cal O}((1/N_c)^0)$~\label{sec:Omega0}}

\begin{figure*}[htp]
\begin{center}
\includegraphics[scale=0.5]{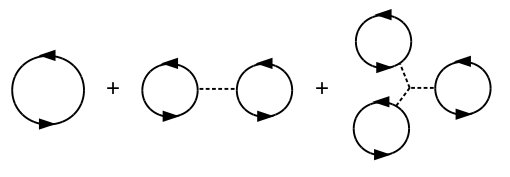}
\caption{\label{fig:Omega1} (From left to right) Noninteracting, Hartree and 't Hooft diagrams contributing at ${O} (N_c)$ to the thermodynamic potential, or $\Omega^{(-1)}_q$. 
We have explicitly introduced a finite range for the interaction (dashed line).}
\end{center} 
\end{figure*}

\begin{figure*}[htp]
\begin{center}
\includegraphics[scale=0.5]{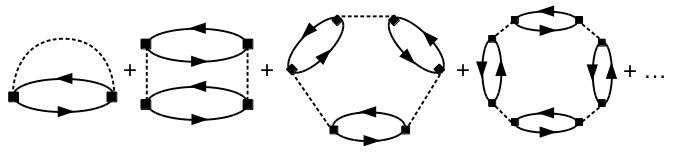}
\caption{\label{fig:Omega0} Fock and ring diagrams contributing to ${\cal O} ((1/N_c)^0)$ of the thermodynamic potential, or $\Omega^{(0)}_q$.
We have explicitly included a finite range for the interaction (dashed lines) and an effective 4-point vertex (squares).}
\end{center} 
\end{figure*}

   At claimed in Sec.~\ref{sec:lagrangian}, the mean-field potential corresponds to the leading order in a $1/N_c$ expansion, $\Omega_q^{(-1)} (T,\mu_i)$~\cite{Quack:1993ie,Hufner:1994ma}. 
This term contains the $O(N_c)$ noninteracting, Hartree and 't Hooft diagrams~\cite{Zhuang:1994dw,Hufner:1994ma}, which we plot in
Fig.~\ref{fig:Omega1}. We have given a spatial range to the contact interaction to distinguish the
different topology between the ${O} (N_c)$ Hartree (second diagram in Fig.~\ref{fig:Omega1}) and the ${O} ((1/N_c)^0)$ Fock (first diagram in Fig.~\ref{fig:Omega0}) interactions.

   This set of diagrams, computed at mean-field level in the imaginary-time formalism, and including the Polyakov loop corrections, provide
the terms contained in Eq.~(\ref{eq:OmegaMF}): The expression of the noninteracting diagram can be found in any textbook~\cite{Kapusta:2006pm} with the standard
Fermi-Dirac factors modified by the coupling to the Polyakov loop, Eqs.~(\ref{eq:fpol1},\ref{eq:fpol2}). At mean-field level, the Hartree diagram contains two powers of the quark condensate and a factor $G$ coming
from the interaction. The 't Hooft diagram contains one quark condensate of each flavor and a factor $H$, cf. Eq.~(\ref{eq:lagPNJL}).

   After incorporating the Hartree diagram into the thermodynamic potential, one is tempted to include the Fock diagram as well (first diagram in Fig.~\ref{fig:Omega0}). In the NJL limit---when the dashed lines 
are contracted into a point---the Hartree and the Fock diagrams are indistinguishable. Nevertheless, in the large-$N_c$ limit the Fock diagram is explicitly $O((1/N_c)^0)$, and it is only one of the
infinite number of terms that contribute at this order. These are depicted in Fig.~\ref{fig:Omega0} and denoted as ``ring diagrams''. A similar resummation of ring diagrams appears in the random-phase approximation~\cite{fetter} of a dense electron gas where electron-hole loops are summed. When the interaction can be taken as local like in the NJL model, it becomes equivalent to the present case.

One has to
consider all these diagrams together to be consistent with $N_c$ counting. 

We now explain the meaning of the filled squares at the vertices of the diagrams in Fig.~\ref{fig:Omega0}. Using mean field we can simplify the original 't Hooft interaction to an effective 4-point vertex which 
is, in addition, combined with the 4-point interaction $G$. The resulting quartic coupling ${\cal K}_M$ depends on the particular flavor channel (denoted by $M$), i.e. depends on the flavor of the incoming and
outgoing quarks. For example, if all the interacting quarks are light, then
\be {\cal K}_M = G - \frac{H}{2} \langle \langle \bar{\psi}_s \psi_s \rangle \rangle \ . \ee
The combinations for the other flavor channels are given in Ref.~\cite{Klevansky:1992qe,Rehberg:1995kh,Torres-Rincon:2015rma}.

Notice that this combination do not spoil $N_c$-counting. The effective vertices ${\cal K}_M (T,\mu_i)$ of the diagrams depicted in Fig.~\ref{fig:Omega0} 
are denoted with a filled square.

  A direct summation of the ring diagram is difficult due to the combinatorial factors in the perturbative expansion of the potential~\cite{abrikosov,fetter}. In practice, it
is more convenient to use the so-called ``coupling-constant integration''~\cite{abrikosov,fetter,Klevansky:1992qe}, which is equivalent to the diagrammatic summation. One computes the 
${\cal O} ((1/N_c)^0)$ potential as,
\be \label{eq:Omegafluc} \Omega^{(0)}_q (T,\mu_i) =  \frac{1}{2} \int_0^1 \frac{d\lambda }{\lambda} \int \frac{d^3p}{(2\pi)^3} T \sum_n \textrm{ Tr } [S^\lambda (i\nu_n, {\bf p}) 
\Sigma^\lambda (i\nu_n, {\bf p} ) ] \ , \ee
where the fermion self-energy $\Sigma^\lambda (i\nu_n,{\bf p})$ and the quark propagator $S^\lambda (i\nu_n,{\bf p})$ admit an
expansion in powers of $1/N_c$ and the coupling constant is rescaled by ${\cal K}_M \rightarrow  \lambda {\cal K}_M$. Finally the trace Tr is taken in color, flavor and spin spaces.

  The full quark propagator $S^\lambda$ contains self energy insertions to all orders in $N_c$. In addition, the quark self-energy $\Sigma^\lambda$ is also computed order by order 
using the full quark propagator. Instead of solving such self-consistent problem, we attach ourselves to a strict $N_c$ expansion keeping the thermodynamic potential at ${\cal O} ((1/N_c)^0)$. We follow the same procedure proposed
in Refs.~\cite{Klevansky:1992qe,Hufner:1994ma}.

The quark propagator is kept at leading order ${\cal O} (1)$, i.e. the Hartree propagator shown in Fig.~\ref{fig:gap},
\be \label{eq:SHartree} S^\lambda (i\nu_n,{\bf p}) = S_{H} (i\nu_n,{\bf p}) \ . \ee
Then, the fermion self-energy $\Sigma^\lambda (i\nu_n,{\bf p})$ is written in terms of the scattering amplitude, $t_M$.
\be \label{eq:selfenergy} \Sigma^\lambda (i\nu_n,{\bf p}) = \sum_M T \sum_m \int \frac{d^3q}{(2\pi)^3} \ \Omega \ S_H (i\omega_m,{\bf q}) \ \bar{\Omega} \ t^\lambda_M (i\nu_n-i\omega_m,{\bf p}-{\bf q}) \ , \ee
with $\Omega = \unit_c \otimes \tau^a \otimes \{ \unit, i\gamma_5 \} $ being factors in color, flavor and spin spaces~\cite{Torres-Rincon:2015rma} depending on the meson channel $M$. 
In this work the meson channel can be scalar or pseudoscalar, $J^\pi=\{0^+,0^-\}$.

The scattering amplitude is strictly ${\cal O} (1/N_c)$,
\be \label{eq:self} t^\lambda_M (i\nu_n-i\omega_m,{\bf p}-{\bf q})   =  \frac{2 \lambda {\cal K}_M}{1-2 \lambda {\cal K}_M \Pi (i\nu_n- i\omega_m,{\bf p}-{\bf q})} \ , \ee
where $\Pi$ is the quark-antiquark correlation function given in Eq.~(\ref{eq:twoprop}) which is ${\cal O} (N_c)$ when the Hartree propagators are used.

  Then, the fermion self-energy in Eq.~(\ref{eq:selfenergy}) is also ${\cal O} (1/N_c)$, and the 
thermodynamical potential is, as desired, ${\cal O} ((1/N_c)^0)$ (the color trace introduce one more power of $N_c$). Had we kept higher powers of $1/N_c$ in the quark propagator of Eq.~(\ref{eq:SHartree}), then we
would have generated ${\cal O} (1/N_c)$ terms in the thermodynamic potential, which are higher-order to the terms considered in this work.

  How mesons appear in this calculation? It is convenient to mention that the same resummation as performed in Eq.~(\ref{eq:self}) for the scattering amplitude has been done for the solution of the Bethe-Salpeter equation for a $q{\bar q}$ 
scattering~\cite{Klevansky:1992qe,Rehberg:1995kh,Buballa:2003qv,Torres-Rincon:2015rma}. These works show that mesons are generated as poles of the
$q\bar{q}$-scattering amplitude, with masses and decay widths given by the location of the poles in the complex-energy plane. Therefore, the presence of mesons is reflected in the increase
of the scattering amplitude (\ref{eq:self}), subsequently affecting the thermodynamic potential in Eq.~(\ref{eq:Omegafluc}).

Combining Eqs.(\ref{eq:Omegafluc}), (\ref{eq:SHartree}), (\ref{eq:selfenergy}) and (\ref{eq:self}) we express the thermodynamic potential as
\be \Omega_q^{(0)} (T,\mu_i) = \sum_{M \in J^\pi=\{0^+,0^-\}} \Omega^{(0)}_M (T,\mu_M (\mu_i))\ , \ee
where $M$ denotes the contribution of each mesonic spin-flavor channel. The individual contribution of a mesonic channel is
\ba \Omega^{(0)}_M (T,\mu_M (\mu_i)) &=&   \frac{g_M}{2} \int_0^1 d\lambda  T \sum_n \int \frac{d^3p}{(2\pi)^3}   T \sum_m  \int \frac{d^3q}{(2\pi)^3} \\ 
&\times&   \textrm{ Tr } \left[  S_{H} (i\nu_n,{\bf p})
  \ \Omega \ S_H (i\omega_m,{\bf q}) \ \bar{\Omega} \  \frac{2  {\cal K}_M}{1-2 \lambda {\cal K}_M \Pi (i\nu_n- i\omega_m,{\bf p}-{\bf q})}  \right] \ , \nn \ea
where $g_M$ is the spin-isospin degeneracy of the meson state. Notice that the meson chemical
potential is a function of the quark chemical potential, $\mu_M=\mu_i - \mu_{i'}$,
with $i$ and $i'$ denoting the valence quark and antiquark of the meson.

We perform a change of variables $i\omega_m \rightarrow  i\nu_n- i\omega_m , {\bf q} ={\bf p} -{\bf q}$ and get
\ba \Omega^{(0)}_M (T,\mu_M) &=&   \frac{g_M}{2} T \sum_m  \int \frac{d^3q}{(2\pi)^3}  \int_0^1 d\lambda \frac{2  {\cal K}_M}{1-2 \lambda {\cal K}_M \Pi (i\omega_m,{\bf q})}  \\ 
&\times&  \ T \sum_n \int \frac{d^3p}{(2\pi)^3}  \textrm{ Tr } \left[  S_{H} (i\nu_n,{\bf p})
  \ \Omega \ S_H (i\nu_n - i\omega_m,{\bf p -\bf q}) \ \bar{\Omega} \   \right] \ , \nn \ea
where in the second line we recognize the quark-antiquark correlation function $\Pi(i\omega_m,{\bf q})$ (cf.~Eq.~(\ref{eq:twoprop})). Therefore,
\be \Omega^{(0)}_M (T,\mu_M) =  \frac{g_M}{2} T \sum_m  \int \frac{d^3q}{(2\pi)^3} \int_0^1 d\lambda \frac{-2  {\cal K}_M \Pi (i\omega_m,{\bf q}) }{1-2 \lambda {\cal K}_M \Pi (i\omega_m,{\bf q})}  \ . \ee
The integration over $\lambda$ can be performed exactly~\cite{Zhuang:1994dw,Hufner:1994ma,Blaschke:2013zaa,Wergieluk:2012gd}
\be \Omega^{(0)}_M (T,\mu_M) = \frac{g_M T}{2} \sum_{n} \int \frac{d^3p}{(2\pi)^3} 
\log [ 1-2 {\cal K}_M \Pi (i\omega_n,{\bf p}) ]  \ . \label{eq:omega0}  \ee

Now we perform the remaining Matsubara summation in Eq.~(\ref{eq:omega0}) by the complex integration around Matsubara frequencies~\cite{Kapusta:2006pm}
\be T \sum_{n} \log [ 1-2 {\cal K}_M \Pi (i\omega_n,{\bf p}) ]  = \frac{1}{2\pi i} \oint_C dz \frac{1}{e^{\beta z}-1} \log [ 1-2 {\cal K}_M \Pi (z,{\bf p}) ]  \ . \ee
We deform the contour of the integral~\cite{Hufner:1994ma} to a contour around the real axis (where the two-quark propagator has analytical cuts, cf. App.~\ref{app:polfunc}). 
We can distinguish two integrations 
\ba T \sum_{n} \log [ 1-2 {\cal K}_M \Pi (i\omega_n,{\bf p}) ]  &=& \frac{1}{2\pi i} \int_{-\infty}^{+\infty} d\omega \frac{1}{e^{\beta \omega}-1} \log [ 1-2 {\cal K}_M \Pi (\omega+i\epsilon,{\bf p}) ] \nn \\
&+& \label{eq:twocuts} \frac{1}{2\pi i} \int_{+\infty}^{-\infty} d\omega \frac{1}{e^{\beta \omega}-1} \log [ 1-2 {\cal K}_M \Pi (\omega-i\epsilon,{\bf p}) ]  \ . \ea

We split the integrals in Eq.~(\ref{eq:twocuts}) ($\int_{-\infty}^{+\infty} = \int_{-\infty}^0 + \int_0^{+\infty}$) and perform the variable shift $\omega \rightarrow \omega-\mu_M$,
\ba T \sum_{n} \log [ 1-2 {\cal K}_M \Pi (i\omega_n,{\bf p}) ]  &=& \frac{1}{2\pi i} \int_{\mu_M}^{+\infty} d\omega \frac{1}{e^{\beta (\omega-\mu_M)}-1} 
\log \frac{   1-2 {\cal K}_M \Pi (\omega-\mu_M+i\epsilon,{\bf p}) }{ 1-2 {\cal K}_M \Pi (\omega-\mu_M-i\epsilon ,{\bf p})}  \nn \\
&+&  \frac{1}{2\pi i} \int_{-\infty}^{\mu_M} d\omega \frac{1}{e^{\beta (\omega-\mu_M)}-1} 
\log \frac{   1-2 {\cal K}_M \Pi (\omega-\mu_M+i\epsilon,{\bf p}) }{ 1-2 {\cal K}_M \Pi (\omega-\mu_M-i\epsilon ,{\bf p})}  \ .  \nn \\
& & \ea
Now we perform the change of variable $\omega \rightarrow - \omega$ in the second integral, and use the property of the polarization function
\be \Pi (\omega-\mu_M-i\epsilon)=\Pi (-\omega -\mu_M+i \epsilon) \ , \ee
where we have used the fact that the polarization function $\Pi(\omega,{\bf p})$ is a function of the energy always through the combination $\omega + \mu_M$ (cf. Eqs.~(\ref{eq:polfunc}, \ref{eq:B0})). We obtain

\ba T \sum_{n} \log [ 1-2 {\cal K}_M \Pi (i\omega_n,{\bf p}) ]  &=& \frac{1}{2\pi i} \int_{\mu_M}^{+\infty} d\omega \frac{1}{e^{\beta (\omega-\mu_M)}-1} 
\log \frac{   1-2 {\cal K}_M \Pi (\omega-\mu_M+i\epsilon,{\bf p}) }{ 1-2 {\cal K}_M \Pi (\omega-\mu_M-i\epsilon ,{\bf p})}  \nn \\
&+&  \frac{1}{2\pi i} \int_{-\mu_M}^{+\infty} d\omega \frac{1}{e^{-\beta (\omega+\mu_M)}-1} 
\log \frac{   1-2 {\cal K}_M \Pi (\omega-\mu_M-i\epsilon,{\bf p}) }{ 1-2 {\cal K}_M \Pi (\omega-\mu_M+i\epsilon ,{\bf p})}  \ .  \nn \\
& & \ea

The integration limits can be again restored to $(0,+\infty)$ by adding and subtracting the same quantity
\ba T \sum_{n} \log [ 1-2 {\cal K}_M \Pi (i\omega_n,{\bf p}) ]  &=& \frac{1}{2\pi i} \int_{0}^{+\infty} d\omega \frac{1}{e^{\beta (\omega-\mu_M)}-1} 
\log \frac{   1-2 {\cal K}_M \Pi (\omega-\mu_M+i\epsilon,{\bf p}) }{ 1-2 {\cal K}_M \Pi (\omega-\mu_M-i\epsilon ,{\bf p})}  \nn \\
&-&  \frac{1}{2\pi i} \int_{0}^{+\infty} d\omega \frac{1}{e^{-\beta (\omega+\mu_M)}-1} 
\log \frac{   1-2 {\cal K}_M \Pi (\omega-\mu_M+i\epsilon,{\bf p}) }{ 1-2 {\cal K}_M \Pi (\omega-\mu_M-i\epsilon ,{\bf p})}  \ .  \nn \\
& & \ea

Then, we use the relation
\be \frac{1}{e^{-\beta (\omega+\mu_M)}-1} = - 1 - \frac{1}{e^{\beta (\omega+\mu_M)} -1}  \ee
to get 
\ba T \sum_{n} \log [ 1-2 {\cal K}_M \Pi (i\omega_n,{\bf p}) ]  &=& \frac{1}{2\pi i} \int_{0}^{+\infty} d\omega \  
\left[ 1 +\frac{1}{e^{\beta (\omega-\mu_M)}-1} +\frac{1}{e^{\beta (\omega+\mu_M)}-1} \right] \nn \\
&\times &\log \frac{  1-2 {\cal K}_M \Pi (\omega-\mu_M+i\epsilon,{\bf p}) }{1-2 {\cal K}_M \Pi (\omega-\mu_M-i\epsilon,{\bf p})}  \ . \ea

Finally, one has 
 \ba \label{eq:big} \Omega^{(0)}_{M} (T,\mu_M) &=& \frac{g_M}{2} \int \dtilde{p}  \frac{1}{2\pi i} \int_{0}^{+\infty} d\omega \  
\left[ 1 +\frac{1}{e^{\beta (\omega-\mu_M)}-1} +\frac{1}{e^{\beta (\omega+\mu_M)}-1} \right] \nn \\
&\times &\log \frac{  1-2 {\cal K}_M \Pi (\omega-\mu_M+i\epsilon,{\bf p}) }{1-2 {\cal K}_M \Pi (\omega-\mu_M-i\epsilon,{\bf p})} \ . \ea

It is interesting to relate this equation to the so-called Beth-Uhlenbeck approach, exploiting the use of the scattering phase shift. We provide a discussion of the scattering phase shift 
in App.~\ref{app:phaseshift}, and refer the reader to Refs.~\cite{Yamazaki:2012ux,Yamazaki:2013yua, Blaschke:2013zaa} for more information. 

Using Eq.~(\ref{eq:help}) we can express $\Omega^{(0)}_M$ in Eq.~(\ref{eq:big}) as an integration over the scattering phase shift,
\be \label{eq:Omega0next} \Omega^{(0)}_{M} (T,\mu_M) = -\frac{g_M}{2\pi} \int \dtilde{p} \int_{0}^{+\infty} d\omega \left[  \frac{1}{e^{\beta (\omega-\mu_M)}-1}
+ \frac{1}{e^{\beta (\omega+\mu_M)}-1}  \right]  \ \delta (\omega, \pvec; T,\mu_M)  \ , \ee
where we have neglected the first term within the brackets of Eq.~(\ref{eq:big}) (this is usually called ``no-sea'' approximation).

An equivalent formula is sometimes used after integrating by parts
\ba \Omega^{(0)}_{M} (T,\mu_M) & =& g_M \int \dtilde{p} \int_{0}^{+\infty} d\omega \left[  \frac{T}{2} \log \left( 1 - e^{-\beta (\omega-\mu_M)} \right) 
+ \frac{T}{2} \log \left( 1 - e^{-\beta (\omega+\mu_M)} \right)  \right] \nn \\
&\times& \frac{1}{\pi} \frac{d \delta (\omega,{\bf p};T,\mu_M) }{d\omega}  \ , \label{eq:Omega0inter} \ea
which resembles the Beth-Uhlenbeck approach to the thermodynamics of an interacting gas~\cite{Blaschke:2013zaa,Beth:1936,Beth:1937zz}.

A simplifying assumption is that the argument of the phase shifts is approximately Lorentz invariant
(we check the validity of this claim in Appendix~\ref{app:phaseshift}). Then, we can introduce the Mandelstam variable $s=\omega^2-p^2$, and approximate
\be \delta_M (\omega,{\bf p};T,\mu_M) \simeq \delta_M (\sqrt{\omega^2-{\bf p}^2},{\bf p =0};T,\mu_M) = \delta_M(\sqrt{s},{\bf p =0};T,\mu_M) \ . \ee
Introducing $s$ in Eq.~(\ref{eq:Omega0next}), we obtain the meson part of the grand-canonical potential,
\ba \label{eq:omegafinal} \Omega^{(0)}_{M} (T,\mu_M)  &=& -\frac{g_M }{8\pi^3} \int dp p^2 \int \frac{ds}{\sqrt{s+p^2}} \ \left[  \frac{1}{e^{\beta (\sqrt{s+p^2}-\mu_M)}-1}
+ \frac{1}{e^{\beta (\sqrt{s+p^2}+\mu_M)}-1}  \right] \nn \\ 
&\times& \ \delta_M (\sqrt{s}; T,\mu_M)    \ , \ea
where the phase shift is computed using Eq.~(\ref{eq:phaseshift}).

 The limits of integration must be chosen in accordance with the UV cutoff of our theory. The momentum integral is limited to $(0,\Lambda)$. However the UV cutoff is
eventually removed in all convergent integrals, for consistency with the mean-field calculation. The integral in $s$ has a lower limit at $0$ (the quark-antiquark propagator is evaluated at real values of $\sqrt{s}$).
The upper limit is taken at $(\Lambda_1+\Lambda_2)^2=(\sqrt{\Lambda^2+m_1^2}+\sqrt{\Lambda^2+m_2^2})^2$, where the unitary cut ends (cf. Appendix~\ref{app:phaseshift} for details of terminology). For low temperatures the phase shift present 
several discontinuities as a function of the energy (see Fig.~\ref{fig:delta3D_vsT}). To avoid systematic numerical uncertainties we use the known limits of the Landau and unitary cuts to perform the integration
over $s$ in a support where the function is nonzero (cf. Fig~\ref{fig:cuts1}).

\subsection{Pressure of a quark---Polyakov loop---meson mixture at $\mu_q=0$}
 
  In this section we finally present our results for the pressure of the PNJL model computed beyond the mean-field approximation at ${\cal O} ((1/N_c)^0)$. For the time being we fix the quark chemical potential to zero
and plot the pressure as a function of the temperature,
\be \label{eq:pressfin} -P(T)=  \Omega_{\rm{PNJL}} (T,\mu_i=0) -\Omega_{\rm{PNJL}} (T=0,\mu_i=0) \ , \ee
where the grand-canonical potential taken at ${\cal O}( (1/N_c)^0)$,
\be \label{eq:potfin} \Omega_{\rm{PNJL}} (T,\mu_i) = \Omega_q^{(-1)} (T,\mu_i) + \sum_{M \in J^\pi = \{ 0^+,0^-\}} \Omega^{(0)}_M (T,\mu_M(\mu_i)) + {\cal U}_{glue} (T) \ , \ee
with $\Omega^{(-1)}, \Omega^{(0)}_M$ and ${\cal U}_{glue}$ given by Eqs.~(\ref{eq:OmegaMF}), (\ref{eq:omegafinal}), and (\ref{eq:paw}), respectively. Notice that all parameters have been fixed 
at mean-field level, so that the mesonic contribution is parameter free for every temperature.

\begin{figure*}[htp]
\begin{center}
\includegraphics[scale=0.36]{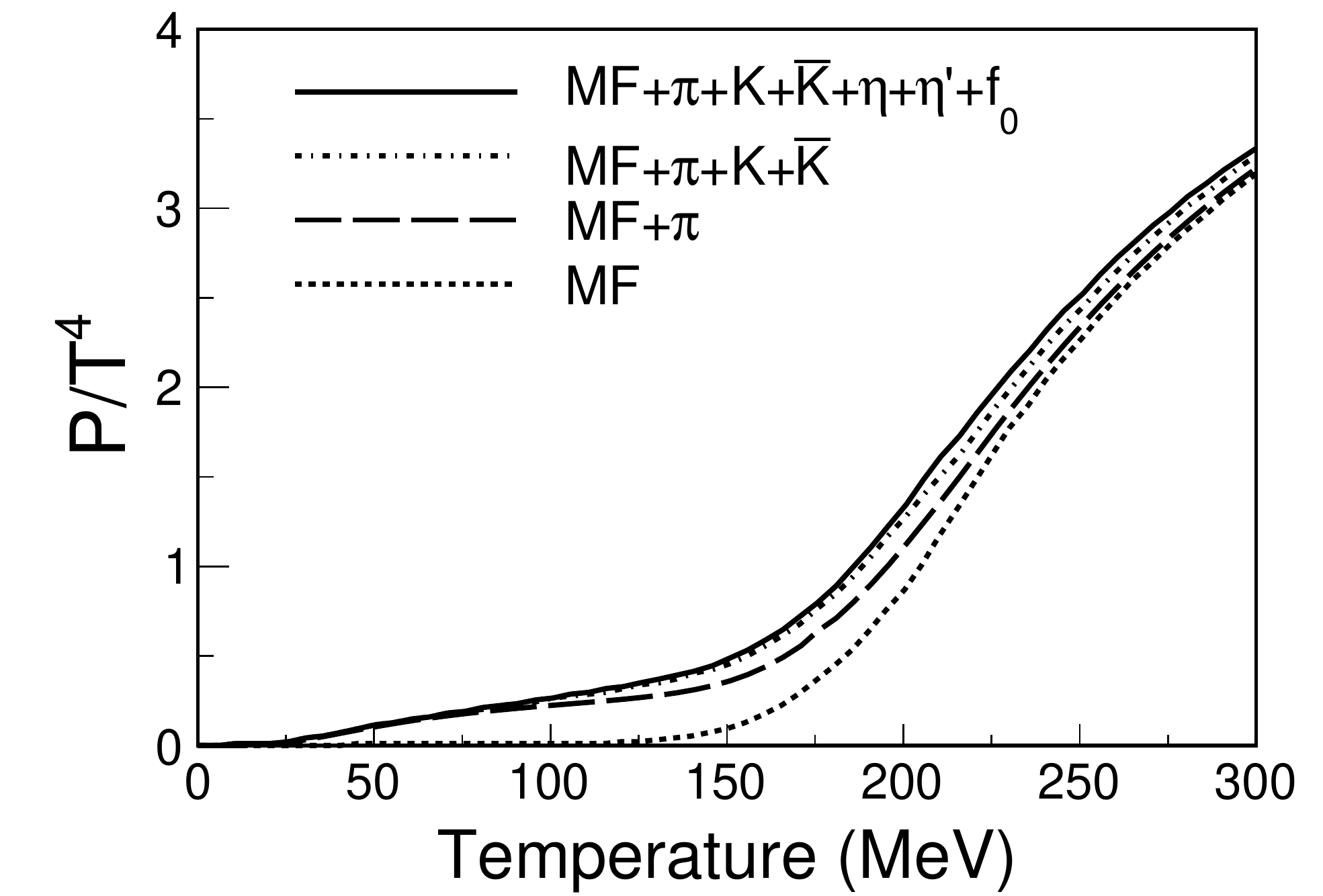}
\includegraphics[width=8cm,height=5cm]{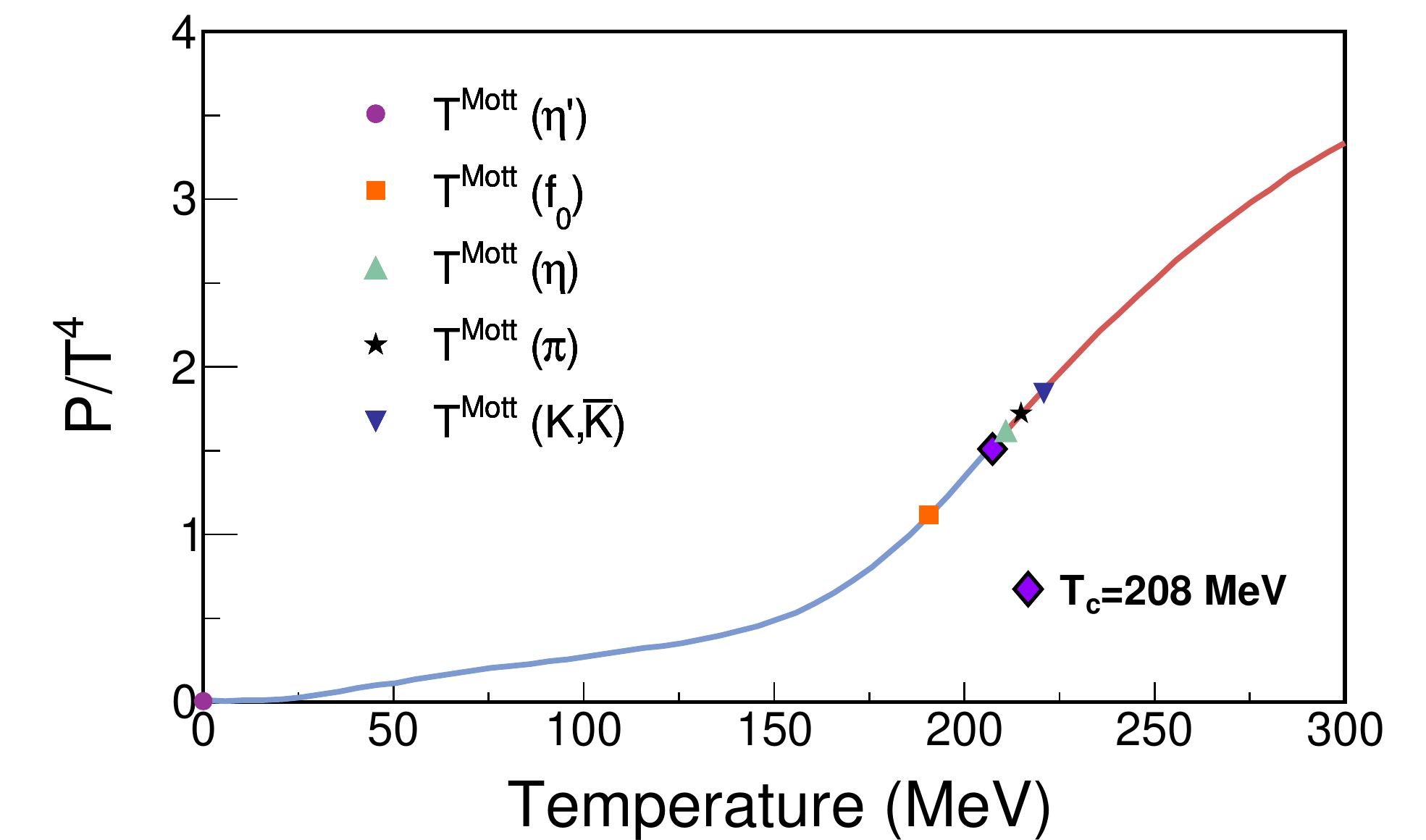}
\caption{\label{fig:Pprogressive} Pressure at $\mu_i=0$ of the PNJL model with the inclusion of the $0^+,0^-$ 
meson states whose mass in vacuum is below 1 GeV: $\pi,K,{\bar K},\eta,\eta',f_0$ (to be identified to the scalar $f_0 (980)$).
Left panel: Sequential addition of mesonic contributions to the pressure. Right panel: Total pressure and Mott temperatures of mesons, indicating where they become unstable against decay into a $\bar{q}q$ pair. 
The blue and red lines are the total pressure below and above $T_c$, respectively.}
\end{center} 
\end{figure*}

  We show in Fig.~\ref{fig:Pprogressive} the total pressure of the system when we add more and more mesonic states on top of the mean-field (MF) result. 
  For the sake of clarity, in this plot we have only included pseudoscalar and scalar mesons with a
vacuum mass below 1 GeV. As expected, the dominant contribution at low temperatures
(up to $T\simeq 100$ MeV) comes from the pions. Then, more massive mesons (in particular the four kaons) start
to affect the thermodynamics. Notice that heavier states make only tiny contribution to the pressure as
they are Boltzmann suppressed due to their higher masses. At high temperature, the mean-field pressure
dominates, and the mesons contribute as resonant states as they melt above their Mott temperature. This temperature is
defined as the one at which the meson pole acquires an imaginary part, thus generating a nonzero decay width. Above this temperature the meson
has nonzero probability to decay into a quark-antiquark pair.

\begin{figure*}[htp]
\begin{center}
\includegraphics[scale=0.4]{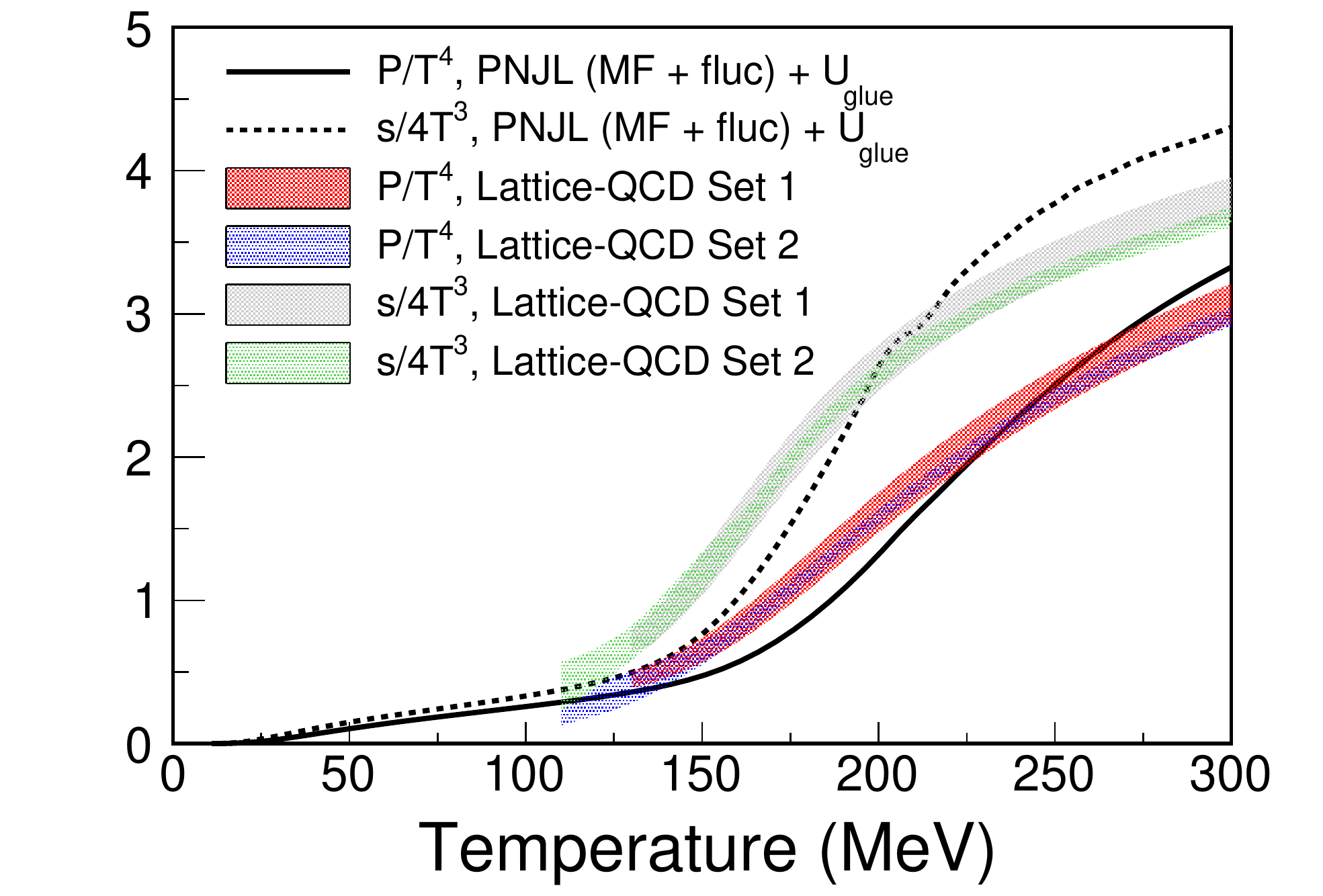}
\includegraphics[scale=0.4]{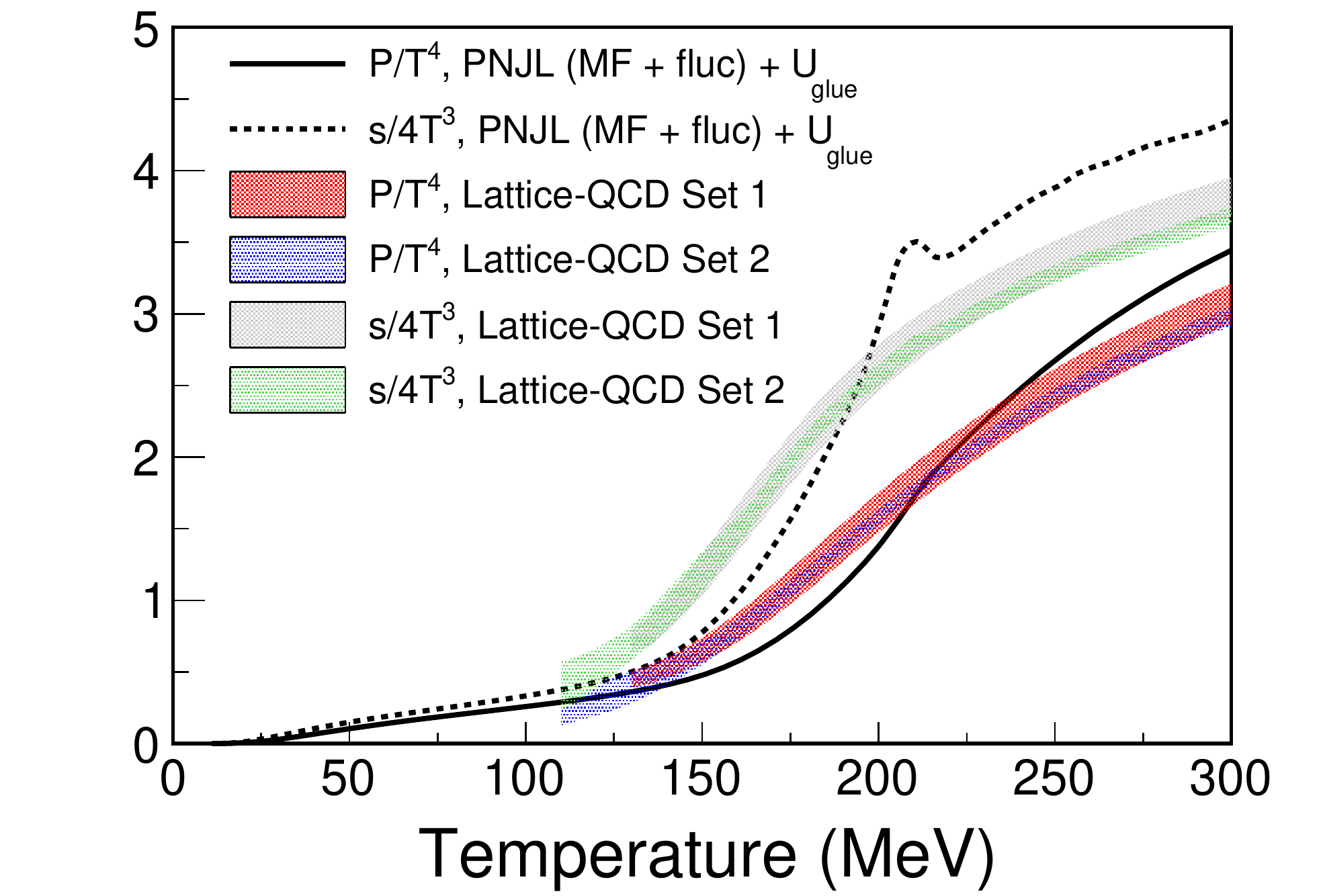}
\caption{\label{fig:Pcomparison} Pressure and entropy density at $\mu_q=0$ of the PNJL model with effect of $0^+,0^-$ 
meson states, compared to lattice-QCD calculations. Left panel: states whose mass in vacuum is below
1 GeV: $\pi,K,{\bar K},\eta,\eta',f_0$ Right panel: all scalar and pseudoscalar states ($\pi,K,\bar{K},\eta,\eta',a_0,K_0^*, \bar{K}_0^*,f_0,f_0'$).}
\end{center} 
\end{figure*}

In Fig.~\ref{fig:Pcomparison} we compare the total pressure and entropy density of the system with the most recent lattice-QCD calculations.
Set 1 corresponds to Ref.~\cite{Borsanyi:2013bia} (Wuppertal-Budapest group) and Set 2 to Ref.~\cite{Bazavov:2014pvz} (HotQCD collaboration). In the left panel we include 
the same mesonic states as accounted in Fig.~\ref{fig:Pprogressive}, i.e., those with a vacuum mass below 1 GeV. We already see 
a large improvement with respect to the pressure at mean field (cf. Fig.~\ref{fig:pressureMF}). The mesonic contribution enhances the low-temperature 
pressure towards the lattice-QCD data. At intermediate temperatures there is still a small deviation which is naturally
associated to the lack of more hadrons. This area can be systematically improved adding more and more states.
For example, if we include all scalar and pseudoscalar states ($\pi,K,\bar{K},\eta,\eta',a_0,K_0^*, \bar{K}_0^*,f_0,f_0'$) we observe in the right 
panel of Fig.~\ref{fig:Pcomparison} that the pressure is closer to the lattice-QCD data. Of course, other missing mesons (and baryons) also
contribute in this region. 

 Finally, we observe that the mesonic contribution overestimates the pressure of lattice-QCD,
which already was correctly reproduced at high temperatures (cf. Fig~\ref{fig:pressureMF}).
  
  Mesons contribute as scattering states to the pressure even at high temperatures. Although the contribution of 
each degree of freedom is tiny (of the order of 1 \% of the total pressure), the addition of many states makes a non negligible 
contribution at large temperatures. This mesonic pressure is nonphysical at such temperatures, and one has to go beyond the present 
approach to cure this problem.  One expects that a reduction of the mesonic pressure 
at high temperatures, the total pressure flattens, thus reducing the entropy density as well.

 A way to remove the non-physical meson pressure at high temperatures might be to incorporate the effect of mesonic fluctuations into the quark propagator. This
idea has been suggested for the NJL model in~\cite{Blaschke:2015bxa}, and applied recently to a simplified version of the PNJL model in~\cite{Blaschke:2016hzu,Blaschke:2016fdh}. In
this approach the meson phase-shift in Eq.~(\ref{eq:Omega0next}) is corrected by an extra factor $-\frac{1}{2} \sin^2 \delta (\omega,{\bf p};T,\mu_M)$, which drastically reduces
the mesonic pressure at high temperatures (see Fig. 3 in Ref.~\cite{Blaschke:2016fdh}).

\begin{figure*}[htp]
\begin{center}
\includegraphics[scale=0.6]{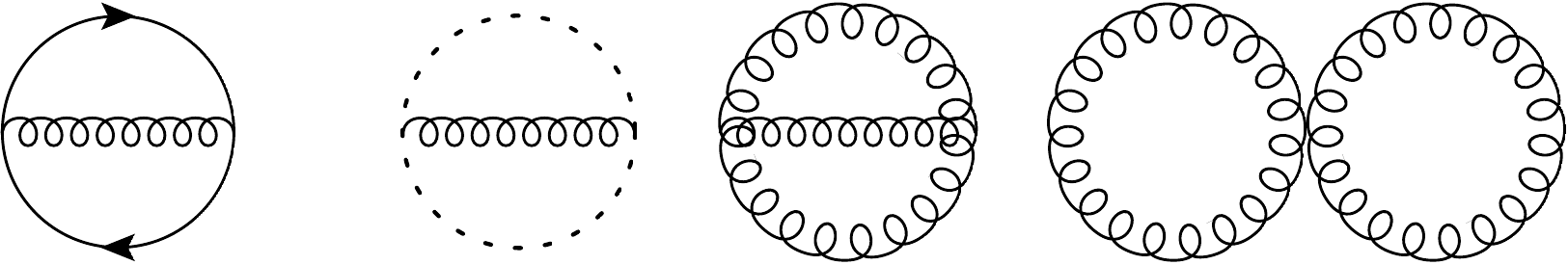}
\caption{\label{fig:pQCDPressure} Diagrams contributing at ${\cal O} (\alpha_s)$ to the QCD pressure~\cite{Kapusta:2006pm} (numerical pre-factors are not displayed).}
\end{center} 
\end{figure*}

In~\cite{Blaschke:2016hzu,Blaschke:2016fdh} a (negative) perturbative correction to the pressure from QCD~\cite{Kapusta:2006pm} is included to lower the pressure at high temperature. We believe that this correction is not appropriate within this scheme. 
It takes into account contributions arising from the diagrams in Fig.~\ref{fig:pQCDPressure} at ${\cal O} (\alpha_s)$ in QCD for the momentum modes between $\Lambda$ and $\infty$ (not included in the original NJL model). Even assuming
that perturbative QCD can be applied to temperatures $T \lesssim 300$ MeV, these diagrams have been already accounted for in the calculation. The diagram involving a quark propagator becomes part
of the Hartree and Fock diagrams. Once removing the UV cutoff in our integrations (to reproduce the Stefan-Boltzmann asymptotic limit) the momentum modes beyond $\Lambda$ have been already integrated. The diagrams
involving gluons and ghosts belong to the Yang-Mills sector, whose pressure is already reproduced by construction thanks to the parametrization of the Yang-Mills effective potential ${\cal U}_{YM}$.
Therefore, these diagrams already participate in the effective description, where gluons with momentum between 0 and $\infty$ are integrated out. As a result, the inclusion of these diagrams would involve a double counting of UV degrees of freedom.

From the results of this manuscript we claim that these corrections are not only not appropriate, but also not needed for the suppression of the mean-field pressure. The back reaction of quarks into the gluon effective 
potential already takes care of this reduction.

\section{\label{sec:mu} Equation of state at finite density}

  We present now some results at finite baryochemical potential by turning on the quark chemical potential. We set the strange chemical potential to zero, $\mu_s=0$, whereas the light quark chemical potential is taken in the
isospin limit, $\mu_q=\mu_u=\mu_d$. In terms of baryon, charge and strangeness chemical potentials we use $\mu_S=\mu_Q=0$, $\mu_B=3\mu_q$.

  The extension to a small chemical potential ---where the transition is still a crossover--- is straightforward. However,
at high densities one faces the problem of multiple solutions in the gap equation (\ref{eq:gap}) for a given temperature. This is the signature of a
coexistence of different phases and the presence of a first-order phase transition. It is important to determine the precise location of the 
phase transition by means of the Maxwell construction. In the next section, we detail how to compute the phase coexistence region and the phase boundary of the model.

\subsection{Phase diagram and boundaries: Maxwell construction}

   We now describe the procedure to determine the phase boundary of the model. To simplify the discussion, we will work with the mean-field thermodynamic potential at ${\cal O} (N_c)$ in this section.
However, we will also incorporate the next-to-leading order terms in our final results.

\begin{figure*}[htp]
\begin{center}
\includegraphics[scale=0.4]{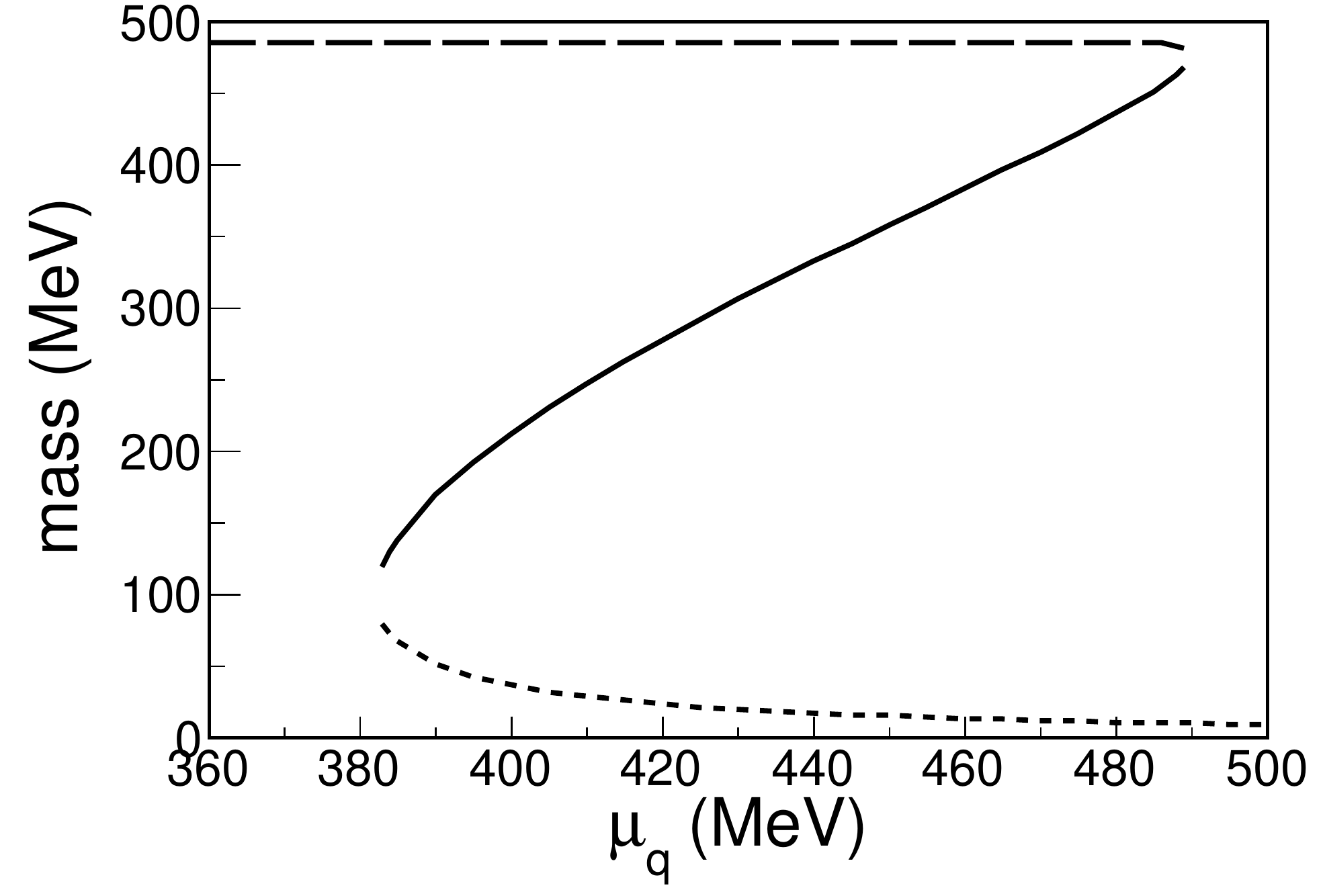}
\includegraphics[scale=0.4]{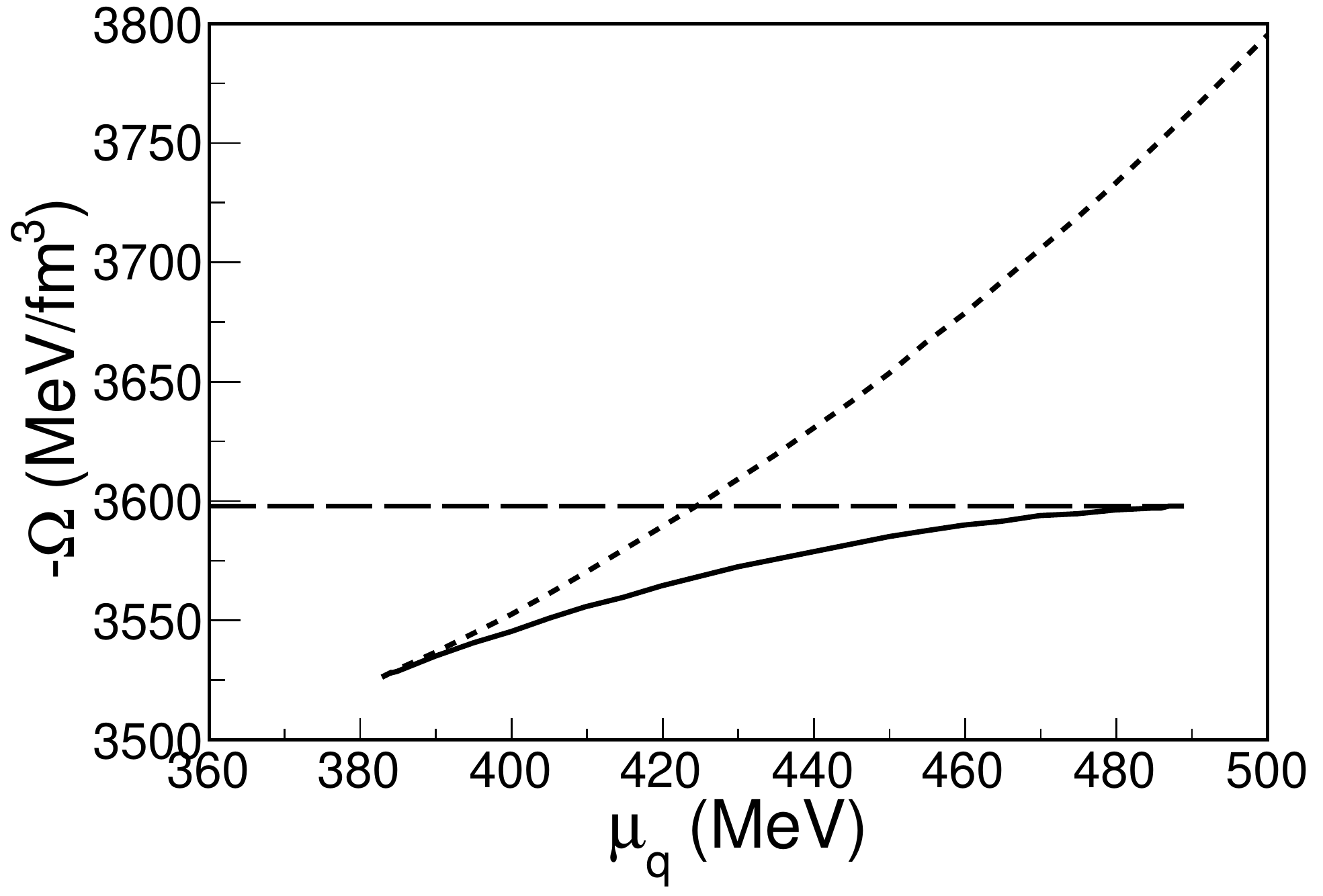}
\caption{\label{fig:maxwell} Left panel: Quark mass as a function of $\mu_q$ at $T=1$ MeV. Right panel: Mean-field thermodynamic potential 
corresponding to the masses of the left panel.}
\end{center} 
\end{figure*}

  For a temperature of $T=1$ MeV, we find a region of the phase transition where several solutions for the quark mass exist for a unique value of the chemical potential. 
In the left panel of Fig.~\ref{fig:maxwell} we show the solutions for the quark mass within a range of $\mu_q$.
  
Between $\mu_q \simeq 380$ MeV and $\mu_q \simeq 490$ MeV, the gap equation has three solutions
for the quark mass. At lower chemical potentials we only find the chirally-broken solution (with
a quark mass of $m_q=480$ MeV), whereas at higher chemical potentials we have the chirally-symmetric phase (with
a quark mass close to the bare one, $m_{q0}$). In between, the three solutions are depicted with different lines styles.

  The phase transition is determined by an analogue of the Maxwell construction. In the presence of different solutions, the energetically favorable
phase is the one with the minimum value of $\Omega (T,\mu_i)$. In the right panel of Fig.~\ref{fig:maxwell} we plot the value of (minus) the thermodynamic potential without vacuum subtraction. At a given
$\mu_q$ the solution with highest $-\Omega(T,\mu_q)$ represents the stable phase. In this example the
chirally-broken phase is the stable one up to $\mu_q=424$ MeV. At this point the quark mass, and hence the order parameter of the chiral transition, present a discontinuity. 
For larger $\mu_q$, the stable phase is the chirally-restored one (with small quark masses).

\begin{figure*}[htp]
\begin{center}
\includegraphics[height=5.4cm]{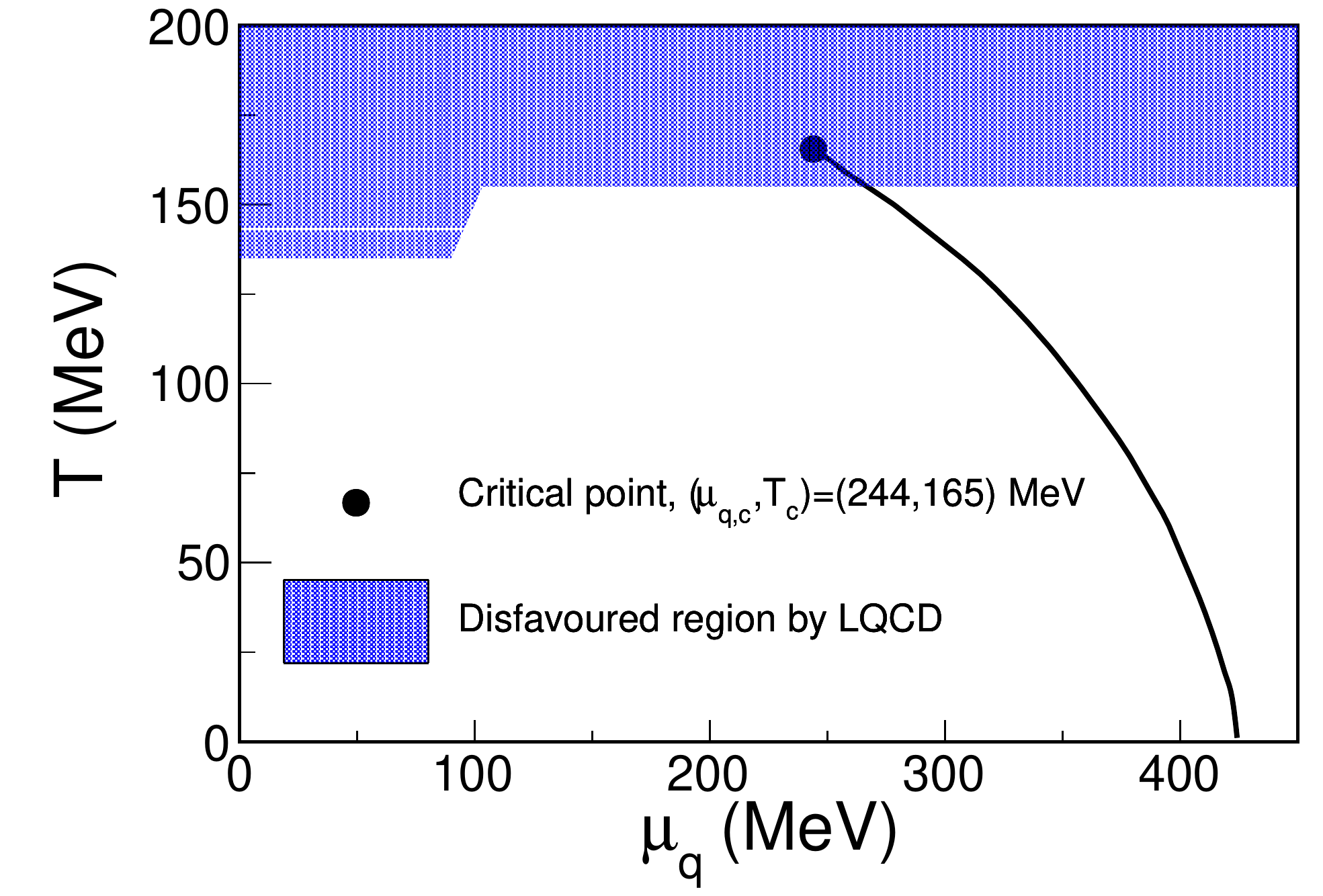}
\includegraphics[height=5.4cm]{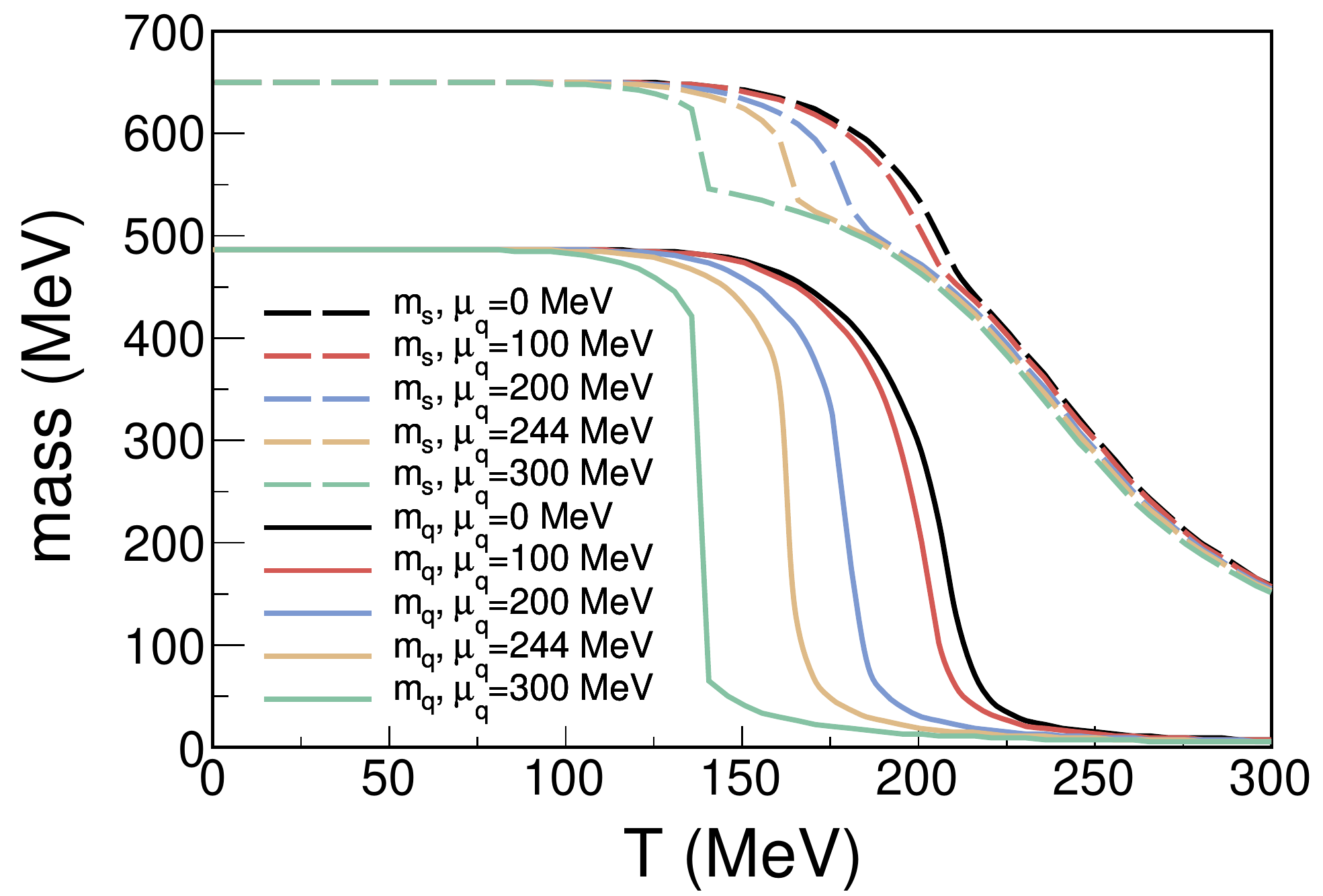}
\caption{\label{fig:boundary} Left panel: Phase boundary of the PNJL model at mean field with the parametrization given in this work. In the shaded region we show the area where the presence of 
a critical point is unlikely according to Ref.~\cite{Bazavov:2017dus}. Right panel: Quark masses as a function of the temperature for several quark chemical potentials.}
\end{center} 
\end{figure*}

  Repeating this procedure for other temperatures we map the first-order critical boundary on the phase diagram, up
to the critical end-point in which the three solutions degenerate into a single one. This signals a second-order
critical point, and the beginning of the crossover regime. The phase boundary is shown in the left panel of Fig.~\ref{fig:boundary}.

  When expressed in terms of the baryochemical potential ($\mu_B=3\mu_q$) the critical point is given by the coordinates $(\mu_{B,c},T_c)=(733,165)$ MeV. For the sake of comparison, the critical point using the ${\cal U}_{YM}$ with $T_0=190$ MeV provides
$(\mu_{B,c},T_c)=(735,169)$ MeV. We also show the area where the presence of a critical point is unlikely according to a recent lattice-QCD calculation~\cite{Bazavov:2017dus}. Notice that vector interactions among quarks---not considered in this work---can modify these values at mean field, and lower the critical temperature~\cite{Fukushima:2008wg}.

  Quark masses are computed as functions of $T$ and $\mu_q$ using Eq.~(\ref{eq:gap}). We present them in the right panel of Fig.~\ref{fig:boundary} for several quark chemical potentials. Notice the discontinuity of
the masses for $\mu_q > 244$ MeV, as a consequence of the first-order phase transition.

\begin{figure*}[htp]
\begin{center}
\includegraphics[scale=0.4]{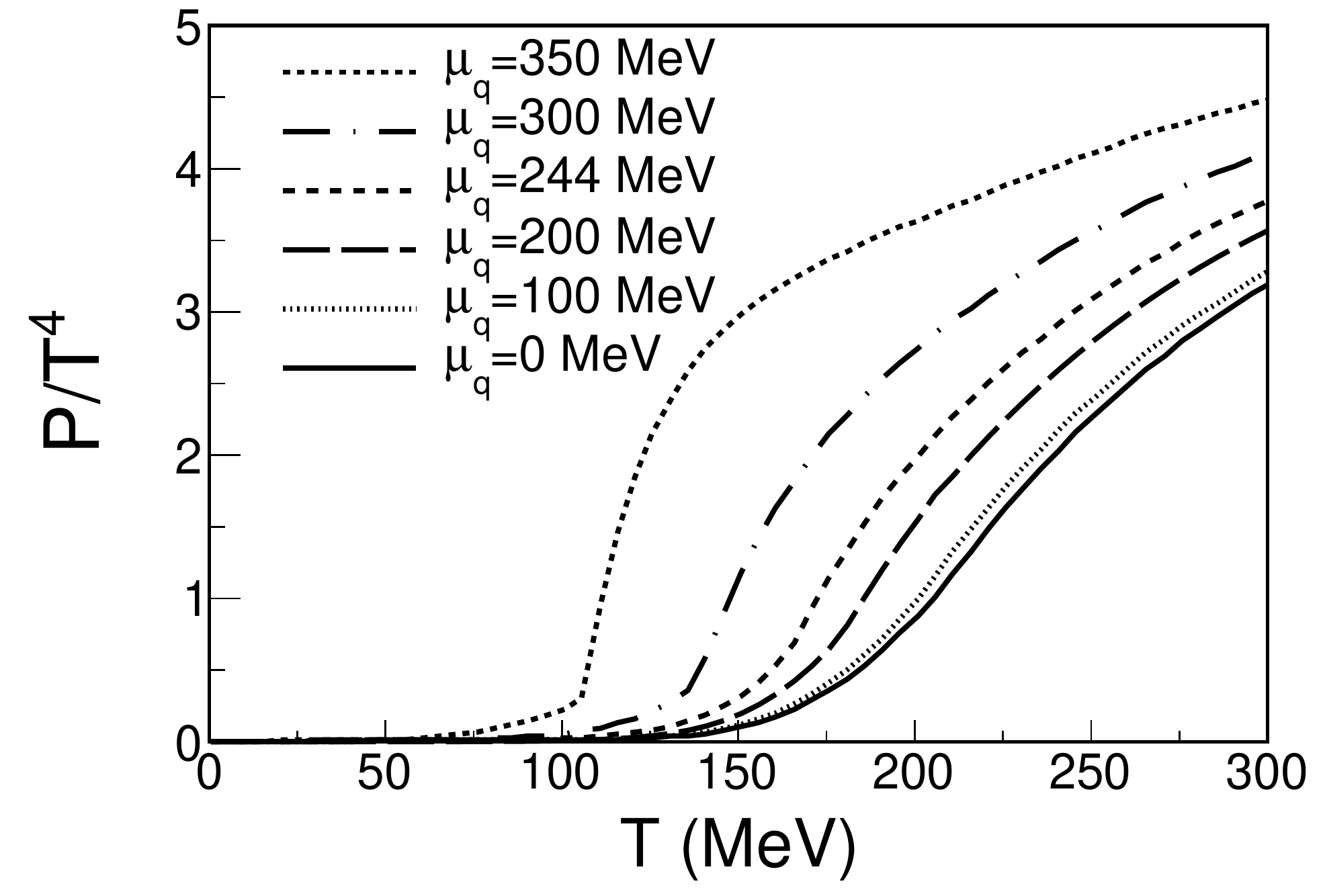}
\caption{\label{fig:pressureMFmu} Pressure in the mean-field approximation of the PNJL model,
for several chemical potentials.}
\end{center} 
\end{figure*}

  Finally in Fig.~\ref{fig:pressureMFmu} we show the mean-field pressure of the system (\ref{eq:pressure}) as a function of $T$ and $\mu_q$. We recall that the glue potential ${\cal U}_{\rm{glue}}$ has no explicit dependence on $\mu_q$ in our scheme.
It only depends on $\mu_q$ through the coupling of the quark field to the EVPL.

  Notice that the pressure is always a monotonically increasing function of the temperature, and continuous at the phase transition by construction,
but with a discontinuous first derivative for the case of a first-order phase transition (see lines with $\mu_q > 244$ MeV).

\subsection{Properties at finite $\mu_q$}

   Following the scheme outlined for $\mu_q=0$ to include mesonic fluctuations in the thermodynamical potential, we will compute the PNJL model pressure
at ${\cal O} ((1/N_c)^0)$ at finite chemical potential. From the results at $\mu_q=0$ we can assume that the main contribution to the 
pressure comes from pions ---lightest mesons, hence easy to create thermally--- and from kaons/antikaons ---more massive, and therefore Boltzmann suppressed, 
but with an isospin degeneracy of 4. For simplicity, we do not consider other mesons due to their expected small contribution.

\subsubsection{Approximate determination of meson masses}

   In order to understand the influence of mesons, we show the dependence of the masses of $\pi$, $K$ and $\bar{K}$ on the temperature and chemical potential. 
The dynamically generated mesons appear as poles of the $q\bar{q}$-scattering amplitude~(\ref{eq:self}),
\be t_M (p_0,{\bf p}) = \frac{2 {\cal K}_M}{1- 2 {\cal K}_M \Pi_M (p_0,{\bf p})} \ee
in the appropriate flavor channel $M$~\cite{Rehberg:1995kh,Torres-Rincon:2015rma}. 
A standard determination of the meson mass (and possible decay width) comes from the pole position $1- 2 {\cal K}_M \Pi (z = M-i\Gamma/2, {\bf p}={\bf 0})=0$ in the complex-energy plane, when the scattering
amplitude is analytically continued to the appropriate Riemann sheet. The polarization function $\Pi (z,{\bf p})$ is shown in Eq.~(\ref{eq:polfunc}).

   If the pole is not very far from the real axis, one can simplify the search by neglecting the imaginary part of the solution in the argument of the $B_0$ function in Eq.~(\ref{eq:polfunc}). In this case, 
one does not need to extend the $B_0$ function to the second Riemann sheet~\cite{Rehberg:1995kh}. To illustrate the meson masses we follow this path, even if at large temperatures/chemical potentials it might not be 
suitable (the decay width increases fast after the Mott temperature, and the pole moves away from the real axis). In our calculation of the pressure we make no such approximation; it is only used in this
section to extract the meson masses.
   
\begin{figure*}[htp]
\begin{center}
\includegraphics[scale=0.4]{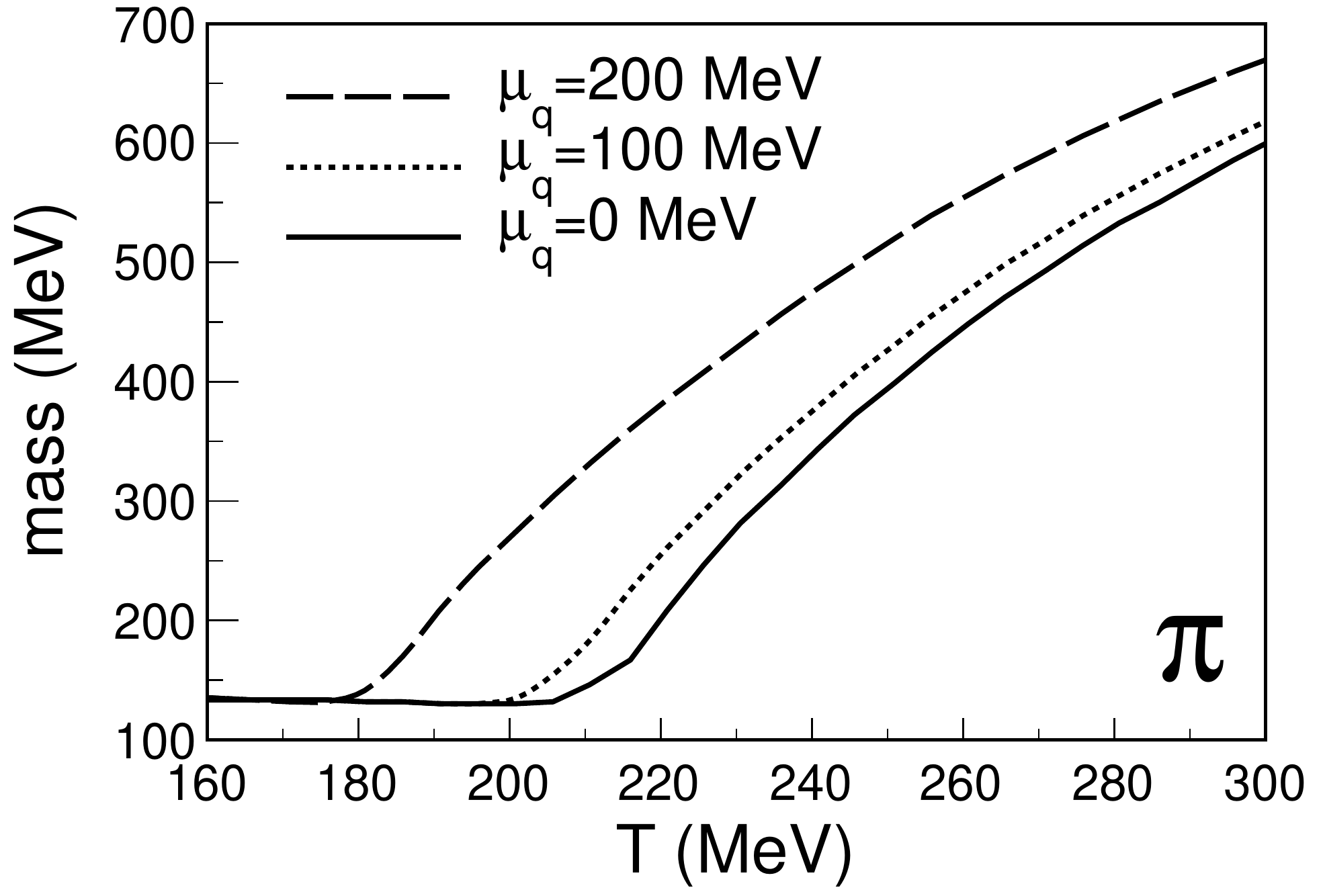}
\includegraphics[scale=0.4]{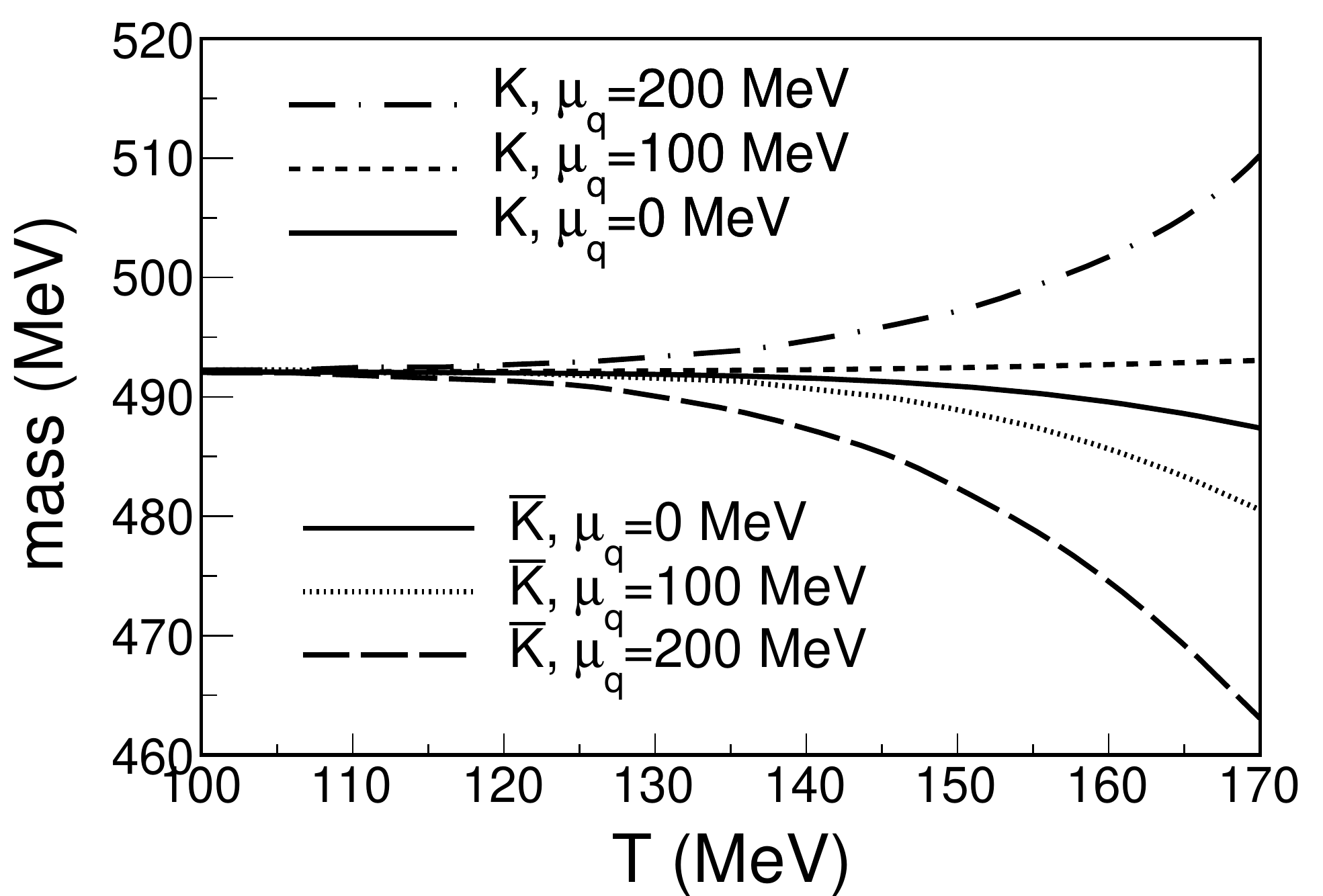}
\caption{\label{fig:pionkaonmu} $\pi$ (left panel) and $K, \bar{K}$ (right panel) masses functions of temperature and chemical potential.}
\end{center} 
\end{figure*}

   In the left panel of Fig.~\ref{fig:pionkaonmu} we show the pion masses, and in the right panel the $K$ and $\bar{K}$ masses, for a given set of chemical potentials.

  The kaon masses show a peculiar behavior due to the fact that isospin degeneracy is broken by the chemical potential. As we take
$\mu_s=0$, the $K$ ($K^+$ and $K^0$ isodoublet) will carry a meson chemical potential $\mu_K=\mu_{q}$, whereas
the ${\bar K}$ ($K^-$, $\bar{K}^0$ doublet) will have $\mu_{{\bar K}}=-\mu_q$. This difference produces non equal masses 
for $K$ and $\bar{K}$ at high temperatures and high baryochemical potentials. In the right panel of Fig.~\ref{fig:pionkaonmu},
we kept a small temperature range for clarity between the curves.

\begin{figure*}[htp]
\begin{center}
\includegraphics[scale=0.5]{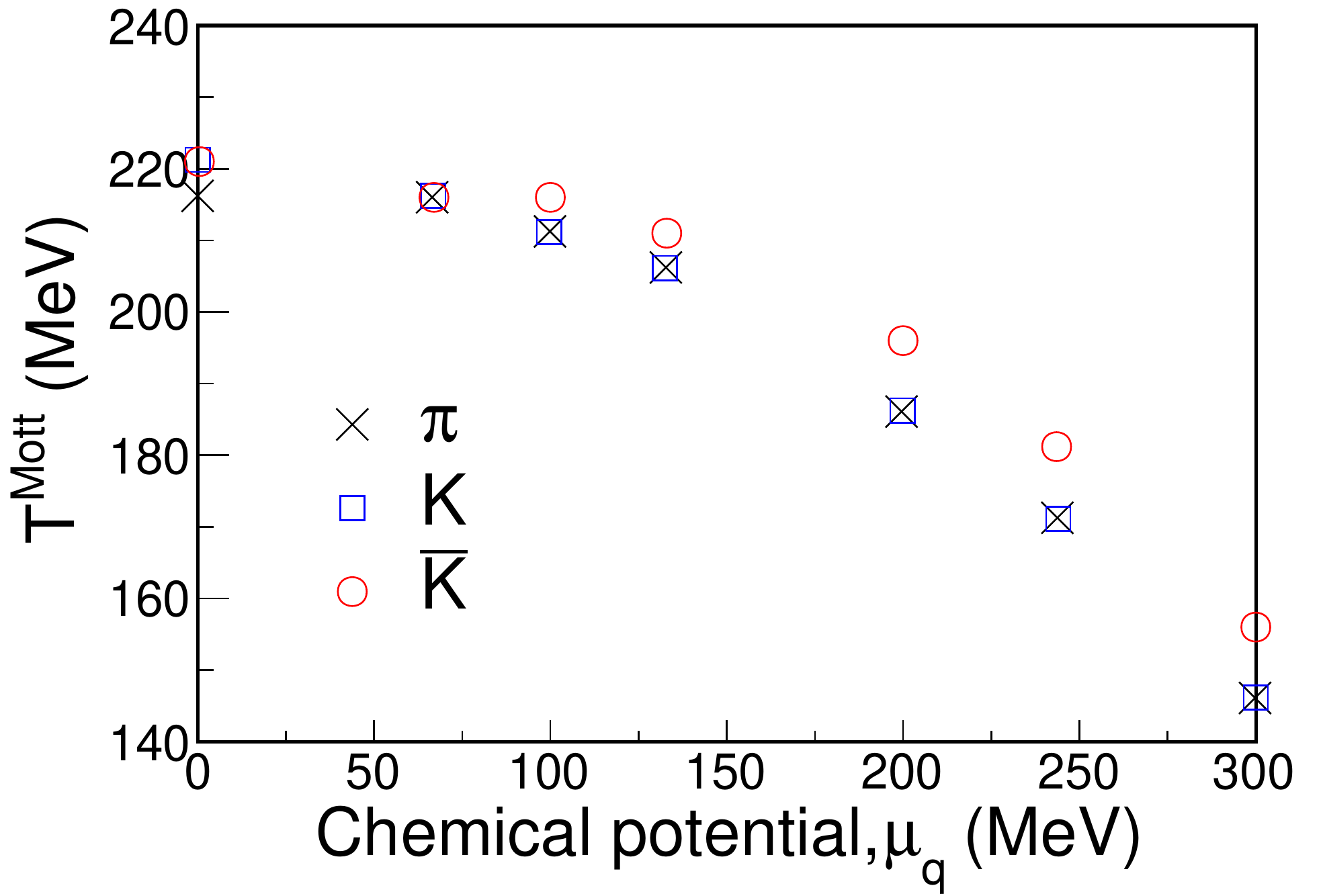}
\caption{\label{fig:Mott} Mott (or melting) temperature of $\pi,K$ and ${\bar K}$ mesons versus chemical potential. The Mott temperature is calculated within an uncertainty of $\pm 3$ MeV.}
\end{center} 
\end{figure*}

  In Fig.~\ref{fig:Mott} we summarize the Mott temperatures $T^{\textrm{Mott}}$ for the lightest mesons versus the quark chemical potential. It might be illustrative to known the temperature at which the different mesons 
acquire a decay width as a function of the chemical potential, and how this temperature decreases as long as $\mu_q$ increases.
  
  It is interesting to see that the ${\bar K}$ states have a slightly larger Mott temperature that the $K$ mesons. This indicates that ${\bar K}$ is more stable in the plasma
than the $K$, especially at high chemical potential.

\subsubsection{Pressure of the PNJL model at ${\cal O} ((1/N_c)^0)$ and $\mu_q \neq 0$}

\begin{figure*}[htp]
\begin{center}
\includegraphics[scale=0.4]{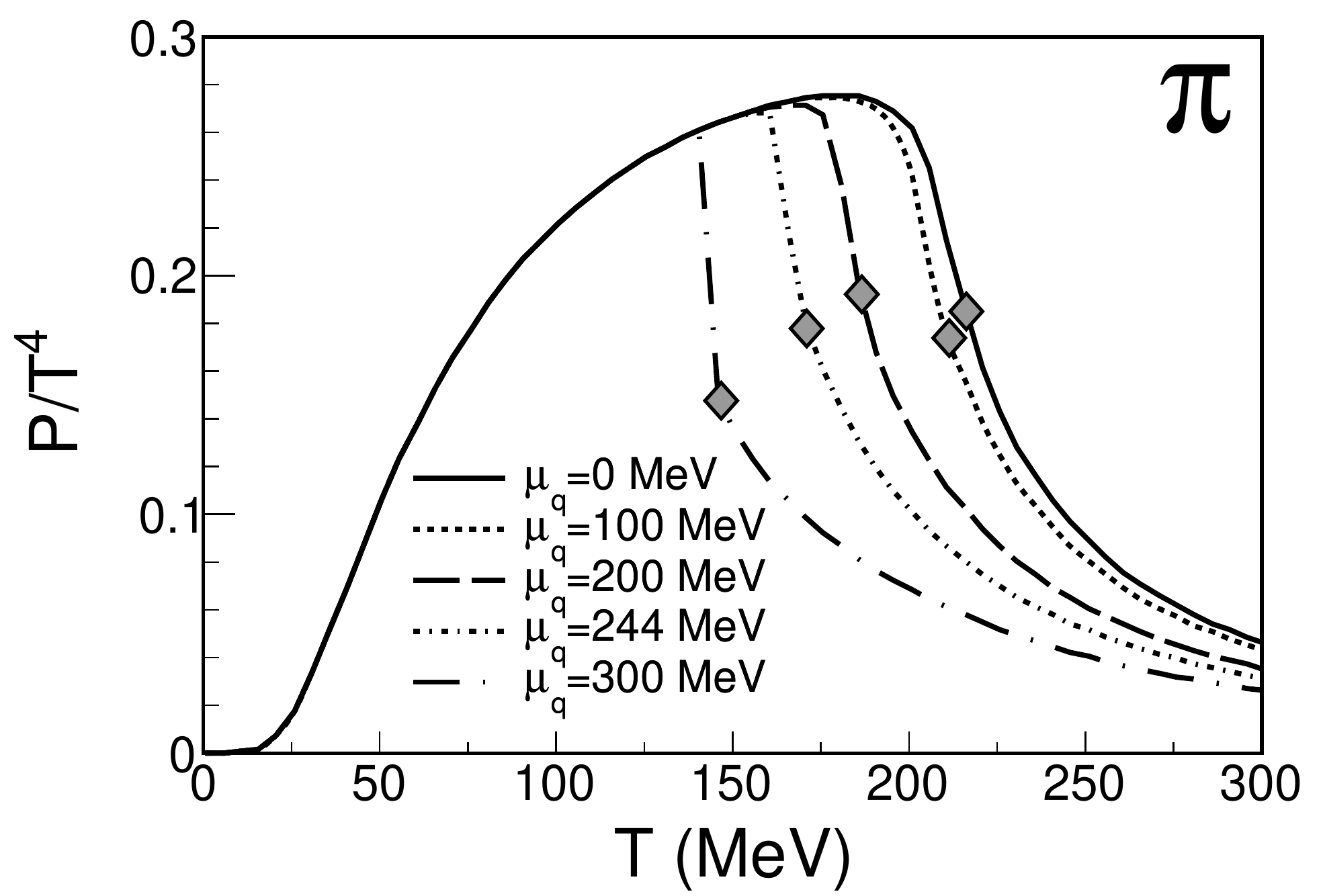}
\includegraphics[scale=0.4]{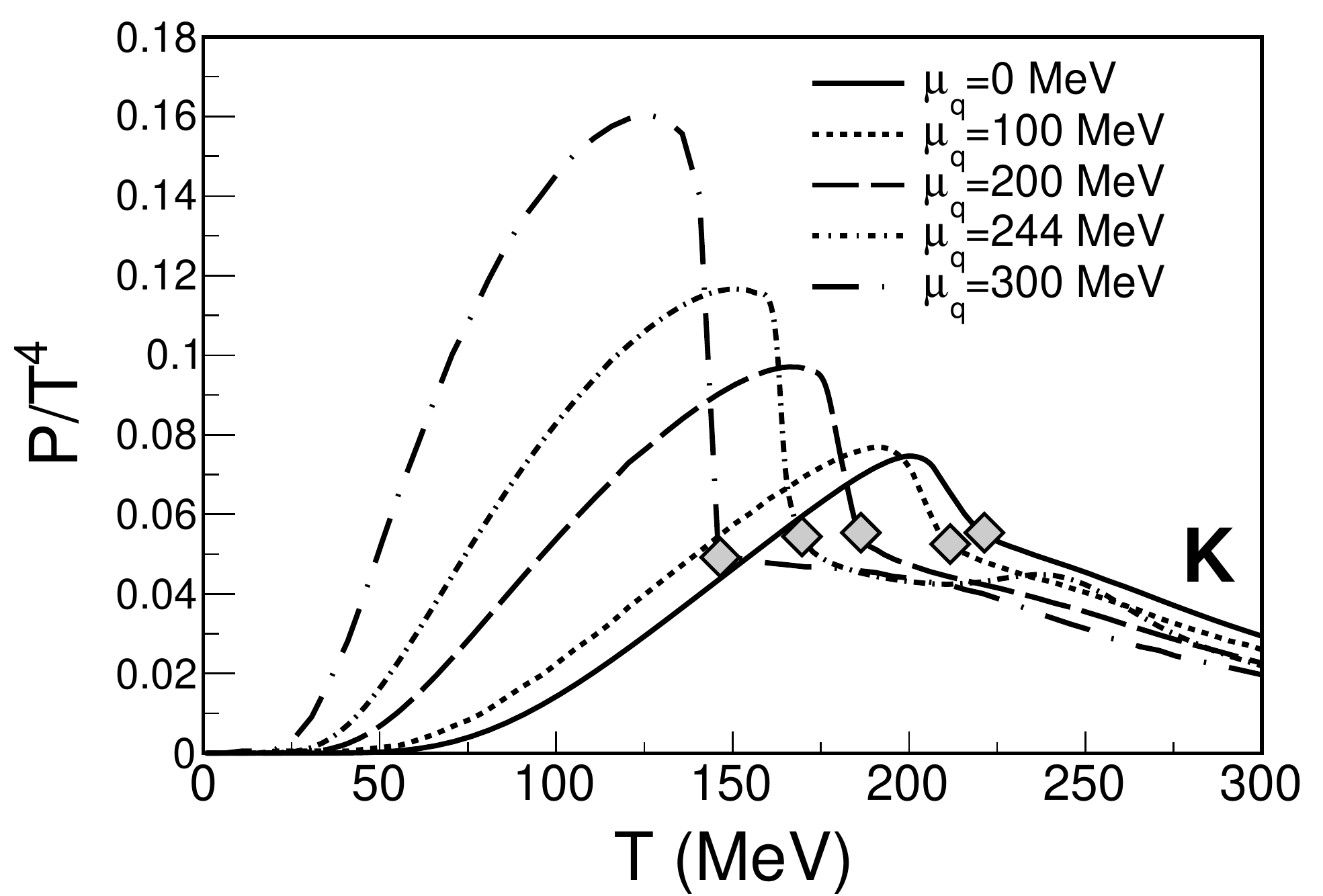}
\caption{\label{fig:press_pion} Partial pressure of $\pi$ and $K = (K^+,K^0)$ as a function of temperature and chemical potential.
The diamonds indicate the Mott temperature for each $\mu_q$.}
\end{center} 
\end{figure*}

  Using Eqs.~(\ref{eq:pressfin},\ref{eq:potfin}) we now present the pressure of the PNJL model for several values of the chemical potential including mesonic fluctuations.
The pressure created by pions and kaons (those states with strangeness = 1) is plotted in Fig.~\ref{fig:press_pion}, as a function of temperature for several chemical potentials.
The Mott temperature is also indicated in the plot for each chemical potential.

  We obtain a jump of the pressure when crossing the first-order phase transition ($\mu_q=300$ MeV). It is more pronounced in the kaon case. 
Although across a first-order transition the total pressure should be continuous, the partial pressure of a given meson is in general discontinuous.

\begin{figure*}[htp]
\begin{center}
\includegraphics[scale=0.4]{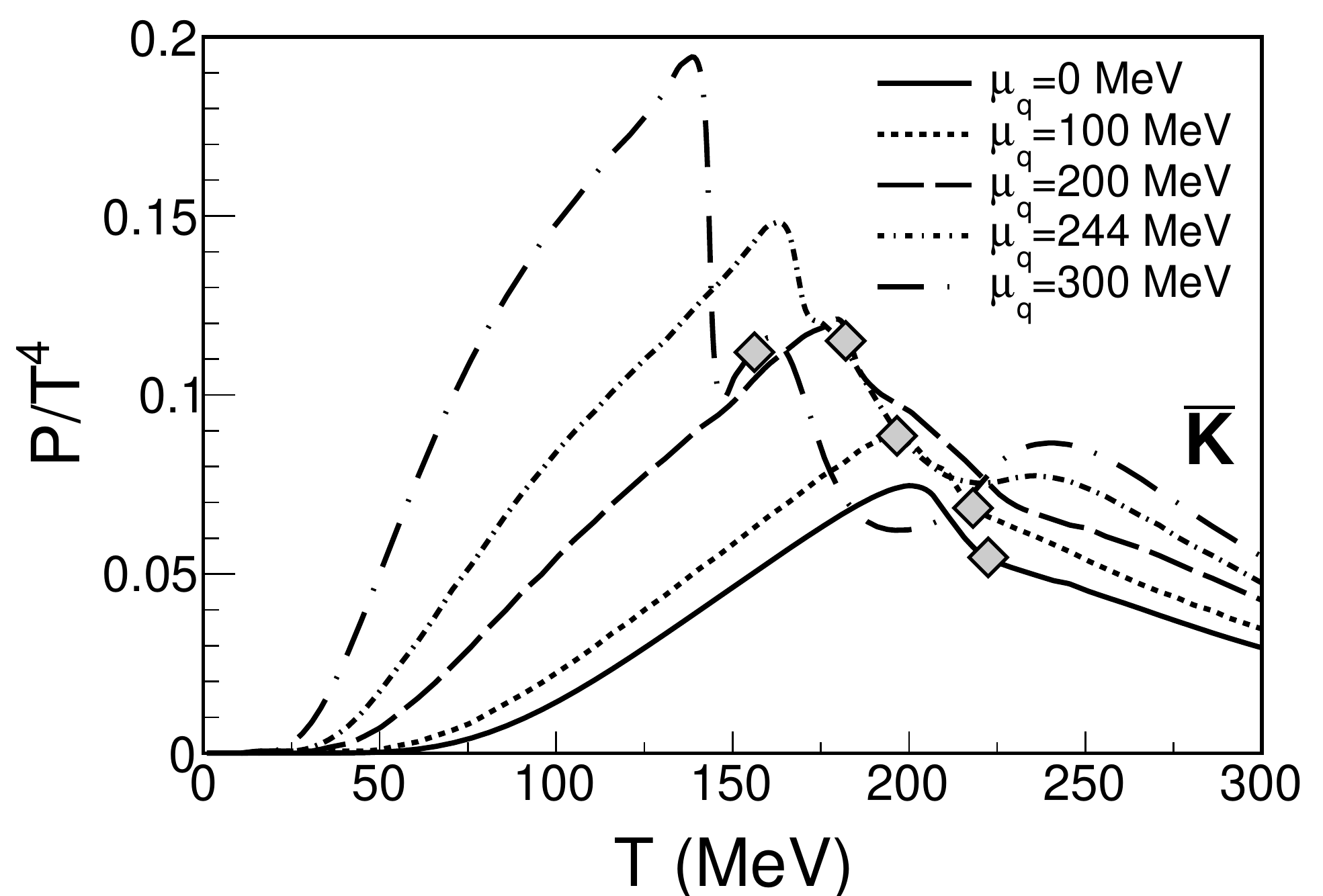}
\includegraphics[scale=0.4]{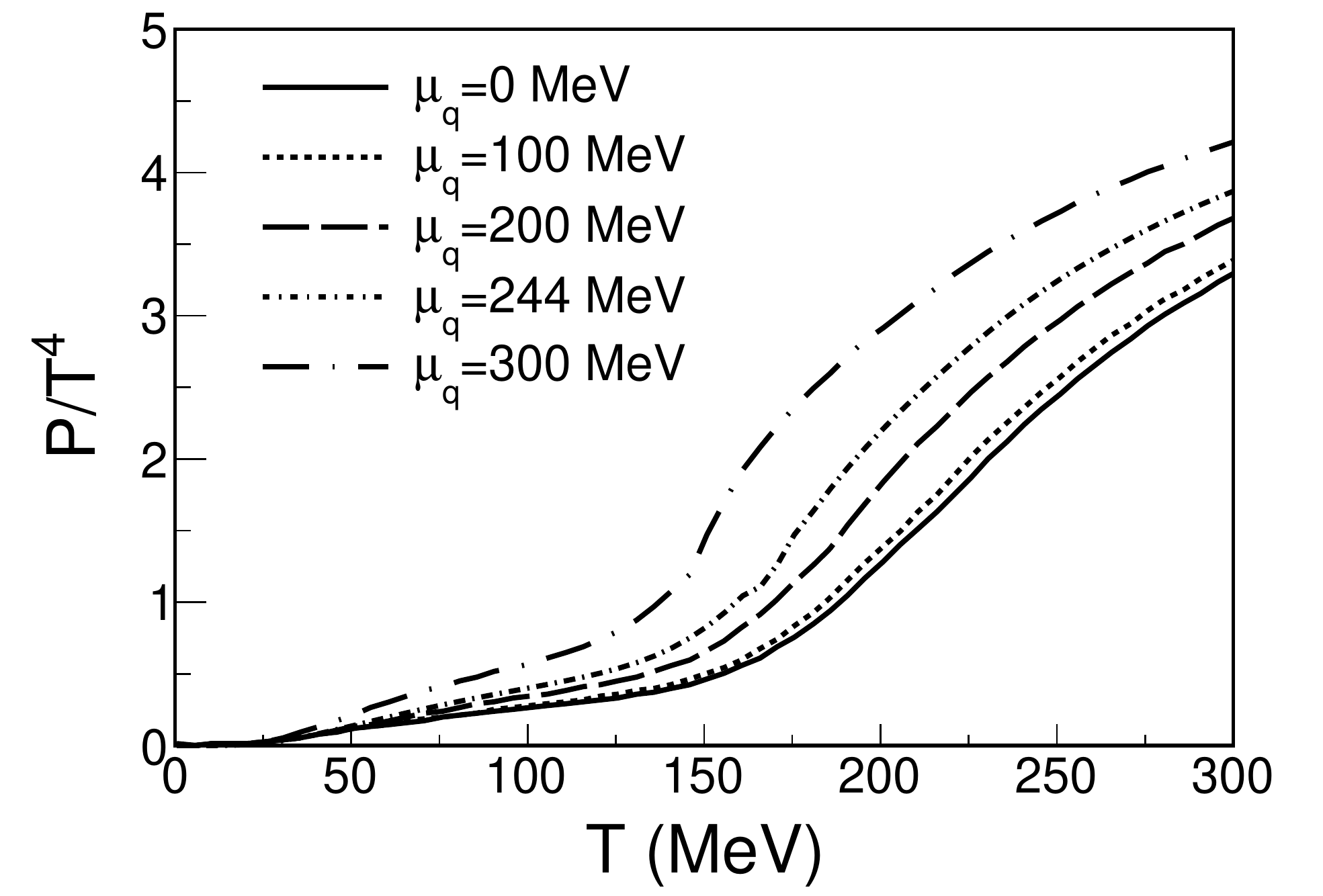}
\caption{\label{fig:press_total} Partial pressure of $\bar{K}=(\bar{K}^0,K^-)$ and total pressure of the system (quarks+Polyakov loop+mesons) as a function of temperature and chemical potential.
The diamonds indicate the Mott temperature for each $\mu_q$.}
\end{center} 
\end{figure*}

  In Fig.~\ref{fig:press_total} we show the pressure of antikaons (those with strangeness = 1) in the left panel, and the total pressure of the system, i.e. a mixture of quarks, the dominant mesons ($\pi,K,\bar{K}$), 
and the effects of the Polyakov loop.

  It is important to remind that in this section we have determined the location of the first-order phase transition by minimizing the grand-canonical potential at ${\cal O} ((1/N_c)^0)$.
Had we used the potential at ${\cal O} (N_c)$, we would have obtained a continuous mean-field pressure (as in Fig.~\ref{fig:pressureMFmu}), but a non-physical discontinuous 
total pressure. In other words, the phase boundary of the model computed at mean field is corrected by the mesonic fluctuations. For example, at $\mu_q=300$ MeV the temperature at which the first-order 
transition takes place is modified by $\sim 5$ MeV.

\begin{figure*}[htp]
\begin{center}
\includegraphics[scale=0.4]{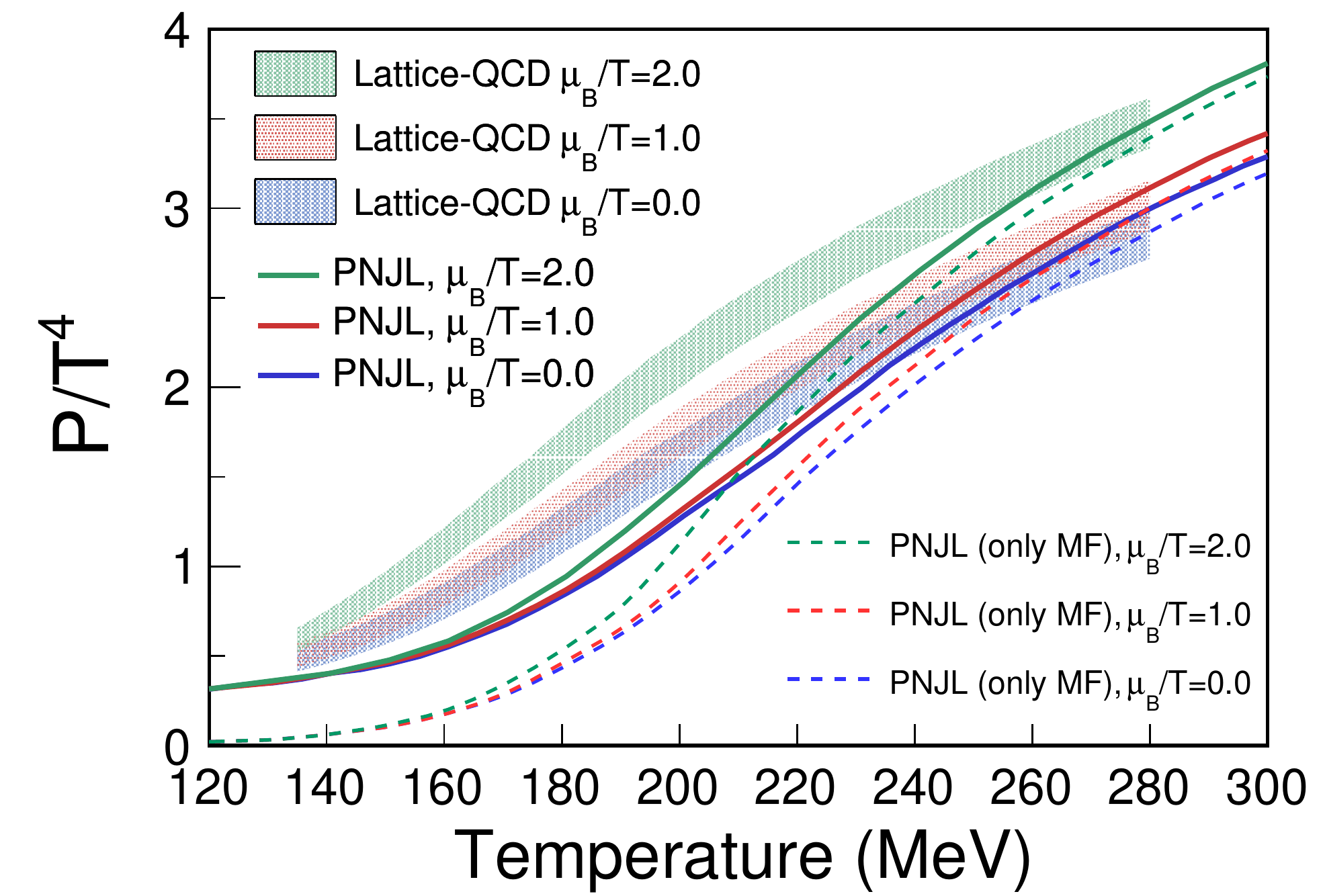}
\caption{\label{fig:press_muB} Pressure of a gas of quarks+Polyakov loop+$\pi+K+\bar{K}$ as a function of temperature and chemical potential. Lattice-QCD data are taken from Ref.~\cite{Bazavov:2017dus}.}
\end{center} 
\end{figure*}

  Finally, we present a calculation of the pressure compared to recent lattice-QCD results at finite chemical potential~\cite{Bazavov:2017dus}. To be consistent with lattice-QCD data we have used $\mu_s=\mu_q=\mu_B/3$, so we have recomputed all our results using a nonzero strange quark chemical potential.
The outcome is shown in Fig.~\ref{fig:press_muB}. 

   The results follow the trend of our Figs.~\ref{fig:pressureMF} and \ref{fig:Pcomparison}. The mean-field results (dashed lines) provide a good description of lattice-QCD data at high temperature thanks to the use of the glue effective potential, but with a poor description of pressure at low temperatures. Addition of meson fluctuations (pions, kaons and antikaons) considerably improves the low-temperature part, still not close to the data due to the absence of additional meson states.

\section{Conclusions and extensions\label{sec:conc}}

    In this work we have revisited the thermodynamic properties of the $N_f=2+1$ PNJL model including ${\cal O} ((1/N_c)^0)$ corrections to the grand-canonical potential to account for physical mesonic excitations in the system. In addition, we have taken into account the modification of the Yang-Mills Polyakov-loop effective potential by the presence of quarks.

    At zero chemical potential we have obtained a pressure which is close to the most recent common lattice-QCD results from several groups in a fairly wide range of temperatures. Then, we have computed the phase diagram of the model, and explored the equation of state at finite chemical potential. The results provide thermodynamical information of regions in the phase diagram away from the $\mu_q=0$ axis. This is the relevant equation of state for the quark-gluon-plasma created in heavy-ion collisions with center-of-mass energies of a couple of GeV, like the low-energy collisions at the Beam Energy Scan at RHIC collider in Brookhaven (U.S.A.), the current NA61/SHINE experiment
from the LHC accelerator at CERN at Geneva, the future CBM experiment at FAIR/GSI facility in Darmstadt (Germany) and NICA experiment at Dubna (Russia). For a final implementation of the PNJL model into numerical simulations~\cite{Marty:2012vs,Marty:2014zka,Marty:2015iwa} at these energies, one needs to extract masses and cross sections of the different species of particles (including quark, mesons, diquarks and baryons) at finite temperatures and densities. This is currently under progress.

    Including the pseudoscalar and scalar mesons, this approach describes reasonably well the low-temperature side of the phase diagram, which is dominated by mesons. We can also describe the phase diagram at moderate temperatures due to the combined contribution of quarks and the Polyakov-loop effective potential. At higher temperatures it is evident that dynamical gluons cannot be neglected and the description if not suitable (although the Stefan-Boltzmann limit is asymptotically recovered). However, the PNJL model is valid for the typical temperatures (and densities) that can be reached presently in heavy-ion reactions (the temperature extracted from direct thermal photons in ALICE Pb+Pb collisions at $\sqrt{s}_{NN}=2.76$ TeV is $T \sim 300$ MeV~\cite{Adam:2015lda}).

    This work can be improved in several ways. A straightforward extension is to consider more mesonic states like vector and axial vector states, by enlarging the effective Lagrangian to include these interaction
channels. Diquarks can be effectively introduced as described in Ref.~\cite{Blaschke:2014zsa}. Their contribution to the pressure is suppressed at low temperatures by the Polyakov loop in a similar way than for quarks.
Baryons~\cite{Blanquier:2011zz, Torres-Rincon:2015rma} should also be introduced, especially at finite chemical potential, where the contribution of nucleons can largely affect the thermodynamics. However, it is not evident
how to perform this extension at the level of the thermodynamic potential. An alternative effective description has been used in Ref.~\cite{Blaschke:2015zwa}.

   As discussed in the text we believe that $\alpha_s$ corrections are not justified to improve the mean-field sector~\cite{Blaschke:2016fdh,Blaschke:2016hzu}. They are not even required for a good matching of lattice-QCD data, because the back reaction 
of quarks into the gluon effective potential~\cite{Haas:2013qwp} already corrects the mean-field pressure at moderate temperatures without the additional hypothesis of the validity of perturbative expansion.
    
    An undesirable effect beyond mean field is the non negligible contribution of mesons to the pressure around $T \sim 300$ MeV. A promising path to suppress this contribution is sketched in
Ref.~\cite{Blaschke:2015bxa}, where the mesonic correction of the quark propagator produces a negative correction to the mesonic pressure. However, this approach cannot be accommodated
into a strict $N_c$ counting, like the one presented here. An alternative solution might be to perform a full self consistent approach, for example, based in a two-particle-irreducible potential like the one in Ref.~\cite{Muller:2010am}. Such a study involves a big
technical effort and it is beyond the scope of this work.

  One could also adjust the parameters of the Polyakov-loop effective potential after having included the mesonic contributions. This would allow us to suppress the total pressure around $T=300$ MeV, and reach lattice-QCD data. However, this would hide the problem of having an unphysical meson pressure at high temperatures. Moreover, if more and more mesons are included in the calculation, then a new parametrization must be fixed each time to reduce the pressure at high temperatures. In our approach the parameter set is fixed at mean field once for all.

\section{Acknowledgements}

We thank D. Blaschke for useful discussions and K. Yamazaki for email exchange. JMTR also acknowledges D. Blaschke 
and the Institute of Theoretical Physics at University of Wroc\l{}aw for hospitality during the realization of this work.
This work has been funded by the programme TOGETHER from R\'egion Pays de la Loire, a Helmholtz Young Investigator Group VH-NG-822 from the Helmholtz
Association and GSI, and EU Integrated Infrastructure Initiative HadronPhysics3 Project under Grant Agreement n. 283286. JMTR also thank support from the Spanish 
Ministerio de Ciencia e Innovaci\'on under contract FPA2013-43425-P.

\appendix

\section{In-medium quark-antiquark propagator~\label{app:polfunc}}

  For a particular meson channel $M$, the quark-antiquark 2-propagator $\Pi^M_{qq'}$ at finite temperature and density is of key importance for the derivation of the grand-canonical potential at ${\cal O}((1/N_c)^0)$
in the large-$N_c$ expansion. In the imaginary-time formalism it is expressed as~\cite{Rehberg:1995nr, Torres-Rincon:2015rma} 
\be \label{eq:twoprop} i\Pi^M_{qq'} (i\omega_m,\pvec)= -i T  \sum_n \int \dtilde{k} \textrm{Tr} \ [ \bar{\Omega}_q S_q(i\nu_n,\kvec)
 \Omega_{q'} S_{q'}(i\nu_n-i\omega_m,\kvec-\pvec)] \ , \ee 
where $S_q(i\nu_n,\kvec)$ is the propagator of quark of flavor $q$ in the Hartree approximation. The trace is to be taken in
color, flavor and spin spaces ($\textrm{Tr} = \textrm{tr}_c \textrm{ tr}_f \textrm{ tr}_\gamma$). The factor $\Omega_q$ is
\be  \Omega_q =  \unit_c \otimes \tau^q \otimes \Gamma_M \ , \ee
where $\unit_c$ is the unit in color space, and $\Gamma_M=\{ \unit, i\gamma_5 \}$ for scalar and pseudoscalar channels, respectively. The flavor trace provides a factor 2 for all mesons 
(already factorized out in the expressions of Ref.~\cite{Rehberg:1995nr}), and the color trace gives a trivial factor $N_c$.
Then, we have
\be \Pi^M_{qq'} (i\omega_m,\pvec)= -2N_c \sum_n \int \dtilde{k} \textrm{tr}_\gamma \ [ \Gamma_M S_q(i\nu_n,\kvec)
 \Gamma_M S_{q'} (i\nu_n-i\omega_m,\kvec-\pvec)] \ . \ee 
After performing the Dirac trace and the Matsubara summation one obtains~\cite{Rehberg:1995kh,Rehberg:1995nr}
\be \label{eq:polfunc} \Pi^M_{qq'} (z,{\bf p})= - \frac{N_c}{4\pi^2} \{ A_q + A_{q'} + \left[ (m_q \mp m_{q'})^2 - (z+\mu_q-\mu_{q'})^2 
+{\bf p}^2 \right] B_0 (z,|{\bf p}|,m_q,\mu_q,m_{q'},\mu_{q'})\} \ , \ee
where we have analytically continued the bosonic Matsubara frequency $i\omega_n$ to an arbitrary complex variable ($i\omega_n \rightarrow z \in \mathbb{C}$). 
The sign $\mp$ corresponds to pseudoscalar and scalar channels, respectively.

The $A_q$ function reads
\be \label{eq:A} A_q = -16 \pi^2 \int \frac{d^3k}{(2\pi)^3} \frac{1}{2E_q} \left[  1 - f_{\Phi}^+ (E_q-\mu_q) - f_{\Phi}^- (E_q+\mu_q) \right] \ , \ee
with the generalized Fermi-Dirac distribution functions of Eqs.~(\ref{eq:fpol1},\ref{eq:fpol2}) accounting for the Polyakov loop effects. 
The $B_0$ function is defined as~\cite{Rehberg:1995kh,Rehberg:1995nr}
\ba \label{eq:B0} B_0 (z,|{\bf p}|,m_q,\mu_q,m_{q'},\mu_{q'})  = 16 \pi^2 \int \dtilde{k}
  & & \left\{ \frac{f_\Phi^+(E_q-\mu_q)}{2E_q} \ \frac{1}{(z+\mu_q-\mu_{q'}-E_q)^2-E_{q'}^2} \ \right. \nn \\
 &-& \left.   \frac{ \left[1 - f_\Phi^-(E_q+\mu_q) \right]}{2E_q} \ \frac{1}{(z+\mu_q-\mu_{q'}+E_q)-E_{q'}^2} \right. \nn \\
 &+& \left. \frac{f_\Phi^+(E_{q'}-\mu_{q'})}{2E_{q'}} \ \frac{1}{(z+\mu_q-\mu_{q'}+E_{q'})^2-E_q^2} \right. \nn \\ 
 &-& \left.  \frac{\left[ 1-f_\Phi^-(E_{q'}+\mu_{q'}) \right]}{2E_{q'}} \ \frac{1}{(z+\mu_q-\mu_{q'}-E_{q'})^2-E_q^2} \right\} \nn \ , \\
 & & \ea
where $E_q=\sqrt{k^2+m_q^2}$ and $E_{q'}=\sqrt{({\bf p}-{\bf k})^2 + m_{q'}^2}$. 
Notice that the $B_0$ function is a complex function, whose first argument, $z$, is complex in general.

To compute meson masses and widths one should look for possible poles of the $q\bar{q}$ scattering amplitude in the complex energy plane, taking ${\bf p}=0$ (meson at rest).
The scattering amplitude in a certain meson channel $M$ reads~\cite{Torres-Rincon:2015rma}
\be t_M (z, \pvec=0) = \frac{2 {\cal K}_M}{1-2 {\cal K}_M \Pi^M (z,\pvec=0)} \ , \ee
where ${\cal K}_M$ is the effective four-fermion coupling introduced in Sec.~\ref{sec:Omega0}.

Following the standard results of scattering theory, any pole of this amplitude along the real axis below any two-body threshold
(physical Riemann sheet) is identified with a bound mesonic state. Alternatively, a pole in the lower half complex plane above 
the two-body threshold (in the second or superior Riemann sheets) is associated to a resonant mesonic state with certain decay width (related to the 
imaginary part of the pole position). The first situation occurs for temperatures below the Mott temperature, and the later above the Mott temperature.
Therefore, the Mott effect represents the effective realization of deconfinement in the context of the NJL/PNJL models.

\section{Scattering phase shift~\label{app:phaseshift}}

The Jost representation of the {\cal S}-matrix element $s_M$ is~\cite{taylor1972scattering}
\be \label{eq:bethu} s_M (\omega, \pvec; T,\mu_M) = \frac{1-2 {\cal K}_M \Pi_M (\omega -\mu_M - i\epsilon, \pvec)}{1-2 {\cal K}_M \Pi_M (\omega -\mu_M  + i\epsilon, \pvec)} \ , \ee
where the denominator plays the role of the Jost function~\cite{taylor1972scattering}. Due to the Schwarz reflection property
\be \label{eq:schwarz} \Pi_M (z^*,{\bf p})=\Pi^*_M(z,{\bf p}) \ , \ee
for arbitrary $z \in {\mathbb C}$, one can deduce that $s_M$ is a pure phase. This phase is what we call the scattering phase shift, defined as in~\cite{taylor1972scattering}
\be \label{eq:smatrix} s_M (\omega, \pvec; T,\mu_M)= \exp \left[ 2i\delta_M(\omega, \pvec; T, \mu_M) \right] \ . \ee

Equating Eq.~(\ref{eq:bethu}) and Eq.~(\ref{eq:smatrix}) one obtains
\be \label{eq:delta} \delta_M(\omega, \pvec; T, \mu_M)  = \frac{1}{2i} \log \frac{1-2 {\cal K}_M \Pi_M (\omega -\mu_M - i\epsilon, \pvec)}{1-2 {\cal K}_M \Pi_M (\omega -\mu_M  + i\epsilon, \pvec)} \ . \ee

The structure in the right-hand side appears in the last line of Eq.~(\ref{eq:big}). The logarithm can be simply expressed as

\be \label{eq:help} \log \frac{1-2 {\cal K}_M \Pi_M (\omega -\mu_M + i\epsilon, \pvec)}{1-2 {\cal K}_M \Pi_M (\omega -\mu_M  - i\epsilon, \pvec)} = -2 i \delta_M(\omega, \pvec; T, \mu_M) \ee
to obtain Eq.~(\ref{eq:Omega0next}).

On the other hand, the scattering phase shift is computed with the help of Eq.~(\ref{eq:delta}) and Eq.~(\ref{eq:schwarz}), 
\ba \label{eq:phaseshift} \delta_M(\omega, \pvec; T, \mu_M) & =& -\frac{1}{2i} \log \frac{ 1-2 {\cal K}_M \Pi_M (\omega -\mu_M + i\epsilon, \pvec) }{[1-2 {\cal K}_M \Pi_M (\omega -\mu_M  + i\epsilon, \pvec)]^*} \nn \\
& = & -\frac{1}{2i} \log |1| - \frac{1}{2}  \textrm{Arg} \frac{1-2 {\cal K}_M \Pi_M (\omega -\mu_M + i\epsilon, \pvec) }{[1-2 {\cal K}_M \Pi_M (\omega -\mu_M  + i\epsilon, \pvec)]^*} \nn \\
&=& - \textrm{Arg} \left[ 1-2 {\cal K}_M \Pi_M (\omega -\mu_M  + i\epsilon, \pvec) \right]    \ . \ea

To exploit Eq.~(\ref{eq:phaseshift}) one needs to evaluate the quark-antiquark propagator (App.~\ref{app:polfunc}) for real energies at $z=\omega-\mu_M+i\epsilon$ with nonzero momentum ${\bf p} \neq 0$. For the particular case ${\bf p}=0$ one can use the formulas presented in
Ref.~\cite{Rehberg:1995nr} to compute meson and diquark masses, as we already did in our previous work~\cite{Torres-Rincon:2015rma}.
However, at ${\bf p} \neq 0$ we have found that the method
in Ref.~\cite{Rehberg:1995nr} does not provide consistent results in the so-called Landau cuts, perhaps due to an inconsistent change of integration variables in that reference. Therefore, to study the case ${\bf p} \neq 0$ we have
evaluated Eq.~(\ref{eq:phaseshift}) by numerical complex integration and taking thereafter the limit $\epsilon \rightarrow 0$.

\begin{figure*}[htp]
\begin{center}
\includegraphics[scale=0.4]{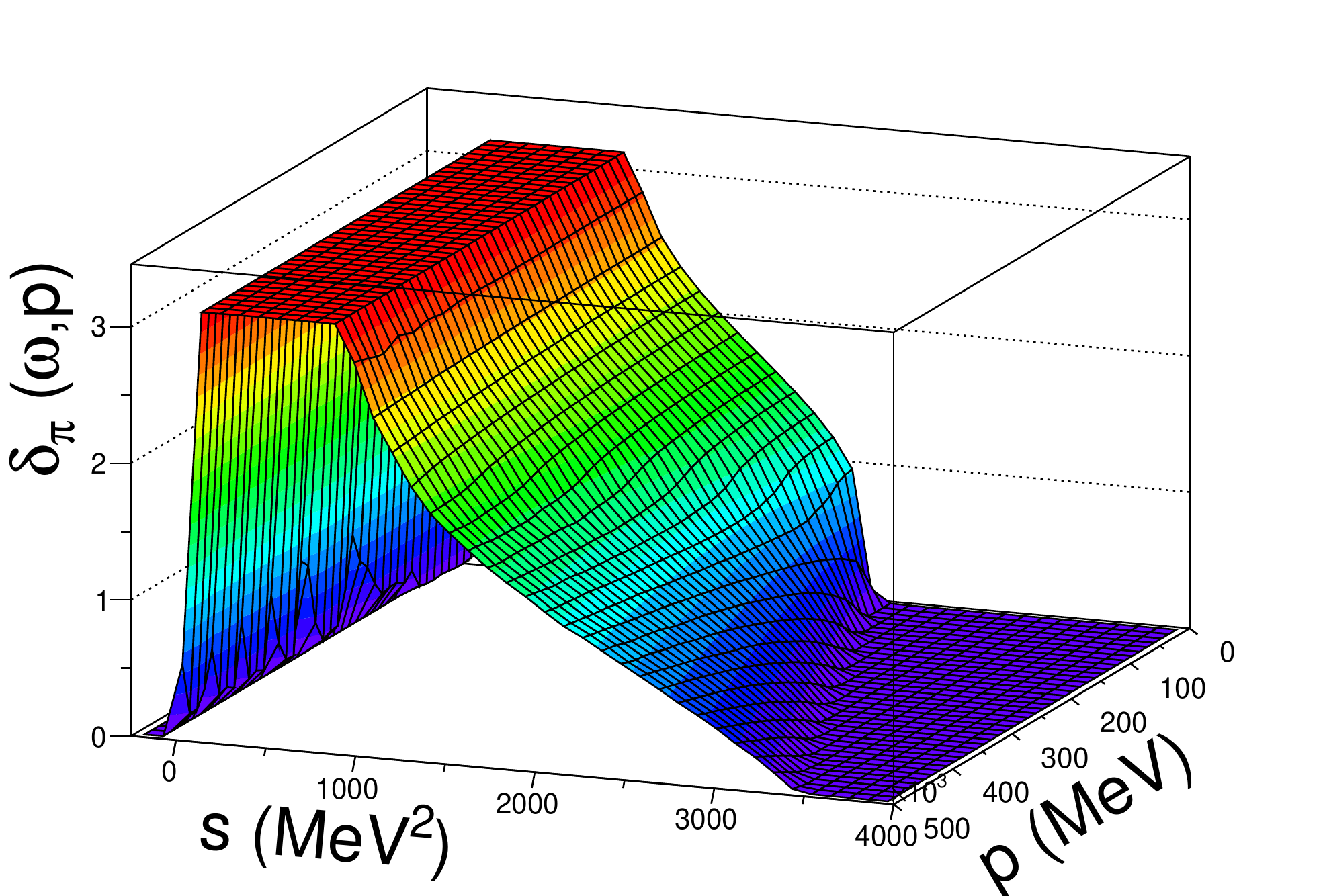}
\includegraphics[scale=0.4]{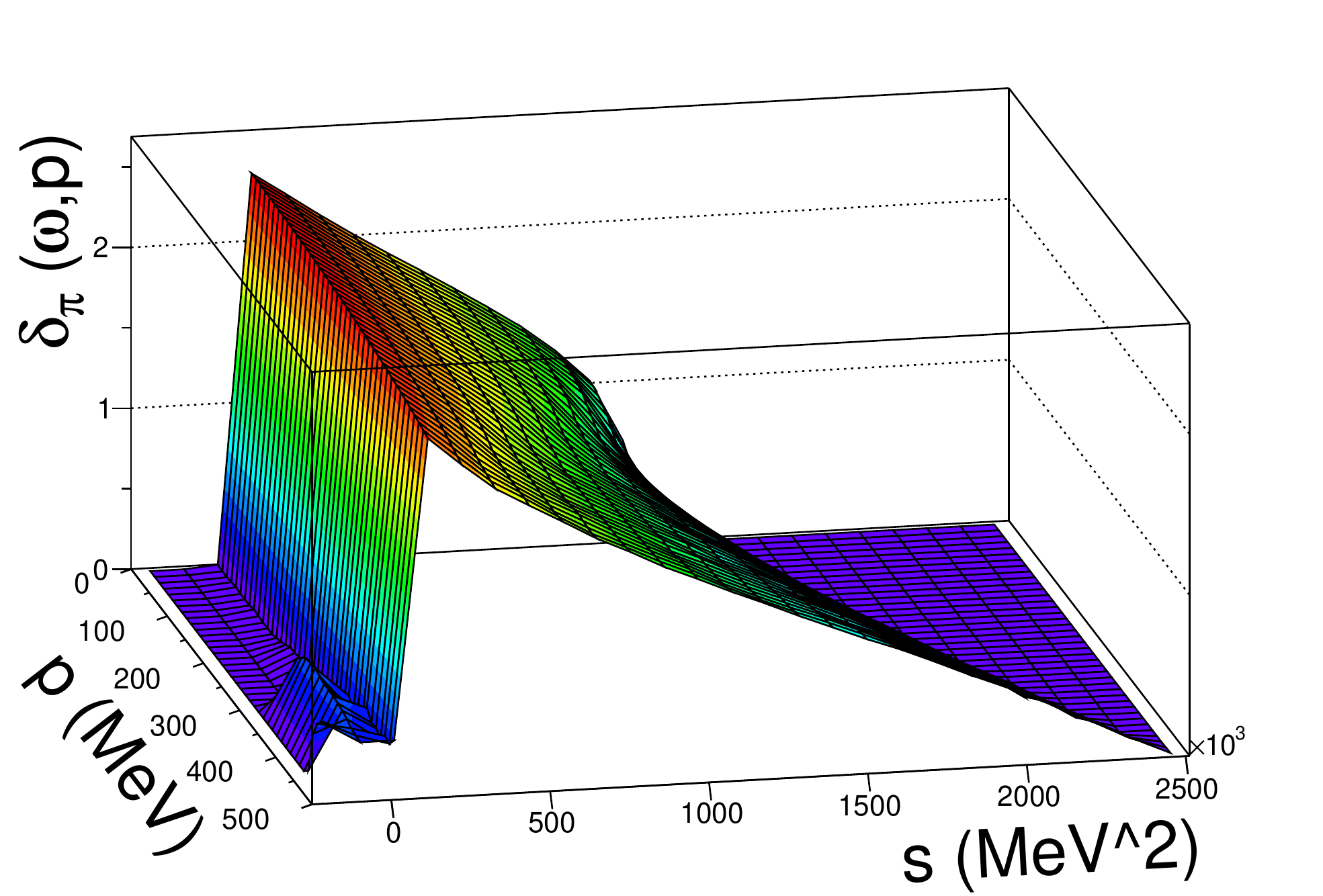}
\caption{\label{fig:delta3D_vsk} Quark-antiquark scattering phase shift corresponding to the pion at $T=1$ MeV (left) and 
$T=300$ MeV (right). The phase shift is plotted as a function of the momentum and $s=\omega^2-p^2$.}
\end{center} 
\end{figure*}

We have numerically checked whether the scattering phase shift at finite momentum behaves approximately in a Lorentz invariant way. This assumption is used in some of the 
previous works~\cite{Zhuang:1994dw, Hufner:1994ma, Blaschke:2013zaa}, arguing that the phase shift can be taken as a function of the Mandelstam variable $s=\omega^2-p^2$ only.
We have computed the scattering phase shifts at finite momentum ${\bf p} \neq 0$ for the pion case. In Fig.~\ref{fig:delta3D_vsk} we present the phase shift as a function of $p$ and $s=\omega^2-p^2$
for two different temperatures: $T=1$ MeV (left panel) and $T=300$ MeV (right panel), and $\mu_q=0$. If the phase shift is Lorentz invariant, then it should be independent on the momentum $p$ when it is written as a function of $s$.

In the left panel we observe that this invariance is slightly broken at high $s$. However, this region is expected 
to be thermally suppressed by the distribution function when the phase shift is integrated in Eq.~(\ref{eq:omegafinal}). In the right panel, we observe an additional violation due to the presence of the Landau cut ($s<0$)
at finite momentum, not present at $T=0$. However, the numerical value of the phase shift at this cut is small, so one can expect that the contribution of that region to the thermodynamical 
potential is negligible. In conclusion, we can safely assume that the assumption of Lorentz-invariance is approximately valid.

\begin{figure*}[htp]
\begin{center}
\includegraphics[scale=0.4]{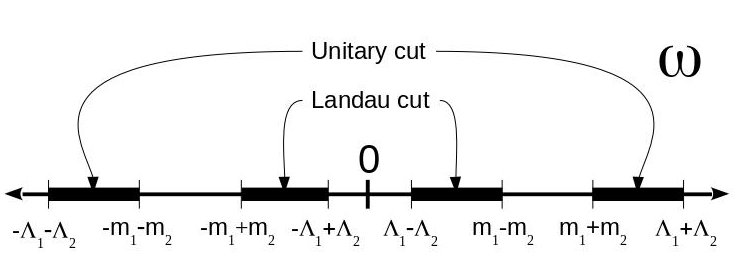}
\caption{\label{fig:cuts1} Landau and unitary cuts along the real axis of the complex-energy plane for the quark-antiquark propagator function $\Pi_{q_1,q_2} (\omega,{\bf p}=0)$.}
\end{center} 
\end{figure*}

 When the quark masses are different, even at ${\bf p}=0$, a Landau cut at finite temperature appears due to an additional cut along the real axis of the $B_0$ function.
This is important, for instance, for the kaon/antikaon case, where one propagates a light quark and a strange antiquark (or vice versa). Therefore, it is convenient to know the general analytical structure 
of the quark-antiquark propagator (or the $B_0$ function). For the case ${\bf p}=0$, with a finite UV cutoff $\Lambda$, and two different quark masses $m_1 \neq m_2$, 
the branch points for the different cuts are shown in Fig.~\ref{fig:cuts1}. Without loss of generality we take $m_1 > m_2$ to avoid a crossing of the Landau cuts, which would make the pictorial representation difficult.
   
   In Fig.~\ref{fig:cuts1} we have denoted $\Lambda_{1,2} = \sqrt{m_{1,2}^2+\Lambda^2}$. We keep the UV cutoff $\Lambda$ in all the momentum integrals in this Appendix. Hence, we have been
able to numerically check the limits described in Fig.~\ref{fig:cuts1}. However, as argued in the text, in our actual calculations we remove the cutoff in all convergent integrals at finite temperature.

   To grasp the meaning of this branch points let us consider the case of a pion at rest. At any temperature below the Mott temperature, the pion appears as a bound state in the scattering amplitude of a light quark and a light
antiquark. It is unable to decay into a quark-antiquark pair because its mass is smaller than the mass of the quark-antiquark pair~\cite{Torres-Rincon:2015rma}, unless it has enough kinetic energy to do so. In this case,
the pion pole lies on the real energy axis at $\omega_{pole}=m_\pi$, below the two quark threshold $\omega_{th}=2m_q$. As a consequence, the decay width (imaginary part of the pole position) is zero.
The polarization function $\Pi_\pi (\omega + i\epsilon, 0)$ is real for $\omega < \omega_{th}$.

  We find that the function $1 - 2 {\cal K}_\pi \Pi_\pi (\omega +i \epsilon, 0)$ is positive for $\omega<\omega_{pole}=m_\pi$. Therefore from Eq.~(\ref{eq:phaseshift}) we expect $\delta_\pi=0$.
At $\omega_{pole}$ this function becomes zero by definition, and changes sign for $\omega>\omega_{pole}$.
The phase shift presents a sudden jump from $\delta_\pi=0$ to $\delta_\pi=\pi$. This jump signals the existence of the bound state at $m_\pi$.

  The phase shift keeps its constant value $\delta_\pi=\pi$ until $\omega_{th}=2 m_q>m_\pi$, where the propagating quark-antiquark pair has enough energy to dissolve. This is the beginning of the
so-called unitary cut, producing an imaginary part of the polarization function, which becomes discontinuous along the real axis. The nonzero imaginary part makes the phase shift to decrease according to Eq.~(\ref{eq:phaseshift}), evolving 
non-trivially with $\omega$. At the upper limit of the unitary cut $\omega_\Lambda=\Lambda_1+\Lambda_2 = 2 \sqrt{\Lambda^2+m_q^2}$, the polarization function takes again real values, and the phase shift
equals again $\delta_\pi=0$, in accordance with the Levinson theorem~\cite{taylor1972scattering},
\be \delta_\pi (\omega = \infty, {\bf p}=0) - \delta_\pi (\omega=0,{\bf p}=0) = 0 \ . \ee

\begin{figure*}[htp]
\begin{center}
\includegraphics[scale=0.4]{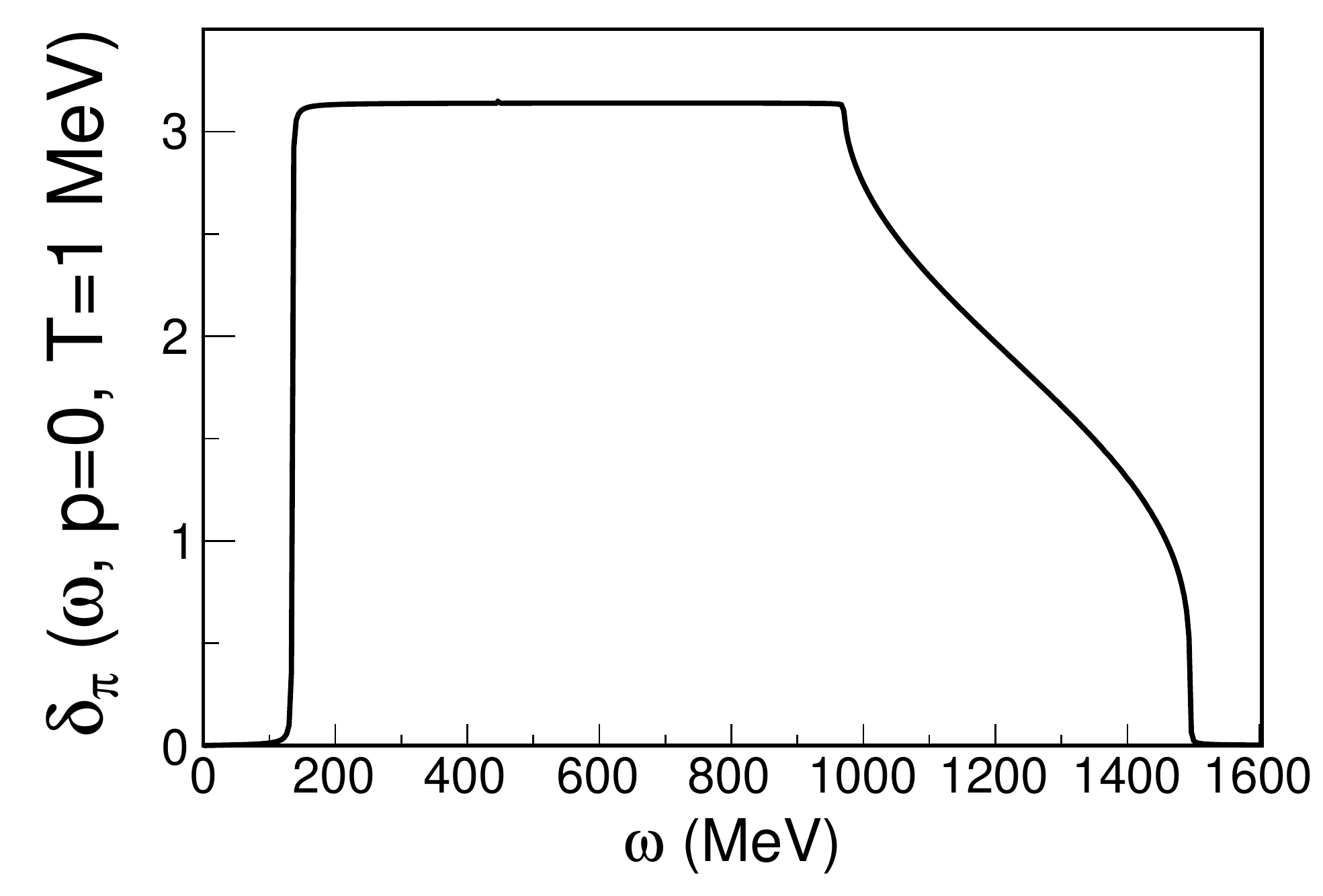}
\caption{\label{fig:deltapi} Pion phase shift as a function of the energy at ${\bf p}=0$ and $T=1$ MeV.}
\end{center} 
\end{figure*}

  We plot the phase shift for $T=1$ MeV in Fig.~\ref{fig:deltapi}, where one can clearly observe the jump from 0 to $\pi$ at $\omega_{pole}=m_\pi$. Then, the phase shift starts to decrease from
$\omega_{th}=2m_q \simeq 970$ MeV until $\omega_\Lambda=\Lambda_1+\Lambda_2=2 \sqrt{\Lambda^2+m_q^2} \simeq 1495$ MeV.

  At temperatures above the Mott transition the pion is a resonant state, whose pole position contains a negative imaginary part. The polarization function is a complex function for all energies 
and the phase shift is nonzero everywhere. If the decay width is small, then the position of the resonance can still be captured by a rapid increase of $\Delta \delta_\pi \sim \pi$ in the scattering phase shift.

  As anticipated, an additional complication occurs when the two quark masses are different, like for kaons and antikaons. In this case the additional Landau cut provides an extra contribution
at $T \neq 0$ to the kaon phase shift~\cite{Yamazaki:2013yua}.

\begin{figure*}[htp]
\begin{center}
\includegraphics[scale=0.4]{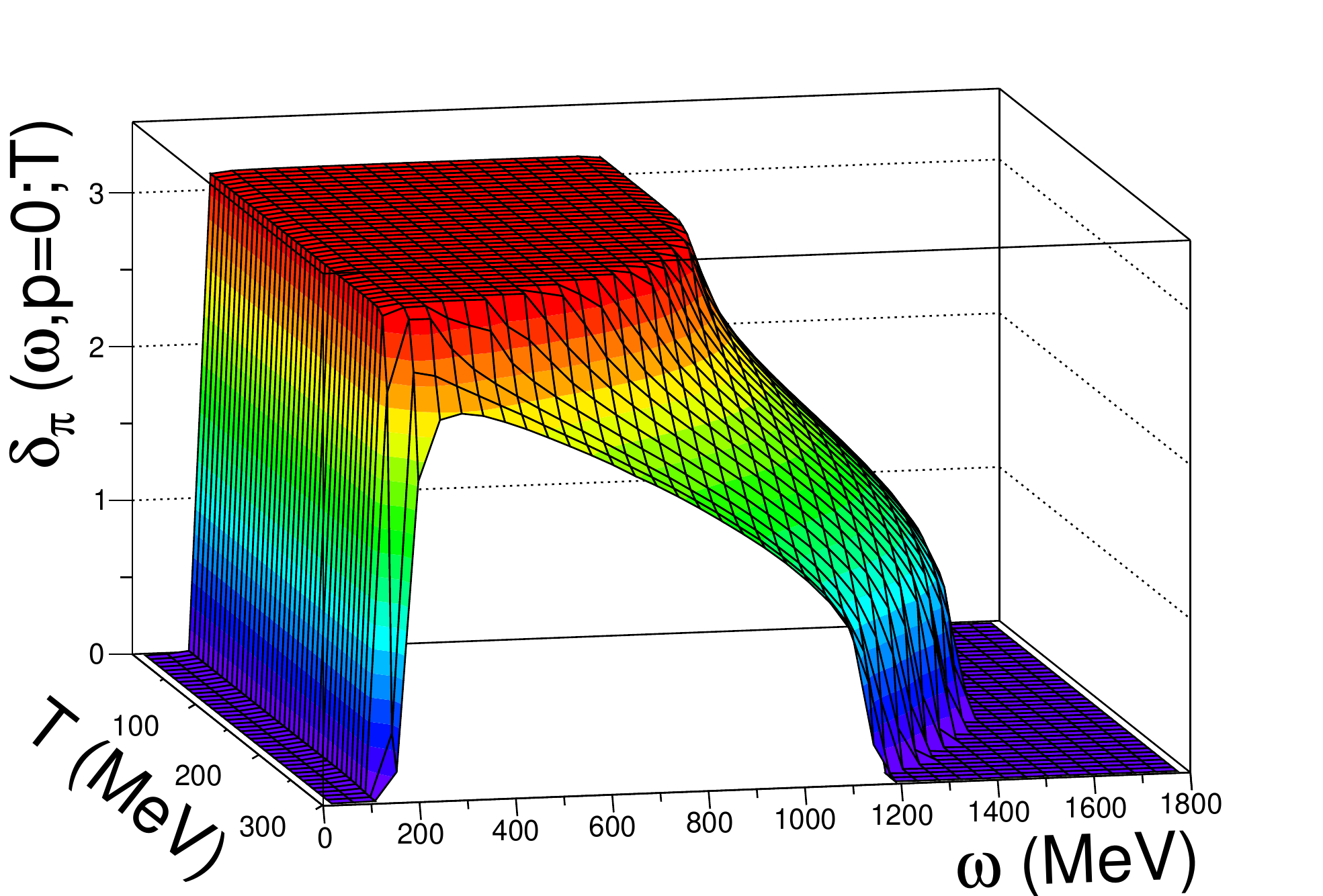}
\includegraphics[scale=0.4]{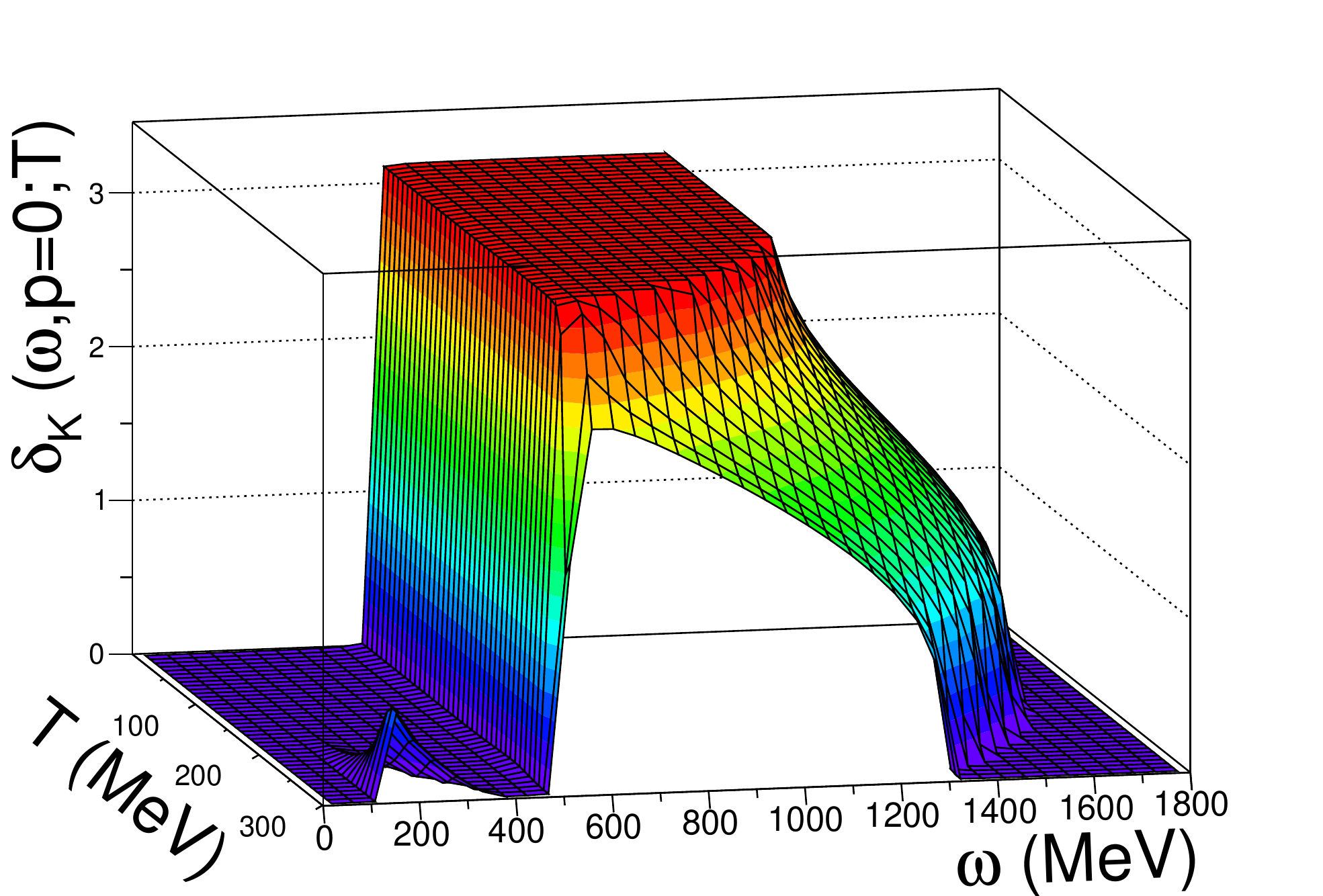}
\caption{\label{fig:delta3D_vsT} Quark-antiquark scattering phase shift corresponding to the pion (left) and 
kaon (right) channels. The phase shift is plotted at $p=0$ as a function of the energy and temperature.}
\end{center} 
\end{figure*}

  In Fig.~\ref{fig:delta3D_vsT} we summarize all these considerations by plotting the scattering phase shifts as functions of the energy and temperature for ${\bf p}=0$. 
We show the pion case in the left panel and the kaon one in the right panel. The Landau cut is visible in the kaonic case at low energies. We have checked that its range coincides 
with the theoretical limits $(\sqrt{\Lambda^2+m_s^2}-\sqrt{\Lambda^2+m_q^2} , m_s-m_u)$. This is the kinematic range where a kaon can decay by absorbing a quark from the thermal bath,
leaving another quark in the final state.

\end{document}